\documentclass[preprint]{elsarticle}
\usepackage[T1]{fontenc}
\usepackage{amsmath}
\usepackage{amssymb}
\usepackage{natbib}
\usepackage{amsthm}
\usepackage{stmaryrd}
\usepackage{tikz}
\usetikzlibrary{trees,arrows,arrows.meta,shapes,decorations.pathmorphing,backgrounds,positioning,fit,petri}
\usepackage{algorithm}
\usepackage[noend]{algpseudocode}
\usepackage{xfrac}
\usepackage{mathtools}
\usepackage{subcaption}
\usepackage{times}
\usepackage{graphicx}
\usepackage{hyperref}
\hypersetup{pdfauthor=author}
\usepackage{listings}
\usepackage{syntax}
\usepackage{geometry}
\usepackage{proof}
\usepackage{imakeidx}
\usepackage[textsize=footnotesize,disable]{todonotes}
\usepackage{bbm}
\usepackage{soul}
\usepackage{labelschanged}

\newcommand{\Lat}{L}                        
\newcommand{\M}{\mbox{\bf M}}               
\newcommand{\Poset}{P}                      

\newcommand{\defsymbol}{\stackrel{{\scriptscriptstyle \textup{\texttt{def}}}}{=}}

\newcommand{\expl}[1]{ \langle {\scriptstyle #1} \rangle}
\newcommand{\func}[3]{#1 : #2 \to #3}

\newcommand{\Kvar}{\mathnormal{K}}
\newcommand{\Dvar}{\mathcal{D}}
\newcommand{\Dform}{\mathnormal{D}}
\newcommand{\Kfun}[1]{\Kvar_{#1}}
\newcommand{\Dkfun}[1]{\Dvar_{#1}}
\newcommand{\Kfunapp}[2]{\Kfun{#1}{\left( #2 \right)}}
\newcommand{\Dkfunapp}[2]{\Dkfun{#1}{\left( #2 \right)}}
\newcommand{\Dkform}[1]{\Dform_{#1}}

\newcommand{\Avar}{\mathsf{K}}
\newcommand{\Gvar}{\mathsf{D}}
\newcommand{\Afun}[1]{\Avar_{#1}}
\newcommand{\Gfun}[1]{\Gvar_{#1}}
\newcommand{\Kop}[2]{\Afun{#1}{\left( #2 \right)}}
\newcommand{\Gop}[2]{\Gfun{#1}{\left( #2 \right)}}

\newcommand{\sfunspace}[1]{\mathcal{S}(#1)}    
\newcommand{\tuplespace}[2]{{#1}^{#2}}         

\newcommand{\sfuntuple}{\mathfrak{s}}  

\newcommand{\pfuntuple}{\pi}
\newcommand{\Pfuntuple}{\Pi}
\newcommand{\Dfuntuple}{\mathbb{D}}
\newcommand{\dfuntuple}{\mathbbm{d}}
\newcommand{\dplusfuntuple}{\overline{\Dfuntuple}}
\newcommand{\dapprox}{\sigma}

\newcommand{\sfun}[1]{\sfuntuple_{#1}}
	
\newcommand{\pfun}[1]{\pfuntuple_{#1}}	
\newcommand{\Pfun}[1]{\Pfuntuple_{#1}}	
\newcommand{\Dfun}[1]{\Dfuntuple_{#1}}	
\newcommand{\dfun}[1]{\dfuntuple_{#1}}
\newcommand{\dplusfun}[1]{\dplusfuntuple_{#1}}
\newcommand{\dapproxfun}[1]{\dapprox_{#1}}

\newcommand{\sfunapp}[2]{\sfun{#1}{\left( #2 \right)}}		

\newcommand{\pfunapp}[2]{\pfun{#1}{\left( #2 \right)}}
\newcommand{\Pfunapp}[2]{\Pfun{#1}{\left( #2 \right)}}
\newcommand{\Dfunapp}[2]{\Dfun{#1}{\left( #2 \right)}}	
\newcommand{\dfunapp}[2]{\dfun{#1}{\left( #2 \right)}}			
\newcommand{\dplusfunapp}[2]{\dplusfun{#1}{\left( #2 \right)}}	
\newcommand{\dapproxfunapp}[2]{\dapproxfun{#1}{\left( #2 \right)}}

\newcommand{\C}{\mathbf{C}}         	 
\newcommand{\Con}{\C}   		         
\newcommand{\true}{\it{true}}			 
\newcommand{\false}{\it{false}}			 
\newcommand{\join}{\sqcup}	 			 
\newcommand{\bigjoin}{\bigsqcup}		 
\newcommand{\meet}{\sqcap}	 			 
\newcommand{\bigmeet}{\bigsqcap}		 
\newcommand{\cleq}{\sqsubseteq} 		 
\newcommand{\cgeq}{\sqsupseteq} 		 
\newcommand{\cl}{\sqsubset} 			 
\newcommand{\cg}{\sqsupset} 			 
\newcommand{\Cs}{{\scal{(\C)}}}  	     
\newcommand{\fleq}{\cleq_{{\texttt s}}}  
\newcommand{\fgeq}{\cgeq_{{\texttt s}}}  
\newcommand{\pleq}{\cleq_{{\scriptscriptstyle \Poset}}}     

\newcommand{\bigjoinp}[1]{\bigjoin\nolimits_{{\scriptscriptstyle #1}}}
\newcommand{\bigmeetp}[1]{\bigmeet\nolimits_{{\scriptscriptstyle #1}}}

\newcommand{\meetp}[1]{\meet_{{\scriptscriptstyle #1}}}

\newcommand{\imp}{\rightarrow}

\newcommand{\conj}{\wedge}
\newcommand{\disj}{\vee}
\newcommand{\ltrue}{\mathtt{T}}
\newcommand{\lfalse}{\mathtt{F}}
\newcommand{\sop}{\ominus}

\newcommand{\tupleset}[2]{T_{#1}^{#2}}
\newcommand{\tuplesetc}[1]{T_{#1}}

\newcommand{\acal}{\mathcal{A}}
\newcommand{\ccal}{\mathcal{C}}

\newcommand{\lcal}{\mathcal{L}}
\newcommand{\pcal}{\mathcal{P}}
\newcommand{\rcal}{\mathcal{R}}
\newcommand{\scal}{\mathcal{S}}
\newcommand{\wcal}{\mathcal{W}}

\newcommand{\rmat}{\mathbb{R}}
\newcommand{\nmat}{\mathbb{N}}
\newcommand{\zmat}{\mathbb{Z}}

\newcommand{\bbold}{\mathbf{B}}
\newcommand{\kbold}{\mathbf{K}}

\newdefinition{example}{Example}
\newdefinition{definition}{Definition}
\newdefinition{remark}{Remark}
\newdefinition{notation}{Notation}

\newtheorem{proposition}{Proposition}
\newtheorem{theorem}{Theorem}
\newtheorem*{theorem*}{Theorem}

\newtheorem{lemma}{Lemma}

\newtheorem*{claim*}{Claim}

\newproof{pot}{Proof of Theorem}

\newcommand{\vect}[1]{\mathbf{#1}}
\newcommand{\diltuple}{\delta}
\newcommand{\erotuple}{\varepsilon}
\newcommand{\dilatn}[1]{\diltuple_{#1}}
\newcommand{\erosn}[1]{\erotuple_{#1}}
\newcommand{\scale}[1]{s_{#1}}
\newcommand{\dilatnapp}[2]{\diltuple_{#1}\left(#2\right)}
\newcommand{\erosnapp}[2]{\erotuple_{#1}\left(#2\right)}
\newcommand{\scaleapp}[2]{s_{#1}\left(#2\right)}

\newcommand{\powerset}[1]{\mathcal{P}(#1)}

\newcommand{\den}[1]{\llbracket #1 \rrbracket} 
\newcommand{\krel}[2]{\overset{#1}{\longrightarrow}_#2} 

\newcommand{\resp}[1]{\todo[bordercolor=blue,color=yellow]{#1}}
\newcommand{\hltext}[1]{{\color{black} #1}}

\makeindex

\makeindex[name=operator,title=Index of Symbols]

\begin{document}


\begin{frontmatter}
\title{Algebraic Structures from Concurrent Constraint Programming Calculi for 
       Distributed Information in Multi-Agent Systems\tnoteref{t1}}

\author[4]{Michell Guzm{\'a}n\fnref{adrs}}
\ead{guzman@fortiss.org}

\author[3]{Sophia Knight}
\ead{sknight@d.umn.edu}

\author[4]{Santiago Quintero}
\ead{squinter@polytechnique.fr}

\author[1]{Sergio Ram{\'i}rez}
\ead{sergio@javerianacali.edu.co}

\author[1]{Camilo Rueda}
\ead{crueda@javerianacali.edu.co}

\author[1,2]{Frank Valencia\corref{cor1}}
\ead{frank.valencia@lix.polytechnique.fr}

\address[1]{Departament of Electronics and Computer Science \\ Pontificia Universidad Javeriana \\ Cali, Colombia}

\address[2]{CNRS, LIX {\'E}cole Polytechnique de Paris, France}

\address[3]{University of Minnesota Duluth, Minnesota, USA}

\address[4]{LIX {\'E}cole Polytechnique de Paris, France}

\fntext[adrs]{Present address: fortiss - Research Institute of the Free State of Bavaria}

\cortext[cor1]{Corresponding author}

\tnotetext[t1]{This work has been partially supported by the ECOS-NORD project FACTS (C19M03)}

\begin{abstract}
Spatial constraint systems (scs) are semantic structures for reasoning about
spatial and epistemic information in concurrent systems. We develop the theory
of scs to reason about the \emph{distributed information} of potentially
\emph{infinite groups}. We  characterize the notion of distributed information
of a group of agents as the infimum of  the set of  join-preserving functions
that represent the spaces of the agents in the group. We provide an
alternative characterization of this notion as the greatest family of
join-preserving functions that satisfy certain basic properties. For
completely distributive lattices, we establish that the distributed
information of $c$ amongst a group is the greatest lower bound of all possible
combinations of information in the spaces of the agents in the group that
derive $c$. We show compositionality results for these characterizations and
conditions under which information that can be obtained by an infinite group
can also be obtained by a finite group. Finally, we provide an application on
mathematical morphology where dilations, one of its fundamental operations,
define an scs on a powerset lattice. We show that distributed information
represents a particular dilation in such scs.
\end{abstract}


\end{frontmatter}


\tableofcontents

\section{Introduction}
\label{sec:intro}

In current distributed systems such as social networks, actors behave more as
members of a certain \emph{group} than as isolated individuals. Information,
opinions, and beliefs of a particular actor are frequently the result of an
evolving process of interchanges with other actors in a group. This suggests a
reified notion of group as a single actor operating within the context of the
collective information of its members. It also conveys two notions of
information, one spatial and the other epistemic. In the former, information
is localized in compartments associated with a user or group. In the latter,
it refers to something known or believed by a single agent or collectively by
a group.

In this paper we pursue the development of a principled account of a reified
notion of group by taking inspiration from the epistemic notion of
\emph{distributed knowledge}~\cite{halpern1990knowledge}. A group has its
information distributed among its member agents. We thus develop a theory
about what exactly is the information available to agents as a group when
considering all that is distributed among its members. 

In our account a group acts itself as an agent carrying the collective
information of its members. We can interrogate, for instance, whether there is
a potential contradiction or unwanted distributed  information that a group
might be involved in among its members or by integrating a certain agent. This
is a fundamental question since it may predict or prevent potentially
dangerous evolutions of the system. 

Furthermore, in many real life multi-agent systems, the agents are unknown in
advance. New agents can subscribe to the system in unpredictable ways. Thus,
there is usually no a-priori bound on the number of agents in the system. It
is then often convenient to model the group of agents as an infinite set. In
fact, in models from economics and epistemic
logic~\cite{Hildenbrand1970,halpern2004reasoning}, groups of agents have been
represented as infinite, even uncountable, sets. In accordance with this fact,
in this paper we consider that groups of agents can also be infinite. This
raises interesting issues about the distributed information of such groups. In
particular, that of group compactness: information that when obtained by an
infinite group can also be obtained by one of its finite subgroups.

\emph{Context.} {Constraint systems} (cs)\footnote{For simplicity we use
\emph{cs} for both \emph{constraint system} and its plural form.} are
algebraic structures for the semantics of process calculi from concurrent
constraint programming (ccp) \cite{saraswat1991semantic}. In this paper we
shall study cs as semantic structures for distributed information of a group
of agents.  

\resp{We added some clarifications about order in the next paragraphs to address {\bf N14}.}
\hltext{A cs is typically formalized as a complete lattice $(\Con,
\cleq)$~\cite{de1995nondeterminism}. The elements of $\Con$ represent partial 
information and we shall think of them as being \emph{assertions}. They are
traditionally referred to as \emph{constraints} since they naturally express
partial information (e.g., $x>42$). The relation  $\cleq$ corresponds to
information order; $c \cleq d$, often written $d \cgeq c$, means that $c$ can
be derived from $d$,  that $d$ represents at least as much information as $c$,
or that if we assume that $d$ holds true then $c$ also holds true (e.g., $x>42
\cgeq x>41$). The join $\join$, the meet $\meet$, the bottom $\true$, and the
top $\false$ of the lattice correspond to conjunction, disjunction, the empty
information and, the join of all (possibly inconsistent) information,
respectively. For example, $(x>42 \join x<42) = \false$ and $(x>42 \meet x<42)
= \true$\footnote{\hltext{The inverse of any boolean algebra is a cs. The
opposite is not true since distributivity of meets over joins is not required
in cs.}}.}

The notion of computational space and the epistemic notion of belief in the
spatial  ccp (sccp) process calculi~\cite{knight:hal-00761116} is represented
as a family of bottom and join-preserving maps $\sfun{i}:\Con \rightarrow \Con$
called \emph{space functions}. A cs equipped with space functions is called a
\emph{spatial constraint system} (scs). From a \emph{computational point of
view}, $\sfunapp{i}{c}$ can be interpreted as an assertion specifying that 
$c$ resides within the space of agent $i$.
\hltext{Thus, given a constraint $s = \sfunapp{i}{c} \join \sfunapp{j}{d} \join e$
we may think of $c$ and $d$ as holding within the spaces of agents $i$ and $j$,
respectively. Similarly, $\sfunapp{i}{\sfunapp{j}{c}}$ can be viewed as a hierarchical
spatial specification stating that $c$ holds within the space the agent $i$ attributes
to agent $j$.}
From an \emph{epistemic point of view}, $\sfunapp{i}{c}$ specifies that  $i$
considers $c$ to be true. An alternative epistemic view is that $i$ interprets
$c$ as $\sfunapp{i}{c}$. All these interpretations convey the idea of $c$
being local or subjective to agent $i$ 
\footnote{\hltext{A polymodal algebra~\cite{chagrov-modalogic-1997} is a boolean
algebra with top and meet preserving functions over it. By duality, its inverse
is a spatial cs.}}.

\emph{This work.} In the spatial ccp process calculus
\emph{sccp}~\cite{knight:hal-00761116},  scs are used to specify the spatial
distribution of information in configurations $\langle P, c \rangle$ where $P$
is a process and $c$ is a constraint, called \emph{the store},  representing
the current partial information. E.g.,  a reduction \( \langle \ P,
{\sfunapp{1}{a}}\join {\sfunapp{2}{b}} \ \rangle \longrightarrow \langle\ Q,
{\sfunapp{1}{a}} \join {\sfunapp{2}{b \join c}} \ \rangle \) means that $P$,
with $a$ in the space of agent $1$ and $b$ in the space of agent $2$, can
evolve to $Q$ while adding $c$ to the space of agent $2$. 

Given the above reduction, assume that $e$ is some piece of information
resulting from the combination (join) of the three constraints above,  i.e.,
$e = a \join b \join c$,  but strictly above the join of any two of them. We
are then in the situation where neither agent has $e$ in their spaces, but as
a group they could potentially have $e$ by combining their information.
Intuitively, $e$ is \emph{distributed} in the spaces of the group $I= \{ 1, 2 \}$.
Being able to predict the information that agents $1$ and $2$ may derive as
group is a relevant issue in multi-agent concurrent systems,  particularly if
$e$ represents unwanted or conflicting information (e.g., $e=\false$). 

In this work we develop the theory of group space functions $\Dfun{I}:\Con
\to \Con$ to reason about information distributed among the members of a
potentially infinite group  $I$. We shall refer to $\Dfun{I}$ as the
\emph{distributed space} of group $I$. In our theory $d \cgeq \Dfunapp{I}{e}$
holds exactly when we can derive from $d$ that $e$ is distributed among the
agents in $I$.  For example, for $e=a\join b \join c$ given above, we will have
$d={\sfunapp{1}{a}} \join {\sfunapp{2}{b \join c}} \cgeq \Dfunapp{\{1,2\}}{e}$ 
meaning that from the information  ${\sfunapp{1}{a}} \join \sfunapp{2}{b \join c}$
we can derive that $e$ is distributed among the group  $I= \{ 1, 2 \}$.
Furthermore, $\Dfunapp{I}{e} \cgeq \Dfunapp{J}{e}$ holds whenever $I \subseteq J$ 
since if $e$ is distributed among a group $I$, it should also be distributed in a
group that includes the agents of $I$.

Distributed information of infinite groups can be used to reason about
multi-agent computations with unboundedly many agents.  For example, a
\emph{computation} in sccp is a possibly infinite reduction sequence $\gamma$
of the form $\langle \ P_0, c_0 \ \rangle \longrightarrow \langle\ P_1, c_1 \
\rangle \longrightarrow  \cdots $ with $c_0 \cleq c_1\cleq \cdots $. The
\emph{result} of $\gamma$ is $\bigjoin_{n\geq 0} c_n$, the join of all the
stores in the computation. In sccp all fair computations from a configuration
have the same result~\cite{knight:hal-00761116}. Thus, the \emph{observable
behaviour} of $P$ with initial store $c$, written $\mathcal{O}(P,c)$, is
defined as the result of any fair computation starting from $\langle P, c
\rangle.$  Now consider a setting where in addition to their sccp capabilities
in~\cite{knight:hal-00761116}, processes can also create new agents. Hence,
unboundedly many agents, say agents $1,2,\ldots$, may be created during an
infinite computation. In this case, $\mathcal{O}(P,c) \cgeq
\Dfunapp{\mathbb{N}}{\false}$, where $\mathbb{N}$ is  the set of natural
numbers, would imply that some (finite or infinite) set of agents in any fair
computation from $\langle P, c \rangle$ may reach contradictory local
information among them. Notice that from the above-mentioned  properties of
distributed spaces, the existence of a finite set of agents $H \subseteq
\mathbb{N}$ such that $\mathcal{O}(P,c) \cgeq \Dfunapp{H}{\false}$ implies
$\mathcal{O}(P,c) \cgeq \Dfunapp{\mathbb{N}}{\false}$. The converse of this
implication will be called \emph{group compactness} and we will provide
meaningful sufficient conditions for it to hold.    

\subsection*{Contributions and Organization.}
The paper starts with some background on lattice theory in
Section~\ref{sec:back} and on spatial constraint systems in
Section~\ref{sec:scs}. The main contributions  are given in
Sections~\ref{sec:dist-info} and~\ref{sec:app} and are listed below:
\begin{enumerate}
\item We characterize the distributed space $\Dfun{I}$ as the \emph{greatest}
space function below the space functions that represent the spaces
(or beliefs) of the agents of a \emph{possibly infinite} group $I$
(Section~\ref{ssec:ds-maxsp}).
\item We provide an alternative characterization of a distributed space as
the greatest function that satisfies certain basic properties
(Section~\ref{ssec:ds-gdc}).
\item We show that distributed spaces have an inherent \emph{compositional}
nature: The information of a group is determined by that of its subgroups
(Section~\ref{DistributedSpaces:section}).
\item We provide a \emph{group compactness} result: Given an infinite
group $I$, we identify a meaningful condition under which  $c \cgeq
\Dfunapp{I}{e}$ implies $c \cgeq \Dfunapp{J}{e} $ for some finite group
$J \subseteq I$ (Section~\ref{ssec:group-compact}).
\item We then show that without this meaningful condition we cannot guarantee
that $c \cgeq \Dfunapp{I}{e}$ implies $c \cgeq \Dfunapp{J}{e}$ for some
finite group $J \subseteq I$ (Section~\ref{ssec:group-noncompact}). 
\item We provide a characterization of distributed spaces for distributive
lattices: Given an infinite group $I$, $\Dfunapp{I}{c}$ can be viewed
as the greatest information below all possible combinations of information in
the spaces of the agents in $I$ that derive $c$
(Section~\ref{DistributedSpaces:section}).
\item Finally, we investigate applications of the theory developed in this
paper to  geometry and mathematical morphology (MM) (Section~\ref{sec:app}).
Below we use $A,B,\ldots$ to denote sets in a vector space. In geometry, the
\emph{Minkowski addition} is given by $A \oplus B = \{ a + b \mid a \in A,
b\in B \}$~\cite{schneider-minkowski-2013}.  It is well-known that the
distribution law $A \oplus (B \cap C)=(A  \cap B) \oplus (A  \cap C)$ holds
for convex sets\footnote{A convex set is a set of points such that, given any
two points in that set, the line segment joining them lies entirely within
that set.} but not in general.  As a simple application of our theory, we
identify a novel and pleasant law for $A \oplus (B \cap C)$: Namely,  $A
\oplus (B \cap C)=\bigcap_{X \subseteq A}(X \oplus B) \cup ((A \setminus X)
\oplus C).$ 
 
In MM, a \emph{dilation} by a \emph{structuring element} $A$ can be seen as a
function $\delta_A$ that transforms every input image $X$ into the image
$\dilatnapp{A}{X} = A \oplus X$~\cite{bloch-mm-2007}.  We show that dilations
are space functions and that the distributed space corresponding to these
dilations is the dilation that arises from the intersection of their
structuring elements: i.e., if $\sfun{1}=\dilatn{A}$ and $\sfun{2}=\dilatn{B}$
then $\Dfun{\{1,2\}}=\dilatn{A \cap B}$.
\end{enumerate}

All in all, in this paper we put forward an algebraic theory for group
reasoning in the context of ccp that can also be applied to other domains. The
theory  here developed can be used in the semantics  of the spatial ccp
process calculus to reason about or prevent  potential unwanted  evolutions of
ccp processes.  One could imagine the incorporation of group reasoning  in a
variety of process algebraic settings and indeed we expect that such
formalisms  will appear in due course. We will also show that our algebraic
theory can be applied to prove new results in other realms such as geometry
and mathematical morphology.

\begin{remark}
This paper is the extended version of the CONCUR'19 paper
in~\cite{guzman-reasondistknowldg-2019} with full proofs and the contributions
described above in the points 5, 6 and 7.  For the sake of the reviewers we
also include contents, index and subject tables and for
Sections~\ref{sec:dist-info} and~\ref{sec:app}, which contain the main
contributions, we include a summary at the end of each section.\qed
\end{remark}

\section{Background}
\label{sec:back}

We presuppose knowledge of basic notions from domain theory and order
theory~\cite{davey2002introduction,abramsky1994domain,gierz2003continuous}.
In this section we present notation and definitions that we use
throughout the paper.

\begin{notation}
\label{not:order}
Let $(\Poset, \pleq)$ be a poset and let $S \subseteq \Poset$. We use
$\bigjoinp{\Poset} S$ to denote the \emph{least upper bound (lub)} (or
\emph{supremum} or \emph{join}) of the elements in $S$, and 
$\bigmeetp{\Poset} S$ is the \emph{greatest lower bound (glb)} (\emph{infimum}
or  \emph{meet}) of the elements in $S$.
We shall often omit the index $\Poset$ from $\pleq$, $\bigjoinp{\Poset}$ and
$\bigmeetp{\Poset}$ when no confusion arises.
As usual, if $S= \{ c,d \}$, $c \sqcup d$ and $c \meet d$ represent
$\bigjoin S$ and $\bigmeet S$, respectively.
If $S = \emptyset$, we denote $\bigjoin S = \true$ and
$\bigmeet S = \false$.

An element $e \in S$ is the \emph{greatest element} of $S$ if and only
if for every element $e' \in S$, $e' \cleq e$.
If such an $e$ exists, we denote it by $\max S$.

\resp{We added the following notation.}
\hltext{
The \emph{dual (or inverse)} of a poset $(\Poset, \cleq)$ is the poset
 $(\Poset,\cleq^{\it op})$ where  $c \cleq^{\it op} d$ holds  if and only if $d \cleq c$ holds.}
\end{notation}
\index[operator]{$(\Poset, \cleq)$, poset}
\index[operator]{$\bigjoinp{\Poset}$, join in poset $\Poset$}
\index[operator]{$\bigmeetp{\Poset}$, meet in poset $\Poset$}
\index[operator]{$\bigjoin$, $\join$, supremum, join, lub}
\index[operator]{$\bigmeet$, $\meet$, infimum, meet, glb}
\index[operator]{$\true$}
\index[operator]{$\false$}
\index[operator]{$\cleq$, entailment}
\index[operator]{$\max$, maximum operator}

The next definition introduces complete lattices and some of their basic
 properties.

\begin{definition}[\cite{birkhoff-lattice-1948}]
\label{def:c-lat}

Let $(\Poset, \cleq)$ be a poset
\begin{enumerate}[(i)]
\item $\Poset$ is said to be a \emph{complete lattice} if and only if
$\bigjoin S$ and $\bigmeet S$ exist for every $S \subseteq \Poset$.
\index{Complete lattice}

\item $\Poset$ is \emph{distributive} if and only if for every
$a,b,c \in \Poset$, $a \join ( b \meet c) = (a \join b) \meet (a \join c)$.
\index{Distributive property}

\item A non-empty set $D \subseteq \Poset$ is \emph{directed} if and only if
for every pair of elements $x, y \in D$, there exists $z \in D$ such that $x
\cleq z$ and $y \cleq z$, or, equivalently, if and only if every \emph{finite}
 subset of $D$ has an upper bound in $D$.
\index{Directed set}

\item An element $c \in \Poset$ is \emph{compact} if and only if for
any directed set $D \subseteq \Poset$, $c \cleq \bigjoin D$ implies
that $c \cleq d$ for some $d \in D$.
\index{Compact element}

\item $\Poset$ is said to be a \emph{completely distributive lattice} if
it is a complete lattice and for any doubly indexed subset
$\{x_{ij}\}_{i \in I, j \in J_i}$ of $\Poset$
\[
\bigjoin_{i \in I}\left(\bigmeet_{j \in J_i} x_{ij} \right)
= \bigmeet_{f \in F} \left(\bigjoin_{i \in I} x_{i f(i)} \right)
\]
where $F$ is the class of choice functions $f$ choosing for each index $i \in
I$ some index $f(i) \in J_i$.
\index{Completely distributive lattice}
\end{enumerate}
\end{definition}

Our space functions will be defined as self-maps with some structural
properties intended to capture our notion of space.

\begin{definition}[\cite{gierz2003continuous}]
Let $(\Lat, \cleq)$ be a complete lattice.
A \emph{self-map} on $\Lat$ is a function $f$ from $\Lat$ to $\Lat$.
\index{Self-map}
Let $f$ be a self-map on $\Lat$.

\begin{enumerate}[(i)]
\item $f$ is \emph{monotonic} if for every $a,b \in \Lat$ such that
$a \cleq b$, then $f(a) \cleq f(b)$.
\index{Self-map!Monotonic}

\item We say that $f$ \emph{preserves} the join of a set
$S \subseteq \Lat$  if and only if $f(\bigjoin S) = \bigjoin \{ f(c)\mid c \in S \}$.

\item We say that $f$ \emph{preserves arbitrary joins} if and only if
it preserves the join of any arbitrary set.
\index{Self-map!Arbitrary join preserving}

\item $f$ is \emph{continuous} if and only if it preserves the join
of any directed set on $\Lat$.
\index{Self-map!Continuous}
\end{enumerate}
\end{definition}

We conclude this background section with a well-known fact about continuous functions.

\begin{proposition}[\cite{abramsky-handbooklogic-1995}]
\label{prop:continuity}

Let $(\Poset,\cleq)$ be a poset where $\Poset$ is a countable set. Let $f$ be a
self-map that preserves the join of increasing chains, i.e., for every $S =
\{c_1,c_2,\ldots\} \subseteq P$ such that $c_1 \cleq c_2 \cleq \cdots$, we have
$f(\bigjoin S) = \bigjoin\{ f(c) \mid c \in S\}$. Then $f$ is continuous.
\end{proposition}
\index{Increasing chain} 


\section{Spatial and Standard Constraint Systems}
\label{sec:scs}

In this section we recall the notion of constraint system and its spatial
extension~\cite{knight:hal-00761116}. Furthermore, we generalize this
extension to allow for infinitely many agents and state some results that will
be used in later sections.

\subsection{Constraint Systems}

\hltext{
\resp{We changed this paragraph.}
{Constraint systems (cs)}~\cite{saraswat1991semantic} are semantic structures
to specify partial information. Following the traditional approach
in~\cite{de1995nondeterminism}, a cs can be formalized as a complete lattice 
$(\Con, \cleq)$. The elements of $\Con$ are called \emph{constraints} and
they represent (partial) information.}

\begin{definition}[Constraint Systems~\cite{de1995nondeterminism}]
\label{def:cs} 
A \emph{constraint system} (cs) is a complete lattice $(\Con, \cleq)$.
The elements of $\Con$ are called \emph{constraints}. 
The symbols $\sqcup$,
$\true$ and $\false$  will be used to denote the join 
operation, the bottom, and the top element of $\Con$.
\end{definition}
\index[operator]{$\true$}
\index[operator]{$\false$}
\index[operator]{$\Con, (\Con,\cleq)$, Constraint system, cs}
\index[operator]{$\cleq$, entailment}
\index{Constraints}

\hltext{
\begin{remark}
\label{rmk:bool-cs}
\resp{We added the next remark.}
Some readers familiar with boolean algebras may feel uneasy about referring to
the bottom and top as $\true$ and $\false$, respectively. Nevertheless, this
is the traditional information order in cs~\cite{vijay}. One can think of cs
as the the dual of boolean algebras without the distributivity requirement.
In fact the Herbrand cs~\cite{de1995nondeterminism} given below is not
distributive.\qed
\end{remark}
}

\begin{figure}
\centering
\includegraphics[scale=1.2]{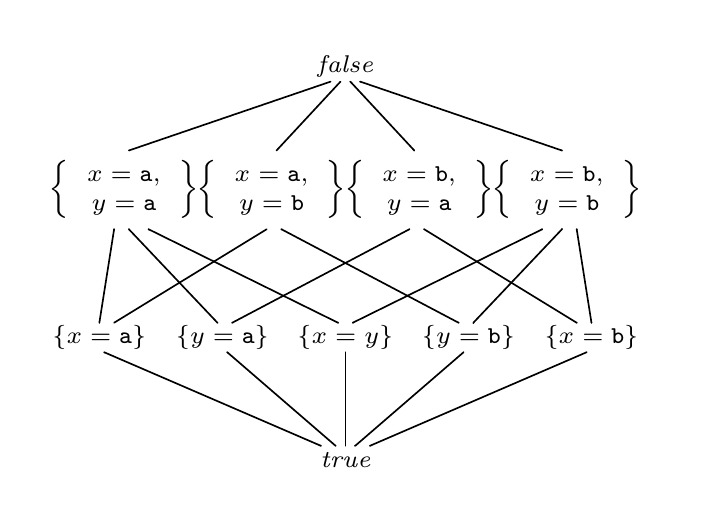}
\caption{Herbrand cs for variables $x$ and $y$ and, constants $\mathtt{a}$ and $\mathtt{b}$.}
\label{fig:herbrand-cs}
\end{figure}

\resp{We added the next four paragraphs to address {\bf N12}, {\bf N13}.}
\hltext{The term \emph{constraints} has been traditionally used in the realm of
concurrent constraint programming to refer to partial information for
historical reasons; they were developed as a generalization of assertions or
restrictions  over variables.  For example, constraints can be instantiated as
arithmetic assertions such as $x > 42$ or as a set of equalities between
Herbrand terms such as $\{x=\mathtt{a}, y=\mathtt{b}\}$ where $x,y$ are
variables and $\mathtt{a,b}$ are constants~\cite{de1995nondeterminism}. In fact,
much of the terminology for the abstract definition of cs is derived from
\emph{concrete} constraint systems such as the Herbrand cs illustrated below
in Ex.~\ref{ex:herbrand-cs}. We will use this concrete cs to give some
intuitions about the intended meaning of cs. 

The relation $\cleq$ represents information order. Thus $c \cleq d$,
alternatively written  $d \cgeq c$, means that the information represented by
$c$ can be derived from the information represented by $d$, that the
assertion $d$ represents at least as much information as $c$, or that if $d$
holds then $c$ must also  hold. For example, we shall see that $\{ x = y \}
\cleq \{x=\mathtt{a}, y=\mathtt{a}\}$ in the Herbrand cs below.  This realizes
the intuition that if $x=\mathtt{a} \wedge y=\mathtt{a}$ holds then $x=y$
holds.

The operator $\sqcup$ represents join of information;  $c \join d$ results
from joining the information from both $c$ and $d$. Notice that $c \join d$ is
the least constraint that allows us to derive \emph{both} $c$ \emph{and} $d$,
i.e., $c \join d \cgeq  c$ and $c \join d \cgeq  d$. The join is typically
interpreted as the conjunction of information and in the concrete case of
Hebrand cs below is obtained by set union. For instance,  in the Herbrand cs
below for $e = \{ x = \mathtt{a}\} \join \{ y = \mathtt{b} \}  = \{ x =
\mathtt{a}, y = \mathtt{b} \}$, we have $e \cgeq \{x = \mathtt{a} \}$ and $e
\cgeq \{ x = \mathtt{b} \}.$  This realizes the intuition that if
$x=\mathtt{a} \wedge y=\mathtt{b}$ holds the both $x=\mathtt{a}$ and
$y=\mathtt{b}$ also hold. 
 
The top element of  the cs represents the join of all, possibly inconsistent,
information, hence it is traditionally refereed to as  $\false$. For example
$\{ x=\mathtt{a} \} \join \{ x= \mathtt{b} \} = \false$ where $\mathtt{a}$
and $\mathtt{b}$ are different constants. This realizes the intuition that 
assertion $x=\mathtt{a} \wedge x=\mathtt{b}$ is inconsistent.  
The bottom element $\true$ represents \emph{empty} or \emph{null} information,
i.e., from $\true$ you cannot derive anything else, $\true \not\cgeq e$ for any
$e\neq \true.$ In the particular case of the Herbrand cs below, $\true$
is the empty set.

The following examples recall two standard concrete cs. The former captures
syntactic equality between terms based on a first-order alphabet and, the
later shows that boolean assignments can be made into a cs and logic
propositions can be interpreted on it.

\resp{The next example is new, to address {\bf N2}.}
\begin{example}[Herbrand Constraint System~\cite{de1995nondeterminism}]
\label{ex:herbrand-cs}
The Herbrand cs captures \emph{syntactic} equality between terms $t,t',\ldots$
built from a first-order alphabet $\mathcal{L}$ with variables $x,y,\ldots$,
function symbols, and equality $=$. The constraints are (equivalence classes
of) sets of equalities over the terms of $\mathcal{L}$: E.g.,  $\{ x = t,  y =
t \}$ is a constraint. The relation ${c}\cleq{d}$ holds if the equalities in
$c$ follow from those in $d$: E.g., ${\{ x = y \}}\cleq{\{  x = t,  y = t
\}}$. The constraint $\false$ is the set of all term equalities in
$\mathcal{L}$ and $\true$ is (the equivalence class of) the empty set. The
compact elements are the (equivalence class of) finite sets of equalities. The
join is the (equivalence class of) set union. Fig.~\ref{fig:herbrand-cs} depicts
the Herbrand cs for variables $x$ and $y$ and, terms (constants) $\mathtt{a}$
and $\mathtt{b}$.\qed
\end{example}
}

\resp{The next paragraph is associated to the above example. We added this also to clarify the order in cs.}

\hltext{In the above example constraints are represented as set of equations and thus
the join  of constraints corresponds to the equivalence class of \emph{union}
of their equations. We can also view a constraint $c$ as a representation of a
set of variable assignments~\cite{aristizabal:hal-00546722}. For instance, a
constraint $x>42$ can be thought of as the set of assignments mapping $x$ to a
value greater than 42; i.e., the solutions to (or models of) $x>42$. In this
case the join of constraints naturally corresponds to the \emph{intersection}
of their assignments, $\false$ as the empty set of assignments, and $\true$ as
the set of all assignments. For example, the interpretation of the constraint
$x>42 \wedge x<50$ corresponds to the intersection, hence the join, of the 
interpretation of the constraints $x>42$ and $x<50$. Below we illustrate this
in more detail for the case of boolean assignments.

\begin{example}[Boolean Constraint System~\cite{guzman:hal-01257113}]
\label{ex:propositional-cs}
\resp{The next example is new, this addresses {\bf N2}.}
Let $\Phi$ be a set of \emph{primitive propositions}. A boolean (or truth)
assignment $\pi$ over $\Phi$ is a total map from $\Phi$ to the set $\{0,1\}.$
We use  $\acal(\Phi)$ to denote the set of all such boolean assignments. We
can now define the boolean cs $\bbold(\Phi)$ as $(\pcal(\acal(\Phi)),
\supseteq)$: The powerset of assignments ordered by $\supseteq$. Thus
constraints in $\bbold(\Phi)$  are sets of assignments, $\sqsubseteq$ is
$\supseteq$, $\false$ is $\emptyset$, $\true$ is $\acal(\Phi)$, the join
operator $\join$ is $\cap$, and the meet operator $\meet$ is $\cup$. A
constraint $c$ in $\bbold(\Phi)$ is compact iff $\acal(\Phi) \setminus c$ is a
finite set. 
 
Notice that logic propositions can be straightforwardly interpreted as
constraints in $\bbold(\Phi)$. Let  $\lcal_0(\Phi)$ be the propositional
language built from $\Phi$ by the grammar
\begin{equation}
\label{propositional-language}
\phi, \psi,\ldots \ := \ p \mid  \phi \land \psi \mid \neg \phi
\end{equation}
where $p \in \Phi$. As usual
$\lfalse \equiv p \conj \neg p$ for some $p \in \Phi$,
$\ltrue \equiv \neg \lfalse$,
$\phi \disj \psi \equiv \neg (\neg \phi \conj \neg \psi)$ and,
$\psi \Rightarrow \phi \equiv \neg \phi \disj \psi$.
A boolean assignment $\pi$ \emph{satisfies} $\phi$ iff $\pi \models \phi$
where $\models$ is defined inductively as follows: $\pi \models p$ iff
$\pi(p)=1$, $\pi \models \phi \wedge \psi$ iff $\pi \models \phi$ and $\pi
\models \psi$, and $\pi \models \neg \phi$ iff $\pi \not\models \ \phi$. We
interpret each formula $\phi$  as the constraint $\bbold\den{\phi} \defsymbol 
\{ \pi \in \acal(\Phi)  \mid \pi \models \phi  \}$ in $\bbold(\Phi)$.
Under this interpretation, we clearly have the following equalities:
\begin{align*}
\bbold\den{\lfalse} &= \false = \emptyset \\
\bbold\den{\ltrue} &= \true = \acal(\Phi) \\
\bbold\den{p} &= \{ \pi \in \acal(\Phi) \mid \pi(p)=1 \} \\
\bbold\den{\phi \wedge \psi} &= \bbold\den{\phi} \join \bbold\den{\psi} = \bbold\den{\phi} \cap \bbold\den{\psi}\\
\bbold\den{\neg \phi} &= \acal(\Phi) \setminus \bbold\den{\phi}.
\end{align*}

Also notice that ${\bbold\den{\phi}}\cleq{\bbold\den{\psi}}$ holds iff
$\psi \Rightarrow \phi $ is \emph{valid}, i.e., satisfied by every truth assignment.
\qed
\end{example}

Other typical examples include constraint system for streams (the Kahn cs),
rational intervals, and first-order theories~\cite{saraswat1991semantic}.
}
\subsection{Distributive Constraint Systems}

Distributivity is ubiquitous in order theory and it plays a fundamental role
in the results of this paper. We consider three forms of distribution.

\begin{definition}[Distributive cs]
\label{def:frames}
A cs $(\Con, \cleq)$ is said to be \emph{distributive (completely
distributive)} iff  it is a distributive (completely distributive) lattice. 
It is said to be a \emph{constraint frame} iff its finite joins distribute over
arbitrary meets. More precisely, $c \join {\bigmeet S} = \bigmeet\{ c \join e
\mid e \in S\}$ for every $c \in \Con$ and $S \subseteq \Con$.
\end{definition}
\index{Constraint frame}

Clearly every completely distributive cs is a constraint frame and every
constraint  frame is also distributive cs. For finite constraint systems all
the three notions of distributivity are equivalent.

Constraint frames allow us to include the subtraction operator from co-Heyting
Algebras, the dual of Heyting algebras~\cite{vickers-topo-1996}.
The \emph{subtraction operator} $d \sop c$ in our setting corresponds to
the \emph{weakest constraint} one needs to join $c$ with to derive $d$.

\resp{The following definition and proposition have changed to address {\bf N26, N27}.}
\hltext{
\begin{definition}[Subtraction~\cite{bellin-catproofth-2014}]
\label{def:ch-sop}
Let $(\Con, \cleq)$ be a cs. Define $d \sop c$ as
$\bigmeet \{ e \in \Con \mid { {c} \join {e}} \cgeq {d} \}$.
\end{definition}
\index{Subtraction operator}
\index[operator]{$\sop$, co-Heyting subtraction}
}

The following properties of subtraction correspond to standard logical properties
if we interpret $\join$ as conjunction, $\cgeq$ as entailment, and $d \sop c$ as 
the logical implication of $d$ by $c$.

\hltext{
\begin{proposition}[\cite{guzman:hal-01257113}]
\label{prop:subtraction}
Let $(\Con, \cleq)$ be a constraint frame. Then for every $c,d \in \Con$ the 
following properties hold:
(1) $c \join (d \sop c) =  c \join d$,
(2) $d \cgeq (d \sop c) $,
(3) $d \sop c = \true$ iff $c \cgeq  d$.
\end{proposition}

Notice that under the above logical interpretation,
Prop.~\ref{prop:subtraction}~(1) corresponds to \emph{Modus Ponens}. Let us
now illustrate a simple application of this proposition.

\begin{example}
\resp{We added the next example.}
Constraint systems of the form $(\pcal(U), \supseteq)$ as $\bbold(\Phi)$ in
Ex.~\ref{ex:propositional-cs} are completely distributive lattices, hence
constraint frames. It is easy to see that $\bbold\den{\phi \Rightarrow \psi}
= \bbold\den{\psi} \sop \bbold\den{\phi}$. From Ex.~\ref{ex:propositional-cs},
we have the following equality $\bbold\den{\phi \conj (\phi \Rightarrow \psi)} 
= \bbold\den{\phi } \join \bbold\den{\phi \Rightarrow \psi}$.
Thus using Prop.~\ref{prop:subtraction}~(1) we obtain 
$\bbold\den{\phi \conj (\phi \Rightarrow \psi)} = \bbold\den{\phi \wedge \psi}$
as expected. \qed
\end{example}
}

The powerset ordered by inclusion is a stereotypical example of completely 
distributive lattice.

\hltext{
\begin{example}[Powerset Constraint System]
\label{ex:powerset}
\resp{The next example has changed according to Prop.~\ref{prop:subtraction} and comment {\bf N27}.}
The power set of any set $S$ ordered by inclusion $(\powerset{S},\subseteq)$
is a cs. In fact it is the stereotypical example of \emph{completely distributive} cs. 
In this case $\cleq \ = \ \subseteq$, $\true=\emptyset$, $\false = S$,
for every $A,B \subseteq S$, $A \join B = A \cup B$,  $A \meet B = A \cap B$,
$B \sop A = B \setminus A$.\qed
\end{example}
}

We conclude this section by discussing the distributivity of the cs of our
previous examples.

\hltext{
\begin{remark}
\resp{We added the next remark.}
Notice that the cs in Ex.~\ref{ex:propositional-cs} and Ex.~\ref{ex:powerset} 
are distributive as they are powersets ordered by (reversed) inclusion, whereas
the Herbrand cs in Ex.~\ref{ex:herbrand-cs} is not. To see this consider the
constraints $c=\{ x=\mathtt{a} \}$, $d = \{ x = \mathtt{a}, y = \mathtt{a} \}$ 
and $e=\{ x=\mathtt{b}\}$ in Fig.~\ref{fig:herbrand-cs}.
We have 
\[ 
c \join (d \meet e) = c \join \true = c \neq (c \join d) \meet (c \join e) = d \meet \false = d.
\]\qed
\end{remark}
}

\subsection{Spatial Constraint Systems}

The authors of~\cite{knight:hal-00761116} extended the notion of cs to account
for  distributed and multi-agent scenarios with a finite number of agents,
each having their own space for local information and their computations. The
extended structures are called spatial constraint systems (scs).  In this
section we adapt scs to reason about  possibly infinite groups of agents.
First we recall the intuition about scs given in~\cite{knight:hal-00761116}. 

\hltext{
\begin{remark}\resp{This remark is new associated to {\bf N9}, {\bf N14}, {\bf N15}, {\bf N23}.}
\label{rmk:modal-scs}
In Remark~\ref{rmk:bool-cs} we pointed out that (distributive) cs are the dual
of boolean algebras, the algebraic generalization of propositional logic. The
extension of cs to scs from~\cite{knight:hal-00761116} is much like the
extension of boolean algebras to \emph{(poly)modal
algebras}~\cite{goldblatt:2000}, the algebraic generalization of the $K_n$
propositional modal logic. A modal algebra is a boolean algebra equipped with
unary operators, called modalities, that preserve meets and top. Dually, a scs
is a cs equipped with  functions, called \emph{space functions}, that preserve
joins and bottom. 

Roughly speaking, the preservation of joins by a space function $f$, i.e.,
$f(c \join d) = f(c) \join f(d)$,  corresponds to the Distribution Axiom
$\mathbf{K}$, of modal logic which states that box modality $\square$
distributes over conjunction. The preservation  of bottom, i.e.,
$f(\true)=\true$, corresponds to Generalization Axiom $\mathbf{N}$, that
states that  if a formula $\phi$ is valid, hence equivalent to true, then the
formula $\square \phi$ is also valid. 

Furthermore, if these space functions are also \emph{closure operators} (i.e.,
idempotent and extensive space functions), the corresponding scs is called
\emph{epistemic scs}~\cite{knight:hal-00761116} and they correspond to the
$S4$ modal logic of knowledge and belief.  In fact epistemic scs are dual to
\emph{closure algebras}~\cite{mckinsey-topo-1944} which are modal algebras
whose modalities are closure operators.  Closure algebras are the
generalization of the $S4$ modal logic for knowledge. The extensiveness of a
space function $f$, i.e., $f(c) \cgeq c$, corresponds to the Truth Axiom
$\mathbf{T}$ of $S4$, $\square \phi \imp \phi$, stating that if $\phi$ is
known then it must be true.   Idempotence of a space function $f$,
$f(c)=f(f(c))$, corresponds to the Axiom $\mathbf{4}$ of $S4$, $\square \phi
\imp \square \square \phi$,  stating that it is known what it is known (the
implication $\square \square \phi \imp \square \phi$ follows from Axiom
$\mathbf{T}$). 
 
Modal logics have been widely used to reason about space, knowledge, and
belief~\cite{belmonte-voxlogica-2019,fagin1995reasoning,hintikka-belief-1962,hendricks-elogic-2008}.
They typically arise as restrictions over the Kripke models of the
above-mentioned modal logic  $K_n$. For this reason we will focus on scs in
general rather than in the more restrictive epistemic scs. This will allow us
to interpret scs as structures for space, belief or knowledge. 

Below we shall often use the term \emph{epistemic} to refer to both belief or
knowledge. When necessary we will use the more specific term \emph{doxastic}
to refer to belief.
\qed
\end{remark}
}

A scs is defined over a set (or group) of agents $G$. Each agent $i \in G$ has
a \emph{space} function $\sfun{i}: \Con \to \Con$ that preserve joins and
bottom. Recall that constraints can be viewed as assertions. Thus given $c \in
\Con$, we can then think of the constraint $\sfunapp{i}{c}$ as an assertion 
stating that $c$ is a piece of information residing \emph{within the space of 
agent} $i$. Some alternative \emph{epistemic} interpretations of
$\sfunapp{i}{c}$ is that it is an assertion stating that agent $i$
\emph{believes} $c$, that $c$ holds within the space of agent $i$, or that
agent $i$ \emph{interprets} $c$ as the constraint $\sfunapp{i}{c}$. All these
interpretations convey the idea that $c$ is local or subjective to agent $i$.
\index{G, group of agents}
\index[operator]{$\sfun{i}$, space function}

\resp{We added the next paragraph to address {\bf N1}, {\bf N14}, {\bf N15}.}
\hltext{The reader familiar with modal logic may notice that the above
intuition about representation of information in scs is fundamentally
different from the representation of information in epistemic/doxastic logic,
or in modal logic in general. In epistemic logic, the statement $w\models
K_i(\phi)$ means ``at world $w$, agent $i$ knows fact $\phi$,'' or ``at world
$w$, agent $i$ possesses information $\phi$.'' Here  $w$ is a world (or
state), part of the model, describing a potential situation or state of
affairs, and $\phi$ and $K_i(\phi)$ are formulas, representing pieces of
information. In scs, on the other hand, the statement $\sfunapp{i}{d} \cleq c$
means that when $c$ is actually true, agent $i$ knows/believes $d$, or
possesses information $d$. In this case, unlike in epistemic logic, one
constraint ($c$) represents a (possibly partial) description of the actual
state of the world, and another constraint ($d$) represents a description of
the information that an agent possesses, and a third constraint,
$\sfunapp{i}{d}$, represents the fact that agent $i$ possesses information
$d$.}

We now introduce the notion of space function.

\begin{definition}[Space Functions]
\label{def:sfunc}
\index{Space function}
A \emph{space function} over a constraint system $(\Con,\cleq)$ is a
\emph{continuous} self-map $f:\Con \to \Con$ such that for every $c,d \in \Con:$

(S.1) $f(\true)=\true$, and
\index{S.1, first space axiom}

(S.2) $f(c \join d) = f(c) \join f(d)$.
\index{S.2, second space axiom}

\noindent We shall use $\sfunspace{\C}$ to denote the \emph{set of all space functions}
over $\C$.
\index[operator]{$\sfunspace{\C}$, set of space functions}
\end{definition}

\resp{We added the following clarification}
\hltext{In Remark~\ref{rmk:modal-scs} we pointed out that S.1 and S.2
correspond to Axioms $\mathbf{N}$ and $\mathbf{K}$ of modal logic. From a
spatial point of view,} the assertion $f(c)$ can be interpreted as saying that
$c$ is in the space represented by $f$.  Property S.1 states that having an
empty local space amounts to nothing. Property S.2 allows us to join and
distribute the information in the space represented by $f$.

A \emph{spatial constraint system} is a constraint system with a possibly
infinite group of agents each one having a space function. We specify such a
group as a  tuple of space functions.

\begin{definition}[Spatial Constraint Systems]
\label{def:scs}
A \emph{spatial constraint system (scs)} is a constraint system
$(\Con, \cleq)$ equipped with a possibly infinite tuple
$\sfuntuple = (\sfun{i})_{i \in G}$ of space functions in
$\sfunspace{\C}$.

We shall use $({\Con},\cleq,(\sfun{i})_{i \in G})$ to denote an scs with a
tuple $(\sfun{i})_{i \in G}$.  We refer to $G$ and $\sfuntuple$ as the
\emph{group of agents} and \emph{space tuple} of $\C$ and to each $\sfun{i}$
as the \emph{space function} in $\C$ of agent $i$. 
Subsets of $G$ are also referred to as groups of agents (or sub-groups of $G$).
\end{definition}
\index[operator]{$({\Con},\cleq,(\sfun{i})_{i \in G})$, spatial constraint system, scs}
\index[operator]{$\sfuntuple = (\sfun{i})_{i \in G}$, tuple of space functions}

\resp{We added the next paragraph to clarify the interpretation of space functions.}
\hltext{As mentioned before we can think of  scs as the dual of  polymodal
 boolean algebras~\cite{goldblatt:2000} without the distributivity requirement, where
 each space function corresponds to a modal operator.
 
In  Remark~\ref{rmk:modal-scs} we pointed out that \emph{epistemic scs} are those
scs whose space functions are idempotent and extensive. They were used
in~\cite{knight:hal-00761116} to reason about knowledge. Here is an example of
such a scs.}

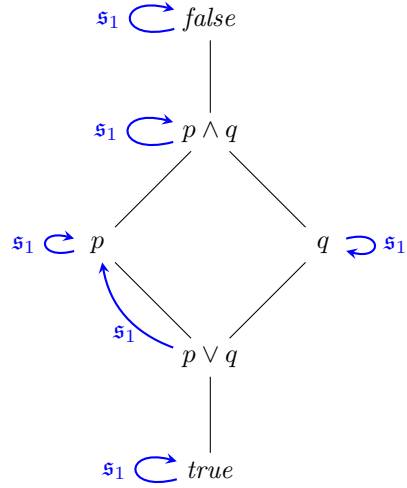
\begin{figure}
\centering
\begin{tikzpicture}[scale=0.6,>=stealth]
  \node (A) at (0,-2.5)   {$p \disj q$};
  \node (B) at (-2.5,0)   {$\; p\;$};
  \node (C) at (2.5,0)    {$\; q\; $};
  \node (D) at (0,2.5)    {$p \conj q$};
  \node (E) at (0,-5)     {$\true$};
  \node (F) at (0,5)      {$\false$};  

  \draw (A) to node {} (B);
  \draw (A) to node {} (C);
  \draw (B) to node {} (D);
  \draw (C) to node {} (D);
  \draw (F) to node {} (D);
  \draw (E) to node {} (A);

  \draw [blue,thick,->,loop left]  		(B) to node {$\sfun{1}$} (B);
  \draw [blue,thick,->,loop right] 		(C) to node {$\sfun{1}$} (C);
  \draw [blue,thick,->,loop left]  		(D) to node {$\sfun{1}$} (D);
  \draw [blue,thick,->,loop left]  		(E) to node {$\sfun{1}$} (E);
  \draw [blue,thick,->,loop left]  		(F) to node {$\sfun{1}$} (F);
  \draw [below,blue,thick,->,bend left] (A) to node {$\sfun{1}$} (B);
\end{tikzpicture}
\caption{Cs ordered by logical implication and space function $\sfun{1}$.}
\label{fig:epist-int}
\end{figure}

\hltext{
\begin{example}[\hltext{Knowledge}]
\resp{We added the next example and Fig.~\ref{fig:epist-int}.}
Consider the cs in Fig.~\ref{fig:epist-int} and the space function $\sfun{1}$
defined on it. It is easy to see that  $\sfun{1}$ is an idempotent and extensive 
space function. If we want to determine what agent 1 knows at $q$, we have no problem because
$q = \sfunapp{1}{q}$, meaning that agent 1 knows $q$. On the other hand,
for $p \disj q$, it is more difficult to determine what agent 1 knows
because there is no constraint $c$ such that $\sfunapp{1}{c} = p \disj q$. We
solve this problem by considering the constraints \emph{below} $p \disj q$. 
Since $c \cleq p \disj q$ means that $p \disj q$ logically implies $c$, if we
can find a constraint $c$ where $c \cleq p\disj q$ and $c = \sfunapp{1}{c'}$
for some constraint $c'$, we know that $p \disj q$ logically implies
$\sfunapp{1}{c'}$, so at $p \disj q$, agent 1 has information $c'$. Finally,
by taking the join (logical conjunction) of all such $c'$, we take into
account all of the information that agent 1 possesses at $p \disj q$.\qed

\end{example}

The idempotence or extensiveness of space functions from epistemic scs allows
for knowledge interpretations but it may be too restrictive. For example, 
notice that extensiveness implies that $\sfunapp{i}{\false} = \false$ and 
$\sfunapp{i}{c} \join \sfunapp{i}{d} = \false$ if $c \join d = \false$. This
would rule out interpretations we wish to allow for general scs as those given
below. 
 
\paragraph{Doxastic and Spatial Interpretations} We could have a scs where
$\sfunapp{i}{\false} \neq \false$ for some agent $i$. This is referred to as
\emph{inconsistency confinement} and, intuitively, can be seen as if
inconsistencies generated within an agent's space are confined to its own
space. We also have the possibility that given two different agents $i$ and
$j$, $\sfunapp{i}{x > 42} \join \sfunapp{j}{x < 42} \neq \false$, despite
$x>42 \join x<42 = \false$. The spatial interpretation is that $x > 42$ is in the space
of $i$ while $x<42$ is in the space of $j$, so there is no conflict or contradiction.
The doxastic interpretation is that agent $i$ believes $x>42$ while agent $j$ believes
that $x<42$.  We refer to this as \emph{freedom of opinion}.
Agents could have different information or perception about the same subject.
Another aspect is \emph{information blindness}. It is possible that an agent
cannot distinguish between two different constraints, e.g., a color blind
agent $i$ that cannot distinguish red from green, i.e.,
$\sfunapp{i}{\textnormal{red}} = \sfunapp{i}{\textnormal{green}}$ as shown in the next
example.

\resp{The next example is new associated to {\bf N1, N3}. We also added Table~\ref{color-sfun:table}.}
\begin{example}[Color Perception]
\label{ex:colors}
Table~\ref{color-sfun:table} describes color perception for four
agents. The underlying cs is based on four colors ordered by color brightness.
  We define space functions $\sfun{1}$, $\sfun{2}$, $\sfun{3}$ and $\sfun{4}$
as shown in the table. In these situations, agent 1 perceives colors
accurately, but agent 2 is colorblind and cannot tell any difference between
yellow, red and green, i.e.,
$\sfunapp{2}{\emph{yellow}}=\sfunapp{2}{\emph{red}}=\sfunapp{2}{\emph{green}}$.
Similarly, agent 3 cannot distinguish between red and yellow, i.e.,
$\sfunapp{3}{\emph{red}}=\sfunapp{3}{\emph{yellow}}$. Finally, agent 4 is
totally blind, he perceives all the colors as black.\qed
\end{example}
}

\begin{table}
\centering
\begin{tabular}{| p{0.4\textwidth} | p{0.4\textwidth} |} 
\hline
\begin{center}
\resizebox{0.3\textwidth}{!}{%
\begin{tikzpicture}[scale=0.7,>=stealth]
  \tikzstyle{every node}=[font=\scriptsize]
  \node[fill=black,shape=circle, draw,text=white]	 (A) at (0,-2.5)  {\emph{black}};
  \node[fill=red!70,shape=circle,draw,inner sep=0.18cm] (B) at (-2.5,0) {\emph{red}};
  \node[fill=green!60!black,shape=circle,draw]		 (C) at (2.5,0)   {\emph{green}};
  \node[fill=yellow,shape = circle, draw] 			 (D) at (0,2.5)  {\emph{yellow}};

  \draw (A) to node {} (B);
  \draw (A) to node {} (C);
  \draw (B) to node {} (D);
  \draw (C) to node {} (D); 

  \draw [thick,->,loop below]  (A) to node {$\sfun{1}$} (A);
  \draw [thick,->,loop above]  (B) to node {$\sfun{1}$} (B);
  \draw [thick,->,loop below]  (C) to node {$\sfun{1}$} (C);
  \draw [thick,->,loop above]  (D) to node {$\sfun{1}$} (D);
\end{tikzpicture}
}%
\end{center}
&
\begin{center}
\resizebox{0.3\textwidth}{!}{%
\begin{tikzpicture}[scale=0.7,>=stealth]
  \tikzstyle{every node}=[font=\scriptsize]
  \node[fill=black,shape=circle, draw,text=white]  	 (A) at (0,-2.5)  {\emph{black}};
  \node[fill=red!70,shape=circle,draw,inner sep=0.18cm] (B) at (-2.5,0) {\emph{red}};
  \node[fill=green!60!black,shape=circle,draw]	 	 (C) at (2.5,0)   {\emph{green}};
  \node[fill=yellow,shape = circle, draw] 			 (D) at (0,2.5)  {\emph{yellow}};

  \draw (A) to node {} (B);
  \draw (A) to node {} (C);
  \draw (B) to node {} (D);
  \draw (C) to node {} (D); 

  \draw [thick,->,loop below]  		(A) to node {$\sfun{2}$} (A);
  \draw [left,thick,->,bend left]   (B) to node {$\sfun{2}$} (D);
  \draw [right,thick,->,bend right] (C) to node {$\sfun{2}$} (D);
  \draw [thick,->,loop above]  		(D) to node {$\sfun{2}$} (D);
\end{tikzpicture}
}%
\end{center}
\\
Agent 1 sees colors accurately.
&
Agent 2 is color blind, he perceives red and green as yellow.
\\
\hline
\begin{center}
\resizebox{0.4\textwidth}{!}{%
\begin{tikzpicture}[scale=0.7,>=stealth]
  \tikzstyle{every node}=[font=\scriptsize]
  \node[fill=black,shape=circle, draw,text=white]	 (A) at (0,-2.5)  {\emph{black}};
  \node[fill=red!70,shape=circle,draw,inner sep=0.18cm] (B) at (-2.5,0) {\emph{red}};
  \node[fill=green!60!black,shape=circle,draw]		 (C) at (2.5,0)   {\emph{green}};
  \node[fill=yellow,shape = circle, draw] 			 (D) at (0,2.5)  {\emph{yellow}};

  \draw (A) to node {} (B);
  \draw (A) to node {} (C);
  \draw (B) to node {} (D);
  \draw (C) to node {} (D); 

  \draw [thick,->,loop below] 		 (A) to node {$\sfun{3}$} (A);
  \draw [left,thick,->,bend left]    (B) to node {$\sfun{3}$} (D);
  \draw [thick,->,loop right]  		 (C) to node {$\sfun{3}$} (C);
  \draw [thick,->,loop above]	     (D) to node {$\sfun{3}$} (D);
  \draw [white,->,loop left]		 (B) to node {$\sfun{3}$} (B);
\end{tikzpicture}
}%
\end{center}
&
\begin{center}
\resizebox{0.3\textwidth}{!}{%
\begin{tikzpicture}[scale=0.7,>=stealth]
  \tikzstyle{every node}=[font=\scriptsize]
  \node[fill=black,shape=circle, draw,text=white]	 (A) at (0,-2.5)  {\emph{black}};
  \node[fill=red!70,shape=circle,draw,inner sep=0.18cm] (B) at (-2.5,0) {\emph{red}};
  \node[fill=green!60!black,shape=circle,draw]		 (C) at (2.5,0)   {\emph{green}};
  \node[fill=yellow,shape = circle, draw] 			 (D) at (0,2.5)  {\emph{yellow}};

  \draw (A) to node {} (B);
  \draw (A) to node {} (C);
  \draw (B) to node {} (D);
  \draw (C) to node {} (D); 

  \draw [thick,->,loop below]  		 (A) to node {$\sfun{4}$} (A);
  \draw [left,thick,->,bend right]   (B) to node {$\sfun{4}$} (A);
  \draw [right,thick,->,bend left]   (C) to node {$\sfun{4}$} (A);
  \draw [left,thick,->]              (D) to node {$\sfun{4}$} (A);
  \draw [white,->,loop above]		 (D) to node {$\sfun{4}$} (D);
\end{tikzpicture}
}%
\end{center}
\\
Agent 3 cannot distinguish between red and yellow.
&
Agent 4 is completely blind.
\\
\hline
\end{tabular}
\caption{Cs (isomorphic to the four-element boolean algebra) of colors ordered by color brightness. Space functions $\sfun{1}$, $\sfun{2}$, $\sfun{3}$ and $\sfun{4}$. Perception of color by agents 1, 2, 3 and 4.}
\label{color-sfun:table}
\end{table}

Our next example illustrate a simple scs that will be used throughout the paper.

\begin{figure}[t]
\centering
\begin{tikzpicture}[scale=0.7,>=stealth]
  \tikzstyle{every node}=[font=\scriptsize]
  \node[shape = circle, draw] (A) at (0,-2.5)   {$p \vee \neg p$};
  \node[shape = circle, draw] (B) at (-2.5,0)   {$\; p\; $};
  \node[shape = circle, draw] (C) at (2.5,0)    {$\neg p$};
  \node[shape = circle, draw] (D) at (0,2.5)    {$p \wedge \neg p$};

  \draw (A) to node {} (B);
  \draw (A) to node {} (C);
  \draw (B) to node {} (D);
  \draw (C) to node {} (D); 

  \draw [above,blue,thick,->,bend left]  (B) to node {$\sfun{1}$} (C);
  \draw [below,blue,thick,->,bend left]  (C) to node {$\sfun{1}$} (B);
  \draw [below,blue,thick,->,loop left]  (A) to node {$\sfun{1}$} (A);
  \draw [below,blue,thick,->,loop left]  (D) to node {$\sfun{1}$} (D);
  \draw [below,red,thick,->,loop right]  (D) to node {$\sfun{2}$} (D);
  \draw [left,red,thick,->,bend left]    (B) to node {$\sfun{2}$} (D);
  \draw [below,red,thick,->,loop right]  (A) to node {$\sfun{2}$} (A);
  \draw [below,red,thick,->,loop right]  (C) to node {$\sfun{2}$} (C);
\end{tikzpicture}
\index{Four-element boolean algebra}
\caption{Cs given by the four-element boolean algebra ordered by logical implication and space functions $\sfun{1}$ and $\sfun{2}$.}
\label{ex:spatial}
\end{figure}
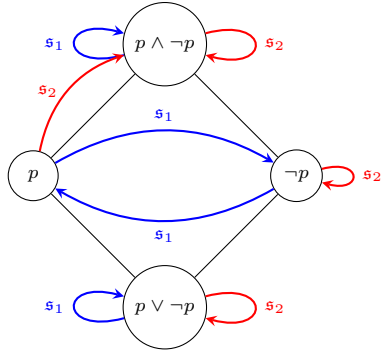

\begin{example}[Proposition Perception]
\label{simple:example}
The scs $({\Con},\cleq,(\sfun{i})_{i \in \{1,2\}})$ in Fig.~\ref{ex:spatial}
is given by the four-element boolean algebra (isomorphic to the complete lattice $\M_2$) and two agents. We have $\Con = \{ p \vee \neg
p,\ p,\ \neg p,\ p \wedge \neg p \}$ and $c \cleq d$ holds if $c$ is a logical
consequence of $d$. The top element $(\false)$ is $p \wedge \neg p$, the
bottom element $(\true)$ is $p \vee \neg p$, and the constraints $p$ and $\neg
p$ are incomparable with each other.

\noindent The set of agents is $\{1,2\}$ with space functions $\sfun{1}$ and
$\sfun{2}$ depicted in Fig.~\ref{ex:spatial}. The intuition is that the agent
$2$  sees no difference between $p$ and $\false$ while agent $1$ interprets
$\neg p$ as $p$ and vice versa. \qed
\end{example}

\hltext{We conclude this section with the Kripke scs from~\cite{guzman:hal-01257113}
which can be used to interpret modal, doxastic and epistemic logics. Other examples of
scs for epistemic reasoning are \emph{Aumann structures} and they will be
illustrated in Sec.~\ref{ssec:ds-aumann}.
\resp{The next subsection is new. We present the Kripke scs.}

\subsection{Kripke Spatial Constraint Systems}
\label{ssec:kripke-scs}
We now extend Ex.~\ref{ex:propositional-cs} by moving from the set boolean
 assignments to the set of (pointed) \emph{Kripke structures}.

\index{KS, Kripke structures}
\begin{definition}[Kripke Structures~\cite{fagin1995reasoning}]
\label{kripke-example}

An $n$-agent Kripke structure (model) (KS) $M$ over a set of atomic
propositions $\Phi$ is a tuple $M = (S, \pi, \rcal_1, \ldots, \rcal_n)$ where

\begin{itemize}
\item $S$ is a nonempty set of states, 
\item $\pi:S \rightarrow (\Phi \rightarrow \{0,1\}) $ is an interpretation 
that associates with each state a truth assignment to the primitive 
propositions in $\Phi$, and 
\item $\rcal_i$ is a binary relation on $S$. 
\end{itemize}
\end{definition}
}
\index[operator]{$\krel{i}{M}$, accessibility relation}
\index[operator]{$(M,s)$, pointed Kripke structure} 
\index[operator]{$\pi$, states interpretation}   
\index[operator]{$\wcal_i(M,s)$, $M$ worlds accessible by $i$ from $s$}

\hltext{
\begin{notation}
\label{kripke:notation}
The states of a KS are often referred to as \emph{worlds}. Each
$\rcal_i$ is referred to as the \emph{accessibility} or
\emph{possibility} relation for agent $i$: $(s,t) \in \rcal_i$ is meant to
capture that agent $i$ considers world $t$ possible given its information in
world $s$. We use  $s \krel{i}{M} t$ to denote $(s,t) \in \rcal_i$ in the KS
$M$. We use $\wcal_i(M,s) = \{ t \mid s \krel{i}{M} t \}$ to denote the
worlds agent $i$ considers possible from a state $s$ of KS $M$. The
interpretation function $\pi$ tells us what primitive propositions are true at
a given world: $p$ holds at state $s$ iff $\pi(s)(p)=1$. We use $\pi_M$ to
denote the interpretation $\pi$ of the KS $M$.
\end{notation}
}
\hltext{Recall that in Ex.~\ref{ex:propositional-cs} constraints are sets of boolean
assignments. This allowed us to interpret each propositional formula as a
constraint; the set of assignments that are models of (or satisfy) the
formula.  Similarly, in the following example (spatial) constraints are sets
of (pointed) KS models. A \emph{pointed KS} is a pair $(M,s)$ where $M$ is a
KS and $s$, called the \emph{actual world}, is a state of $M$.  This will
allows us to interpret each modal formula as its set of pointed KS models;
i.e., a spatial constraint. }

\index[operator]{$\kbold(\scal_n(\cdot))$, Kripke scs}
\index[operator]{$\scal_n(\cdot)$, set of Kripke structures} 
\index[operator]{$i(\cdot)$, space function for Kripke structures} 
\index[operator]{$\Delta$, set of pointed Kripke structures} 
\hltext{
\begin{definition}[Kripke scs~\cite{guzman:hal-01257113}]
\label{def:kripke-scs}
Let $\scal_n(\Phi)$ be a non-empty set of  $n$-agent Kripke structures over
$\Phi$. Let $\Delta$ be the set of all pointed Kripke structures $(M,s)$ such
that $M \in { \scal_n(\Phi) }$.
We define the Kripke $n$-scs for $\scal_n(\Phi)$ as the tuple 
\[
\kbold(\scal_n(\Phi)) = (\C,\sqsubseteq,\Kfun{1},\ldots,\Kfun{n})
\]
where $\Con=\pcal(\Delta)$, and for every $ X,Y \in \C:$ ${X}\cleq{Y}$ iff
$Y\subseteq X$, and
\begin{equation} 
\Kfunapp{i}{X} \defsymbol \{(M,s) \in
\Delta \mid \forall t: s \krel{i}{M} t \mbox{ implies } (M,t) \in X \}
\end{equation}
for every agent $i \in \{ 1,\ldots, n\}$.
\end{definition}
}

\hltext{The Kripke $n$-scs $\kbold (\scal_n(\Phi))$ is a complete lattice given by a powerset
ordered by $\supseteq$. The $\join$ is set intersection, the top element
$\false$ is $\emptyset$, and the bottom $\true$ is the set $\Delta$ of all pointed
Kripke structures $(M,s)$ with $M \in \scal_{n}(\Phi)$. It is easy to verify
that $\Kfunapp{i}{\true}=\true$ and $\Kfunapp{i}{{c_1}\join{c_2}} =
{\Kfunapp{i}{c_1}}\join{\Kfunapp{i}{c_2}}$. Similar to
Ex.~\ref{ex:propositional-cs}, a constraint $c$ in $\kbold(\scal_n(\Phi))$ is
compact iff $\Delta \setminus c $ is a finite set~\cite{guzman:hal-01257113}.} 

\hltext{
\begin{proposition}[\cite{guzman:hal-01257113}]
\label{prop:kripke-scs}
Let $\kbold(\scal_n(\Phi))=(\C,\sqsubseteq,\Kfun{1},\ldots,\Kfun{n})$ be as in
Definition~\ref{def:kripke-scs}. Then $\kbold(\scal_n(\Phi))$ is a scs.
\end{proposition}
}

\hltext{
\paragraph{A modal language} Modal formulae can be interpreted as constraints
in the scs $\kbold(\scal_n(\Phi))$.

The modal language $\lcal_n(\Phi)$ is obtained by extending the grammar for
the propositional language $\lcal_0(\Phi)$ in Eq.~\ref{propositional-language}
with modalities $\square_i \phi$ in the standard way.}
\index[operator]{$\lcal_n(\cdot)$, modal language}

\hltext{
\begin{definition}[Modal Language]
\label{modal-language}
Let $\Phi$ be a set of primitive propositions. The language $\lcal_n(\Phi)$ is
given by the following grammar:
\begin{equation}
\phi, \psi,\ldots \ := \ p \mid  \phi \wedge \psi \mid \neg \phi \mid \square_i \phi
\end{equation}
where $p \in \Phi$ and  $i \in \{1,\ldots,n \}.$
\end{definition}
}

\hltext{The semantics of modal logics is typically given using KS's. We say that a
pointed KS $(M,s)$ \emph{satisfies} $\phi$ iff $(M,s)\models \phi$ where
$\models$ is defined inductively as follows: $(M,s) \models p$ iff
$\pi_M(s)(p)=1$, $(M,s) \models \phi \wedge \psi$ iff $(M,s) \models \phi$ and
$(M,s) \models \psi$, $(M,s) \models \neg \phi$ iff $(M,s) \not\models \
\phi$, and  $(M,s)\models \square_i \phi$ iff $(M,t) \models \phi$  for every
$t$ such that $s \krel{i}{M} t$.
 
As in Ex.~\ref{ex:propositional-cs} we can interpret each formula $\phi$ as
constraints in Kripke constraint systems.}
\index[operator]{$\ccal\den{\phi}$, Kripke scs interpretation of $\phi$} 

\hltext{
\begin{definition}[Kripke Constraint Interpretation]
\label{modal-interpretation}
Let $\ccal$ be a Kripke scs $\kbold(\scal_n(\Phi)).$  Given a modal formula
$\phi$ in the  language $\lcal_n(\Phi)$, its interpretation in the Kripke scs
$\ccal$ is the constraint $\ccal\den{\phi}$  inductively defined as follows:
  \begin{eqnarray*}
   \ccal \den{ p } &  = & \{ (M,s) \in \Delta  | \  \pi_M(s)(p) = 1   \ \} \\
   \ccal \den{\phi \wedge \psi} & = & \ccal \den{\phi} \sqcup \ccal \den{\psi} \\
   \ccal \den{\neg\phi} & = & \Delta \setminus { \ccal \den{\phi} } \\
   \ccal \den{\square_i \phi} & = & \Kfunapp{i}{\ \ccal \den{\phi}\ } 
 \end{eqnarray*}
where  $\Delta$ is the set of all pointed Kripke structures $(M,s)$ such that
$M \in { \scal_n(\Phi) }.$ 
\end{definition}
}

\hltext{
\begin{notation}
\label{semantic:notation}
Notice that the interpretation of $\square_i(\phi)$,
$\ccal\den{\square_i(\phi)}$, is equal to the constraint
$\Kfunapp{i}{\,\ccal\den{\phi} \,}$ in $\kbold(\scal_n(\Phi)).$  Often,
by abuse of notation,  we shall suppress the semantic symbols
$\ccal\den{ \ }$ from formulae--e.g.,  we write $\Kfunapp{i}{\phi}$ for
the constraint $\Kfunapp{i}{\,\ccal\den{\phi} \,}.$
\end{notation}
}

\hltext{Following our intended meaning of constraints, we think of $\Kfunapp{i}{\phi}$
as stating that  $\phi$ holds in the space of agent $i$, or as an epistemic
assertion stating that agent $i$ considers/believes $\phi$ to be true.}

\hltext{
\subsection{Continuity}
\resp{We reorganized the next two paragraphs.}
In~\cite{knight:hal-00761116} space functions were not required to be
continuous. Nevertheless, we will argue later, in Remark~\ref{remark:cont},
that continuity  comes naturally in the intended phenomena we wish to capture:
modeling information of possibly \emph{infinite} groups.  In fact,
in~\cite{knight:hal-00761116} scs could only have finitely many agents. 
In this work we also extend scs to allow  arbitrary, \emph{possibly infinite}, sets
of agents.  We illustrate scs with infinite groups in the next section.

Notice that the continuity and preservation of finite joins by space functions
will provide us with their preservation of arbitrary joins.  The following
proposition  gives us sufficient conditions for the existence of the join of
an arbitrary set on a given poset.  A sketch of its proof is presented
in~\cite{johnstone1982stone}, for the sake of completeness we present our own
proof of it. }

\begin{proposition}[\cite{johnstone1982stone}]
\label{prop:arb-join}
Let  $(\Poset,\cleq)$ be a poset. Suppose that  $\bigjoin F$ and $\bigjoin D$
exist for every finite set $F\subseteq \Poset$ and for every directed set
$D\subseteq \Poset.$
Then  $\bigjoin A$ also exists for every $A\subseteq \Poset.$
\end{proposition}

\begin{proof}
Let $A \subseteq \Poset$ be an arbitrary set and let $D =
\left\{\bigjoin F \mid F \subseteq A \text{ and } F \text{ is finite}
\right\}$. First we prove that $D$ is directed.
For $F_1, F_2 \subseteq A$, both finite sets, the elements
$\bigjoin F_1, \bigjoin F_2 \in D$ and the set $F_3 = F_1 \cup F_2
\subseteq A$ is finite. Then $\bigjoin F_3 \in D$ and, both $\bigjoin 
F_1 \cleq \bigjoin F_3$ and $\bigjoin F_2 \cleq \bigjoin F_3$ hold.

To complete the proof we show that for every $c \in \Poset$:
\[
c  \text{ is an upper bound of } D \text{ if and only if $c$ is 
an upper bound of } A.
\]
\noindent For the ``only if'' direction, we prove its contrapositive. 
Let $c \in \Poset$. If $c$ is not an upper bound of $A$, there is an
$a \in A$ such that $a \not\cleq c$. Notice that $\{a\} \subseteq A$ and thus
$\bigjoin \{a\} = a \in D$ but $a \not\cleq c$. Hence $c$ is not an upper bound of $D$.

\noindent  For the other direction, assume that $c \in \Poset$ is an 
upper bound of $A$. Let $F$ be any finite subset of $A$. Since
$e \cleq c$ for every $e \in F$, then $\bigjoin F \cleq c$. Therefore,
$c$ is an upper bound of $D$. 
 
 We then conclude  $\bigjoin A = \bigjoin D$ as wanted.
\end{proof}

The following proposition states two useful properties of space functions:
monotonicity and preservation of arbitrary joins. 

\begin{proposition}[\cite{guzman-algebraic-2020}]
\label{prop:join-preservation}
Let $f: \C \to \C$ be a function over a cs $(\Con, \cleq)$.  Then
\begin{enumerate}[1.]
\item  If $f$ is space function then $f$ is monotonic.
\item  $f$ is space function if and only if it preserves arbitrary joins.
\end{enumerate}
\index{Arbitrary join preservation}
\index{Space function!Monotonic}
\index{Space function!Arbitrary join preservation}
\end{proposition}

\section{Distributed Information}
\label{sec:dist-info}

This section contains the main technical contributions of this paper. In
particular, we will  characterize the notion of collective information of a
group of agents. Roughly speaking,  the \emph{distributed (or collective)
information} of a group $I$ is the join of each piece of information that
resides in the space of an agent $i \in I$.  For each constraint $c$, the
distributed information of $I$ w.r.t $c$ is the distributive information of
$I$ that can be derived from $c$. We wish to formalize whether a given
constraint $e$ can be derived from the collective information of the group $I$
w.r.t $c$. 

The following examples, which we will use throughout this paper, illustrate
the above intuition. 

\begin{example}
\label{ds-example}
Consider an scs $({\Con},\cleq, (\sfun{i})_{i \in G})$ where $G = \nmat$ and
$(\Con, \cleq)$ is a constraint frame. Let $c = \sfunapp{1}{a} \join
\sfunapp{2}{b \sop a} \join \sfunapp{3}{e \sop b}$. The constraint $c$ specifies
the situation where $a, b \sop a$ and $e \sop b$ are in the spaces of agent $1$,
$2$ and $3$, respectively. Neither agent necessarily holds $e$ in their space
w.r.t $c$. Nevertheless, the information $e$ can be derived from the
collective information of the three agents w.r.t $c$, since from
Prop.~\ref{prop:subtraction} we have $a \join (b \sop a) \join (e \sop b) \cgeq
e$.
\hltext{Let us now consider an example with infinitely many agents.
\resp{We changed this paragraph.}

\emph{Infinitely Many Agents.}
Suppose that there exists an element $e' \in \Con$ and an increasing chain $a_0 \cleq a_1 \cleq \cdots$ 
such that  $\bigjoin_{i \in \nmat} {a_i}\cgeq e'$ and $e' \not\cleq a_i $ for every $i \in \nmat$. Therefore $e'$  is not \emph{compact} (see
Def.~\ref{def:c-lat}).  Let $c'
\defsymbol \bigjoin_{i \in \nmat} \sfunapp{i}{a_i}$. Notice that for no agent $i \in \nmat$ holds (or can derive)
$e'$ in their space since $e' \not\cleq a_i$. Yet, from
our assumption, $e'$ can be derived from the collective information w.r.t
$c'$ of all the agents in $\nmat,$ i.e., $\bigjoin_{i \in \nmat} {a_i} \cgeq e'$.} \qed
\end{example}

The above example may suggest that distributed information can be obtained by
joining individual local information derived from $c$. Such information can be
characterized as the $i$-projection of agent $i$ w.r.t $c$.

\begin{definition}[Agent and Join Projections]
\label{def:agent-proj}

Let $({\Con},\cleq,(\sfun{i})_{i \in G})$ be an scs. Given $i \in G$, the
\emph{$i$-agent projection} of $c \in \Con$ is defined as $\pfunapp{i}{c}
\defsymbol \bigjoin \{ e \mid c  \cgeq \sfunapp{i}{e} \}$. We say that $e$ is
\emph{$i$-agent derivable} from $c$ if and only if $\pfunapp{i}{c} \cgeq e$.
Given $I \subseteq G$ the \emph{$I$-join projection} of a group $I$ of $c$ is
defined as $\pfunapp{I}{c} \defsymbol \bigjoin\{ \pfunapp{i}{c} \mid i\in I
\}$. Similarly, we say that $e$ is \emph{$I$-join derivable} from $c$ if and
only if $\pfunapp{I}{c} \cgeq e$.
\end{definition}
\index{Agent projection}
\index{Join projection}
\index{$i$-agent derivable}
\index{$I$-join projection}
\index[operator]{$\pfunapp{i}{c}$, $i$-agent projection of $c$}
\index[operator]{$\pfunapp{I}{c}$, $I$-join projection of a group $I$ of $c$}

The $i$-agent projection of $i \in G$ of $c$ naturally represents the join of 
all the information that agent $i$ has in $c$. The $I$-join projection of
group $I$ joins individual $i$-agent projections of $c$ for $i \in I$. This
projection can be used as a sound mechanism for reasoning about
distributed-information: If $e$ is $I$-join derivable from $c$ then it follows
from the  distributed-information of $I$ w.r.t $c$.

\begin{example}
\label{ds-example-join}

Let $c$ be as in Ex.~\ref{ds-example}. We have $\pfunapp{1}{c} \cgeq a$,
$\pfunapp{2}{c} \cgeq (b \sop a)$ and $\pfunapp{3}{c} \cgeq (e \sop b)$.
Indeed $e$ is $I$-join derivable from $c$ since $\pfunapp{\{1,2,3\}}{c} =
\pfunapp{1}{c} \join \pfunapp{2}{c}\join  \pfunapp{3}{c} \cgeq e$. Similarly,
we conclude that $e'$ is $I$-join derivable from $c'$ in Ex.~\ref{ds-example}
since $\pfunapp{\nmat}{c'} = \bigjoin_{i \in \nmat}\pfunapp{i}{c}\cgeq
\bigjoin_{i \in \nmat} {a_i} \cgeq e'$.\qed
\end{example}

Nevertheless, $I$-join projections do not provide a complete mechanism for
reasoning about distributed information as illustrated below.

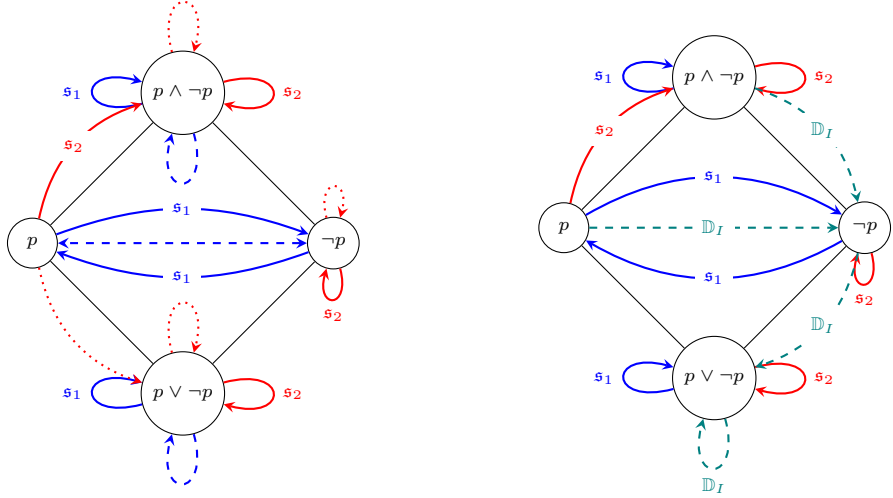
\begin{figure}[t]
\centering
\begin{subfigure}[t]{0.45\textwidth}
\centering
\begin{tikzpicture}[scale=0.5,>=stealth]
  \tikzstyle{every node}=[font=\scriptsize]
  \node[shape = circle, draw] (A) at (0,-4)   {$p \vee \neg p$};
  \node[shape = circle, draw] (B) at (-4,0)   {$\; p\; $};
  \node[shape = circle, draw] (C) at (4,0)    {$\neg p$};
  \node[shape = circle, draw] (D) at (0,4)    {$p \wedge \neg p$};

  \draw (A) to node {} (B);
  \draw (A) to node {} (C);
  \draw (B) to node {} (D);
  \draw (C) to node {} (D);

  \draw [blue,thick,->,bend left,out=20,in=160] (B) to node[fill=white] {$\sfun{1}$} (C);
  \draw [blue,thick,->,bend left,out=20,in=160] (C) to node[fill=white] {$\sfun{1}$} (B);
  \draw [blue,thick,->,loop left]    (A) to node             {$\sfun{1}$} (A);
  \draw [blue,thick,->,loop left]    (D) to node 	         {$\sfun{1}$} (D);

  \draw [red,thick,->,loop right]    (D) to node             {$\sfun{2}$} (D);
  \draw [red,thick,->,bend left]     (B) to node[fill=white] {$\sfun{2}$} (D);
  \draw [red,thick,->,loop right]    (A) to node             {$\sfun{2}$} (A);
  \draw [red,thick,->,loop below]    (C) to node             {$\sfun{2}$} (C);

  \draw [dashed,blue,loop below,->,thick]     (A) to node    {} (A);
  \draw [dashed,blue,loop below,->,thick]     (D) to node    {} (D);
  \draw [dashed,blue,<->,thick]       		  (C) to node    {} (B);

  \draw [dotted,loop above,red,->,thick]      (A) to node    {} (A);
  \draw [dotted,loop above,red,->,thick]      (D) to node    {} (D);
  \draw [dotted,loop above,red,->,thick]      (C) to node    {} (C);
  \draw [dotted,bend right,red,->,thick]      (B) to node    {} (A);
\end{tikzpicture}
\caption{Projections $\pfun{1}$ (dashed) and $\pfun{2}$ (dotted) given $\sfun{1}$ and $\sfun{2}$.}
\label{projection:counter-example}
\end{subfigure}
~
\begin{subfigure}[t]{0.45\textwidth}
\centering
\begin{tikzpicture}[scale=0.5,>=stealth]
  \tikzstyle{every node}=[font=\scriptsize]
  \node[shape = circle, draw] (A) at (0,-4)   {$p \vee \neg p$};
  \node[shape = circle, draw] (B) at (-4,0)   {$\; p\; $};
  \node[shape = circle, draw] (C) at (4,0)    {$\neg p$};
  \node[shape = circle, draw] (D) at (0,4)    {$p \wedge \neg p$};

  \draw (A) to node {} (B);
  \draw (A) to node {} (C);
  \draw (B) to node {} (D);
  \draw (C) to node {} (D);
  \draw [blue,thick,->,bend left]  (B) to node[fill=white] {$\sfun{1}$} (C);
  \draw [blue,thick,->,bend left]  (C) to node[fill=white] {$\sfun{1}$} (B);
  \draw [blue,thick,->,loop left]  (A) to node             {$\sfun{1}$} (A);
  \draw [blue,thick,->,loop left]  (D) to node             {$\sfun{1}$} (D);

  \draw [red,thick,->,loop right] (D) to node              {$\sfun{2}$} (D);
  \draw [red,thick,->,bend left]  (B) to node[fill=white]  {$\sfun{2}$} (D);
  \draw [red,thick,->,loop right] (A) to node              {$\sfun{2}$} (A);
  \draw [red,thick,->,loop below] (C) to node              {$\sfun{2}$} (C);

  \draw [teal,dashed,loop below,->,thick]  (A) to node   	        {$\Dfun{I}$} (A);
  \draw [teal,dashed,bend left,->,thick]   (D) to node[fill=white]  {$\Dfun{I}$} (C);
  \draw [teal,dashed,bend left,->,thick]   (C) to node[fill=white]  {$\Dfun{I}$} (A);
  \draw [teal,dashed,->,thick]             (B) to node[fill=white]  {$\Dfun{I}$} (C);
\end{tikzpicture}
\caption{$\Dfun{I}$ with $I=\{1,2\}$ given 
$\sfun{1}$ and $\sfun{2}$.}
\label{fig:difficult-small}
\end{subfigure}
\caption{Projections (see Def.~\ref{def:agent-proj})
(a) and distributed space function (see Def.~\ref{ds:def})
(b) over four-element boolean algebra.}
\label{fig:projections-delta}
\end{figure}

\begin{example}
\label{ds:counter-example}
Let $d \defsymbol \sfunapp{1}{b} \meet \sfunapp{2}{b}$. Recall that we think
of $\join$ and $\meet$ as conjunction and disjunction of assertions: $d$
specifies that $b$ is present in the space of agent $1$ or in the space of
agent $2$ though not exactly in which one. Thus from $d$ we should be able to
conclude that $b$ belongs to the space of \emph{some} agent in $\{ 1, 2 \}$.
Nevertheless,  $b$ is not necessarily  $I$-join derivable from $d$ since  from
$\pfunapp{\{1,2\}}{d} = \pfunapp{1}{d} \join \pfunapp{2}{d}$ we cannot, in
general, derive $b$.  To see this consider the scs in
Fig.~\ref{projection:counter-example}, taking $b = \neg p$. We have
$\pfunapp{\{1,2\}}{d} = \pfunapp{1}{d} \join \pfunapp{2}{d}=\true \join
\true=\true \not\cgeq b$.  One can generalize this example to infinitely many
agents. 
\resp{We changed this paragraph.}
\hltext{
\emph{Infinite Many Agents.} Consider the scs in Ex.~\ref{ds-example} and let $d' \defsymbol
\bigmeet_{i \in \nmat}\sfunapp{i}{b'}$. We should be able to conclude from
$d'$ that $b'$ is in  the space of \emph{some} agent in $\nmat$ but, in
general, $b'$ is not $\nmat$-join derivable from $d'$.}\qed
\end{example}

\subsection{Distributed Spaces}
\label{ssec:ds}

We have just illustrated in Ex.~\ref{ds:counter-example} that the $I$-join
projection of $c$, $\pfunapp{I}{c}$,  the join of individual projections, may
not project all distributed information of a group $I$.  To solve this problem
we  develop the notion of $I$-group projection of  $c$, written as
$\Pfunapp{I}{c}$. We will first define a space function $\Dfun{I}$ called the
distributed space of group $I$. The function $\Dfun{I}$ can be thought of as a
virtual space including all the information that  can be in  the space of a
member of $I.$  We will then define an $I$-projection, $\Pfun{I}$, in terms of
$\Dfun{I}$ much like $\pfun{i}$ is defined in terms of $\sfun{i}$.

\subsubsection*{Set of Space Functions}

We now introduce a new partial order induced by $\C$: The set of space
functions ordered point-wise. Recall that $\scal(\C)$ denotes the set of all
space functions over a cs $\C$ (Def.~\ref{def:sfunc}). For notational
convenience, we shall use $(f_I)_{I \subseteq G}$ to denote the tuple
$(f_I)_{I \in \pcal(G)}$ of elements of $\scal(\C)$. 

\begin{definition}[Function Order]
\label{def:sfunc-order}

Let $({\Con},\cleq)$ be a cs. Given $f,g: \C \to \C$ define $f \fleq g$ iff
$f(c) \cleq g(c)$ for every $c \in \Con$.
\end{definition}
\index{Space function order}
\index[operator]{$\fleq$, space function order}
\index[operator]{$(\scal(\C),\fleq)$, set of space functions ordered by $\fleq$}

An important design aspect of our structure is that the set of space
functions $\scal(\C)$ can be made into a complete lattice. 

\begin{lemma}[\cite{gratzer-latjoinend-1958}]
\label{lemma:closure-space}
Let $({\Con},\cleq)$ be a cs.
Then $(\scal(\C),\fleq)$ is a complete lattice. 
\end{lemma}

In the next section we use the properties of $(\scal(\C),\fleq)$ to formalize
the distributed space of a group $I$ as the greatest space function below
every space function $\sfun{i}$ with $i \in I$.

\subsection{Distributed Spaces as Max Spaces}
\label{ssec:ds-maxsp}
We can now give the definition of distributed spaces.  It is convenient to
give the following intuition first.

\hltext{
\begin{remark}
\label{rmk:sf-notion}
\resp{We changed the next remark according to comment {\bf N38}.}
Suppose that $f$ and $g$ are space functions in $\Cs$. Intuitively, $f(c)$
means $c$ is within the space represented by $f$. By definition, $f \fleq g$
means $f(c) \cleq g(c)$ for every $c \in \Con$. Intuitively, from the assertion  
that $c$ is in the space represented by $g$ we can derive that  $c$ is also in the space
represented by $f$. This can be interpreted as saying that the space
represented by $g$ is included in the space represented by $f$; in other words
\emph{the bigger the space, the smaller the function that represents it}.
Thus every $c$ in $g$ is also in $f$, hence $f$ is a bigger space.\qed
\end{remark}
}

Following the above intuition, the order relation $\fleq $ of $\Cs$ represents
(reverse) space inclusion and  the join and meet operations in $\Cs$ represent
intersection and union of spaces. The biggest and the smallest spaces are
represented by the bottom  and the top  elements  of the lattice $\Cs$, here
called $\lambda_\bot$ and $\lambda_\top$, respectively, and defined as
follows.

\begin{definition}[Top and Bottom Spaces]
\label{lambda:def}
Let $\Cs$ be the lattice of space functions. Define $\lambda_\bot$ and
$\lambda_\top$ in $\Cs$ as follows:
$\lambda_\bot(c) \defsymbol \true$ for every $c \in \Con$; and 
$\lambda_\top(c)\defsymbol \true  \mbox{ if } c = \true$ and
$\lambda_\top(c) \defsymbol \false  \mbox{ if } c \neq \true$.
\index{Top space}
\index{Bottom space}
\index[operator]{$\lambda_\top$, top space}
\index[operator]{$\lambda_\bot$, bottom space}
\end{definition}

The distributed space $\Dfun{I}$ of a group $I$ can be viewed as the function
that represents the smallest space that includes all the local information of
the agents in $I$. From Remark~\ref{rmk:sf-notion}, $\Dfun{I}$ should be the
\emph{greatest space function} below the space functions of the agents in $I$.
The existence of such a function follows from completeness of
$(\scal(\C),\fleq )$ stated in Lemma~\ref{lemma:closure-space}.  

\begin{definition}[Distributed Space]
\label{ds:def}
\index{Distributed space}
Let  $({\Con},\cleq,(\sfun{i})_{i \in G})$ be an scs. The
\emph{distributed spaces} of $\C$ are given by $\Dfuntuple =
(\Dfun{I})_{I \subseteq G}$ where
\[
\Dfun{I} \defsymbol \max \left\{\  f \in \scal(\C) \mid f \fleq \sfun{i} 
\mbox{ for every } i \in I \ \right\}.
\]

\noindent We shall say that \emph{$e$ is distributed  among $I\subseteq G$ w.r.t $c$} 
if and only if $c \cgeq \Dfunapp{I}{e}$.
We shall refer to each $\Dfun{I}$ as the \emph{(distributed) space}
of the group $I$.
\index[operator]{$\Dfuntuple = (\Dfun{I})_{I \subseteq G}$, distributed spaces}
\index[operator]{$\Dfun{I}$, distributed space of the group $I$}
\end{definition}

\begin{remark}
\label{rmk:ds-meet}
From Lemma~\ref{lemma:closure-space}, 
$\Dfun{I}=\bigjoin_{\scal(\C)} \left\{\  f \in \scal(\C) \mid f \fleq \sfun{i} 
\mbox{ for each } i \in I \ \right\} = \bigmeet_{\scal(\C)}\{ \sfun{i} \mid i \in I \}$ where $\bigjoin_{\scal(\C)}$  and $\bigmeet_{\scal(\C)}$ are the join and meet
in the complete lattice  $(\scal(\C),\fleq )$.\qed
\end{remark}

Let us consider a concrete example.

\begin{example}
Fig.~\ref{fig:difficult-small} illustrates an scs with space  functions
$\sfun{1}$ and $\sfun{2}$, and their distributed space $\Dfun{\{1,2\}}$. The
reader can verify that $\Dfun{\{1,2\}}$ is indeed the greatest function such
that $\Dfun{\{1,2\}}\fleq \sfun{1}$ and $\Dfun{\{1,2\}}\fleq \sfun{2}$. 
Notice that $\sfunapp{1}{p} \join \sfunapp{2}{\neg p}\cgeq \Dfunapp{\{1,2\}}{p
\join \neg p} = \Dfunapp{\{1,2\}}{\false}$ meaning that if agents $1$ and $2$
had $p$ and $\neg p$ in their corresponding spaces, as a group they could
derive an inconsistency.\qed 
\end{example}

\resp{We added the next paragraph and example to address {\bf N41}.}
\hltext{
In~\cite{knight:hal-00761116}, shared information of a group $I$ specifies the
fact that a given constraint $e$ resides within the space of every agent in a
group $I$. In contrast, distributed information specifies that $e$ is
distributed among the members of $I$. The next example illustrates this
difference. 

\begin{example}
\label{ex:shared-dist}
Let $I = \{1,\ldots,n\}$ and let $\sfunapp{I}{e} \defsymbol \bigjoin_{i \in I}
\sfunapp{i}{e}$ as in~\cite{knight:hal-00761116}. Here $\sfunapp{I}{e}$ specifies that  $e$ is shared
by the agents in $I$: $e$ is present in the space of $i$ for every $i \in I$. Instead, $\Dfunapp{I}{e}$
means that the information $e$ is distributed among the spaces of the members
in $I$. To see the difference, let $I =\{1,2\}$ and consider the scs in
Fig.~\ref{fig:difficult-small}. Let $c = \sfunapp{1}{p} \join \sfunapp{2}{\neg
p}$. Notice that $p \wedge \neg p$ is distributed in $I$ w.r.t $c$: $c \cgeq
\Dfunapp{I}{p \wedge \neg p}$. However, $p \wedge \neg p$ is not shared
information of $I$ w.r.t $c$, namely, $c \not\cgeq \sfunapp{I}{p \wedge \neg p}$.
\qed
\end{example}
}

\subsection{Compositionality of Distributed Spaces}
\label{ssec:comp-ds}

Distributed spaces have pleasant compositional
properties. They capture the intuition that the \emph{distributed information}
of a  group $I$ can be obtained from the the distributive information of its
subgroups.  

\begin{proposition}
\label{prop:comp}
Let $(\Dfun{I})_{I \subseteq G}$ be the distributed spaces of an scs
$(\Con,\cleq,(\sfun{i})_{i\in G})$. Suppose that $K,J \subseteq I\subseteq G$.
\begin{enumerate}
\item $\Dfun{I} = \lambda_\top $ if  $I = \emptyset$.

\item $\Dfun{I}=\sfun{i}\ $  if $I=\{ i \}$.

\item $\Dfunapp{J}{a} \join \Dfunapp{K}{b} \cgeq  \Dfunapp{I}{a \join b}$.

\item $\Dfunapp{J}{a} \join \Dfunapp{K}{c \sop a} \cgeq  \Dfunapp{I}{c}$ if
$({\Con},\cleq)$ is a constraint frame. 
\end{enumerate}
\end{proposition}

\begin{proof}
\begin{enumerate}
\item It follows directly from Def.~\ref{lambda:def} and Def.~\ref{ds:def}.

\item Let $I = \{i\}$, from Def.~\ref{ds:def},
$\Dfun{I} = \max\{f \in \scal(\C) \mid f \fleq \sfun{i}\} = \sfun{i}$.

\item Assume $K,J \subseteq I.$ From Def.~\ref{ds:def} we conclude $\Dfun{I} \fleq
\Dfun{J}$ and  $\Dfun{I} \fleq \Dfun{K}$. Thus $\Dfunapp{J}{a} \cgeq
\Dfunapp{I}{a}, \Dfunapp{K}{b} \cgeq \Dfunapp{I}{ b}$ and therefore $\Dfunapp{J}{a}
\join \Dfunapp{K}{b} \cgeq \Dfunapp{I}{a} \join \Dfunapp{I}{b}$.  Since $\Dfun{I}$
is a space function, $ \Dfunapp{I}{a} \join \Dfunapp{I}{b} = \Dfunapp{I}{a \join b}$,
then we obtain $\Dfunapp{J}{a} \join \Dfunapp{K}{b} \cgeq  \Dfunapp{I}{a \join b}$
as wanted.

\item It follows from part (3) with $a = a$ and $b = c \sop a$, and
Prop.~\ref{prop:subtraction}.
\end{enumerate}
\end{proof}

Recall that $\lambda_\top$  corresponds to the empty space (see
Def.~\ref{lambda:def}). The first property realizes the intuition that the
empty subgroup $\emptyset $ \emph{does not} have any information whatsoever
distributed w.r.t a consistent $c$: for if $c \cgeq \Dfunapp{\emptyset}{e}$
and $c \neq \false$  then $e = \true$. Intuitively, the second property says
that the function $\Dfun{I}$ for the group of one agent must be the agent's
space function. The third property states that a group can join the
information of its subgroups. The last property uses subtraction, hence the
constraint frame condition, to express that by joining the information $a$ and
$c \sop a$ of their subgroups, the group $I$ can obtain $c$.

Let us illustrate how to derive information of a group from smaller ones using
Prop.~\ref{prop:comp}. 

\begin{example}
\label{ds-example-2}
Let $c = \sfunapp{1}{a} \join \sfunapp{2}{b \sop a} \join \sfunapp{3}{e \sop
b}$ as in Ex.~\ref{ds-example}. We want to prove that $e$ is distributed among
$I=\{1,2,3\}$ w.r.t $c$, i.e., $c \cgeq \Dfunapp{\{1,2,3\}}{e}$. Using
Properties (2) and (4) in Prop.~\ref{prop:comp} we obtain $ c \cgeq
\sfunapp{1}{a} \join \sfunapp{2}{b \sop a} = \Dfunapp{\{1\}}{a} \join
\Dfunapp{\{2\}}{b \sop a} \cgeq \Dfunapp{\{1,2\}}{b}$, and then $c \cgeq
\Dfunapp{\{1,2\}}{b} \join \sfunapp{3}{e \sop b} = \Dfunapp{\{1,2\}}{b} \join 
\Dfunapp{\{3\}}{e \sop b} \cgeq \Dfunapp{\{1,2,3\}}{e}$ as wanted. 
\end{example}

\begin{remark}[Continuity and Infinitely Many Agents]
\label{remark:cont}
\index{Continuity}
The example with infinitely many agents in Ex.~\ref{ds-example} illustrates
well why we require our spaces to be continuous in the presence of possibly
infinite groups. Clearly $c' = \bigjoin_{i \in \nmat} \sfunapp{i}{a_i} \cgeq
\bigjoin_{i \in \nmat}\Dfunapp{\nmat}{a_i}$. By continuity, $ \bigjoin_{i \in
\nmat}\Dfunapp{\nmat}{a_i} = \Dfunapp{\nmat}{\bigjoin_{i \in {\nmat}}a_i}$
which indeed captures the idea that each $a_i$ is in the distributed space
$\Dfun{\nmat}$.\qed
\end{remark}

\subsection{Distributed Spaces in Aumann and Kripke Structures}
\label{ssec:ds-aumann}

We now consider an important structure from mathematical economics used for
group epistemic reasoning: Aumann structures~\cite{halpern2004reasoning}. We
illustrate that the notion of  distributed knowledge in these structures is an
instance of a distributed space.

\begin{example}
\label{aumann:example}
\resp{We added a footnote in this example to include a comment in {\bf N43}}
\emph{Aumann Constraint Systems~\cite{knight:hal-00761116}.}
Aumann structures are an \emph{event-based}
approach to modelling knowledge. An Aumann structure is a tuple
$\acal = (S, \pcal_1, \dots, \pcal_n)$ where $S$ is a set of states and
each $\pcal_i$ is a partition on $S$ for agent $i$~\cite{halpern2004reasoning}.
The sets of each partition $\pcal_i$ are called \emph{information sets}. If two states $t$ and $u$
are in the same information set for agent $i$, it means that in state $t$
agent $i$ considers state $u$ possible, and vice versa.
An \emph{event} in an Aumann structure is any subset of $S$. Event $e$ holds
at state $t$ if $t\in e$. The set $\pcal_i(s)$ denotes the information set
of $\pcal_i$ containing $s$.
The event of \emph{agent $i$ knowing $e$} is defined as
\[
\Kop{i}{e} = \left\{ s \in S \mid \pcal_i(s) \subseteq e\right\},
\]
and the \emph{distributed knowledge of an event $e$ among the agents in a
group $I  \subseteq \{ 1,\ldots, n\}$} is defined as
\[
\Gop{I}{e} = \left\{s \in S \mid \bigcap_{i\in I} \pcal_i(s) \subseteq e\right\}.
\]
An Aumann structure induces the spatial constraint system
$\C(\acal)$ with events as constraints, i.e., $\Con = \{e \mid e
\mbox{ is an event in } \acal\}$, and for every $e_1, e_2 \in \Con$, $e_1
\cleq e_2$ iff $e_2 \subseteq e_1$. The operators join $(\join)$ and meet
$(\meet)$ are intersection $(\cap)$ and union $(\cup)$ of events,
respectively; $\true = S$ and $\false = \emptyset$\footnote{\hltext{Notice that in
this cs we use the reverse inclusion which is the dual of the powerset example in  Ex.~\ref{ex:powerset}. Then, the join and meet operators and, the top and
bottom elements, are swapped.}}. The space functions are the knowledge operators, i.e.,
$\sfunapp{i}{c}= \Kop{i}{c}$. 

\index{Aumann constraint system}
\index[operator]{$\Kop{i}{e}$, agent $i$ knows $e$}
\index[operator]{$\Gop{I}{e}$, distributed knowledge of $e$ in group $I$}
\index[operator]{$\acal = (S, \pcal_1, \dots, \pcal_n)$, Aumann structure}
\index[operator]{$\pcal_i$, information set}

\hltext{The next proposition states that in fact distributed knowledge and
distributed information coincide.}
\resp{We added the next proposition to address comment {\bf N43}.\\
IMPORTANT: Its proof uses one of the main results of this paper
--Th.~\ref{thm:delta-ast}.}

\hltext{
\begin{proposition}[Distributed Spaces in Aumman Structures]
\label{prop:aumann-dist}
Let $\acal = (S, \pcal_1, \dots, \pcal_n)$  be an Aumann structure and let 
$\mathbf{C}(\acal)$ be its induced scs. Then $\Dfun{I} =
\Gfun{I}$ for every $I \subseteq \{ 1,\ldots, n\}$.
\end{proposition}
}

\hltext{Therefore distributed knowledge in Aumann structures is indeed an instance of 
distributed information. The proof uses one of the main results of this paper
(Th.~\ref{thm:delta-ast}). For the sake of the presentation, we postpone it
to Sec.~\ref{ssec:proof-aumann-dist}.}\qed 
\end{example}

\hltext{The next example shows that distributed knowledge in epistemic logic is also
an instance of distributed information.}
\resp{We added the next example.}

\hltext{
\begin{example}[Distributed Spaces in Kripke Structures]
\label{ex:kripke-ds}
Distributed knowledge in logic is expressed with the formula
$\Dkform{I}{(\phi)}$ whose intended meaning is that the knowledge $\phi$ is
distributed among the members of $I$. Its semantics is expressed in terms of
its Kripke models; i.e., the models of $\Dkform{I}{(\phi)}$ are given by
$\Dkfunapp{I}{X}$ defined next where $X$ are the models of
$\phi$~\cite{fagin1995reasoning}.

Let ${\bf K}(\mathcal{S}_n(\Phi)) = (\C,\sqsubseteq,\Kfun{1},\ldots,\Kfun{n})$
be a Kripke $n$-scs as in Def.~\ref{def:kripke-scs}. The function
$\Dvar : \C \to \C$ is defined as
\[
\Dkfunapp{I}{X} \defsymbol \left\{ (M,s) \in \Delta \mid \forall t: (s,t) \in
\bigcap_{i \in I} \rcal_i \mbox{ implies } (M,t) \in X \right\}.
\]
The following proposition states that $\Dfun{I} = \Dkfun{I}$.
As for Prop.~\ref{prop:aumann-dist}, we use Th.~\ref{thm:delta-ast} to
prove it. The proof is given in Sec.~\ref{ssec:proof-kripke-dist}.
\resp{We added the next proposition.\\
IMPORTANT: Its proof uses one of the main results of this paper
--Th.~\ref{thm:delta-ast}.}

\begin{proposition}[Distributed Spaces in Kripke Structures]
\label{prop:kripke-dist}
Let $\kbold(\scal_n(\Phi)) = (\Con,\sqsubseteq,\Kfun{1},\ldots,\Kfun{n})$ be
a Kripke $n$-scs as in Def.~\ref{def:kripke-scs}. Then $\Dfun{I} =
\Dkfun{I}$ for every $I \subseteq \{1,\ldots,n\}$.
\end{proposition}

Hence  we can extend the semantics in Def.\ref{modal-interpretation} to distributed knowledge formulae $\Dkform{I}{(\phi)}$ as 
$\ccal \den{ \Dkform{I}{(\phi)} }  =  \Dfunapp{I}{\ \ccal \den{\phi}\ }.$
\qed 
\end{example}
}

In Prop.~\ref{prop:comp} we listed some useful properties about 
$(\Dfun{I})_{I \subseteq G}$. In the next section we shall see that
$(\Dfun{I})_{I \subseteq G}$  is the greatest solution of three  basic
properties.  

\subsection{Distributed Spaces as Group Distributions Candidates.}
\label{ssec:ds-gdc}

We now wish to single out a few fundamental properties on tuples  of self-maps
that can be used to characterize distributed spaces. 

\begin{definition}[Distribution Candidates]
\label{def:gdc} Let $({\Con},\cleq,(\sfun{i})_{i \in G})$ an scs.
A \emph{group distribution candidate} (gdc) of $\C$ is a tuple
$\dfuntuple = (\dfuntuple_I)_{I\subseteq G}$ of self-maps on $\Con$ such that
for each $I,J \subseteq G$:

(D.1) $\dfuntuple_I$ is a space function in $\C$,
\index{D.1, first group distribution candidate axiom}

(D.2) $\dfun{I}  = \sfun{i}$ if $I = \{ i\}$,
\index{D.2, second group distribution candidate axiom}

(D.3) $\dfun{I} \fgeq  \dfun{J}$ if $ I \subseteq J$.
\index{D.3, third group distribution candidate axiom}
\end{definition}
\index{Distribution candidate}
\index[operator]{$\dfuntuple = (\dfuntuple_I)_{I\subseteq G}$, group distribution candidate, gdc}

Property D.1 requires each $\dfun{I}$ to be a space function. This is
trivially met for $\dfun{I} = \Dfun{I}$. Property D.2 says that the function
$\dfun{I}$ for a group of one agent must be the agent's space function. 
Clearly, $\dfun{\{i \}} = \Dfun{\{ i \}}$ satisfies D.2; indeed the
distributed space of a single agent is their own space. Finally, Property D.3
states that $\dfun{I}(c) \cgeq  \dfun{J}(c)$, if $ I \subseteq J$. This is
also trivially satisfied if we take $\dfun{I} = \Dfun{I}$ and $\dfun{J} =
\Dfun{J}$. Indeed if a group $I$ has some distributed information $c$ then
any group $J$, that includes $I$, should also have $c$. This realizes
the intuition in Remark~\ref{rmk:sf-notion}: The bigger the group, the bigger
the space and thus the smaller the space function that represents it. 

Properties D.1-D.3, however,  do not determine $\Dfuntuple$ uniquely. In fact,
there could be infinitely-many tuples of  space functions that satisfy them.
For example,  if we were to chose $\dfun{\emptyset} = \lambda_\top,$ $\dfun{\{
i \}}= \sfun{i}$ for every $i \in G$, and $\dfun{I}= \lambda_\bot$ whenever $|
I | > 1 $ then D.1-D.3 would be trivially met. But these space
functions would not capture our intended meaning of distributed spaces: E.g.,
we would have $\true \cgeq  \dfunapp{I}{e}$ for every $e$ thus implying that
any $e$ would be distributed in the empty information $\true$ amongst the
agents in $I \neq \emptyset$.

Nevertheless, we prove that $(\Dfun{I})_{I \subseteq G}$
is the greatest solution satisfying D.1-D.3.

\begin{theorem}[Max gdc]
\label{th:max-gdc}
\index{Maximum group distribution candidate, max gdc}
Let  $(\Dfun{I})_{I \subseteq G}$ be the distributed spaces of
$({\Con},\cleq,(\sfun{i})_{i \in G})$.
Then
\begin{enumerate}
\item $(\Dfun{I})_{I \subseteq G}$ is a gdc of $\C$.
\item If $(\dfuntuple_I)_{I\subseteq G}$ is a gdc of $\C$ then
$\dfuntuple_I  \fleq  \Dfun{I}$ for each $I \subseteq G$.
\end{enumerate}
\end{theorem}

\begin{proof}
Let  $(\Dfun{I})_{I \subseteq G}$ be the distributed spaces of $\C$.

\begin{enumerate}
\item We need to prove that $(\Dfun{I})_{I \subseteq G}$ satisfies  properties
D.1-D.3 in Def.~\ref{def:gdc}.

Property D.1 follows from definition of $\Dfun{I}$ (see Def.~\ref{ds:def}).
Property D.2 is proven in Prop.~\ref{prop:comp} part (2). For property D.3,
let $I,J \subseteq G$ such that $I \subseteq J$. Notice that $\{ f \in
\scal(\C) \mid f \fleq \sfun{i} \mbox{ for every } i \in I \} \subseteq \{ f
\in \scal(\C) \mid f \fleq \sfun{j} \mbox{ for every } j \in J \}$.
Then $\Dfun{J} \fleq \Dfun{I}$.

\item Since $(\dfuntuple_I)_{I\subseteq G}$ is a gdc, we have $\dfuntuple_I
\fleq \dfuntuple_{\{i\}} = \sfun{i}$, for every $i \in I$.
Then $\dfuntuple_I \in \{ f \in \scal(\C) \mid f \fleq \sfun{i} \mbox{ for every }
i \in I \}$ which implies $\dfuntuple_I \fleq \Dfun{I}$.
\end{enumerate}
\end{proof}

Therefore, Th.~\ref{th:max-gdc} tells us that distributed spaces could have
been equivalently defined as the \emph{greatest space functions} satisfying
Properties D.1-D.3. 
We shall use the characterization of distributed spaces in
Th.~\ref{th:max-gdc} in the proofs of Prop.\ref{projection:prop} and
Th.\ref{thm:comp-algo} in the next sections.  Let us first illustrate the
use of such properties in the following example.
\resp{We updated the next example according to {\bf N44}.}

\hltext{
\begin{example}
\label{ex:delta-func2}

Let $(\C,\cleq,(\sfun{i})_{i \in G})$ and $c = \sfunapp{1}{a} \join
\sfunapp{2}{b \sop a} \join \sfunapp{3}{e \sop b}$ as in Ex.~\ref{ds-example}. 
Here we shall show that $e$ can be derived from the distributed
information among $I = \{1,2,3\}$.

\noindent We want to prove $c \cgeq \Dfunapp{I}{ e }$ for $I = \{ 1,2,3 \}$.
Since $\C$ is a constraint frame, by Def.~\ref{def:gdc} and
Prop.~\ref{prop:subtraction} we have
\begin{align*}
c &= \Dfunapp{{\{1\}}}{a} \join \Dfunapp{{\{2\}}} {b \sop a} \join \Dfunapp{{\{3\}}}{e \sop b} \tag{Def.~\ref{def:gdc}~D.2}\\
&\cgeq \Dfunapp{I}{a} \join \Dfunapp{I}{b \sop a} \join \Dfunapp{I}{e \sop b}  \tag{Def.~\ref{def:gdc}~D.3}\\
&= \Dfunapp{I}{ a \join (b \sop a) \join (e \sop b) } \tag{Def.~\ref{def:gdc}~D.1}\\
&\cgeq \Dfunapp{I}{ e } \tag{Prop.~\ref{prop:subtraction}}.
\end{align*}
Thus $c \cgeq \Dfunapp{I}{e}$ as wanted.

\emph{Infinitely Many Agents.} Recall the case with infinitely many agents from
Ex.~\ref{ds-example}.
Similarly, we can show that  $e'$ is distributed in $\nmat$ w.r.t $c'$:

\[
c'
= \bigjoin_{i \in \nmat} \Dfunapp{\{i\}}{a_i}
\cgeq \bigjoin_{i \in \nmat} \Dfunapp{I}{a_i}
= \Dfunapp{I}{\bigjoin_{i \in \nmat} a_i}
\cgeq \Dfunapp{I}{e'}.
\]

Now consider our counter-example in Ex.~\ref{ds:counter-example} where we have
$d = \sfunapp{1}{b} \meet \sfunapp{2}{b}$. Recall that $d$ specifies that $b$
resides either within the space of agent 1 or in the space of agent 2,
although we do not know exactly in which one. Here we wish to prove that $b$
is distributed in the group $\{1,2\}$, i.e., $b$ can be derived from $d$ as
being in a space of a member of $\{1,2\}$.
We want to prove $d \cgeq \Dfunapp{I}{b}$ for $I = \{ 1,2 \}$:
\begin{align*}
d = \sfunapp{1}{b} \meet \sfunapp{2}{b} &= \Dfunapp{\{1\}}{b} \meet \Dfunapp{\{2\}}{b} \tag{Def.~\ref{def:gdc}~D.2}\\
&\cgeq \Dfunapp{\{1,2\}}{b} \meet \Dfunapp{\{1,2\}}{b} \tag{Def.~\ref{def:gdc}~D.3}\\
&=  \Dfunapp{\{1,2\}}{b} \tag{Prop. of $\meet$}
\end{align*}
Then $d \cgeq \Dfunapp{\{1,2\}}{b}$ as wanted. 

\emph{Infinitely Many Agents.} If we have $d' \defsymbol \bigmeet_{i \in \nmat}\sfunapp{i}{b'}$ we can show that $b'$
is distributed in the group $\nmat$ w.r.t $d'$:  $d' = \bigmeet_{i \in
\nmat}\Dfunapp{\{i\}}{b'} \cgeq \bigmeet_{i \in \nmat}\Dfunapp{\nmat}{b'} =
\Dfunapp{\nmat}{b'}$. \qed 
\end{example}
}

\hltext{The above example shows the capability of $\Dfun{I}$ to express the
distributed information of a group $I$ in different scenarios.
In the later sections we shall provide properties of $\Dfun{I}$ 
and its characterization for completely distributive lattices.}

\subsection{Group Projections}
\label{ssec:group-proj}

As promised at the beginning of Section~\ref{ssec:ds-maxsp} we now give a 
definition of \emph{Group Projection}. The function $\Pfunapp{I}{c}$ extracts
exactly all information that the group $I$ may have distributed w.r.t $c$.

\begin{definition}[Group Projection]
\label{group-projection}
\index{Group projection}

Let $(\Dfun{I})_{I \subseteq G}$ be the distributed spaces of an scs
$({\Con},\cleq,(\sfun{i})_{i \in G}).$  Given the set $I \subseteq G$, the
$I$-group projection of $c \in \Con$ is defined as $\Pfunapp{I}{c} \defsymbol
\bigsqcup \{ e \mid c  \cgeq \Dfunapp{I}{e} \}$.
We say that $e$ is $I$-group derivable from $c$ if and only if
$\Pfunapp{I}{c}\cgeq e$.
\index{$I$-group derivable}
\index{$I$-group projection}
\index[operator]{$\Pfunapp{I}{c}$, $I$-group projection of $c$}
\end{definition}

Much like space functions and agent projections, group projections and
distributed spaces  also form a pleasant correspondence:
a \emph{Galois connection}~\cite{davey2002introduction}. 

\begin{proposition}
\label{projection:prop}

Let  $(\Dfun{I})_{I \subseteq G}$ be the distributed spaces of an scs
$({\Con},\cleq,(\sfun{i})_{i \in G})$. For every $c,e\in \Con$,
\begin{enumerate}
\item $ c \cgeq \Dfunapp{I}{e} \mbox{ if and only if } \Pfunapp{I}{c} \cgeq e $.
\index{Galois connection}

\item $\Pfunapp{I}{c}\cgeq  \Pfunapp{J}{c}$ if $J \subseteq I$.

\item $\Pfunapp{I}{c}\cgeq  \pfunapp{I}{c}$.
\end{enumerate}
\end{proposition}

\begin{proof}

Let $(\Dfun{I})_{I \subseteq G}$ be the distributed spaces of $\C$
and let $c,e \in \Con$.

\begin{enumerate}
\item Let $S = \{d \mid c \cgeq \Dfunapp{I}{d}\}$. First, assume that
$c \cgeq \Dfunapp{I}{e}$. Since $e \in S$, by Def.~\ref{group-projection},
$\Pfunapp{I}{c} = \bigjoin S \cgeq e$.
Second, assume $\Pfunapp{I}{c} \cgeq e$. Then by monotonicity, 
$\Dfunapp{I}{\Pfunapp{I}{c}} \cgeq \Dfunapp{I}{e}$.
From continuity of $\Dfun{I}$, we know that $\Dfunapp{I}{\Pfunapp{I}{c}} =
\Dfunapp{I}{\bigjoin S} = \bigjoin\{ \Dfunapp{I}{d} \mid d \in S\}$ and by
definition of $S$, for every $d \in S$, we have $c \cgeq\Dfunapp{I}{d}$, then $c \cgeq \bigjoin\{ \Dfunapp{I}{d} \mid d \in S\}$.
Therefore, $c \cgeq \Dfunapp{I}{e}$.

\item Given that $(\Dfun{I})_{I \subseteq G}$ is a gdc (see
Th.~\ref{th:max-gdc}), if $J \subseteq I$, then
$\Dfun{J} \fgeq \Dfun{I}$. Hence $\{d\ |\ c \cgeq \Dfunapp{J}{d}\}
\subseteq \{d \mid c \cgeq \Dfunapp{I}{d}\}$ and thus
$\Pfunapp{I}{c} \cgeq \Pfunapp{J}{c}$ for every $c \in \Con$.

\item By part (2), for every $\{i\} \subseteq I$ and every $c \in \Con$,
we have $\Pfunapp{I}{c} \cgeq \Pfunapp{\{i\}}{c}$. It implies,
$\Pfunapp{I}{c} \cgeq \bigjoin_{i \in I} \Pfunapp{\{i\}}{c}$, for every
$c \in \Con$. Then
$\bigjoin_{i \in I}\Pfunapp{\{i\}}{c}=
\bigjoin_{i \in I} \{\bigjoin\{d\ |\ c \cgeq \Dfunapp{\{i\}}{d}\}\}=
\bigjoin_{i \in I} \{\bigjoin\{d\ |\ c \cgeq \sfunapp{i}{d}\}\}=
\bigjoin_{i \in I} \{\pfunapp{i}{c}\}=\pfunapp{I}{c}$. Therefore,
$\Pfunapp{I}{c} \cgeq  \pfunapp{I}{c}$, for every $c \in \Con$.
\end{enumerate}
\end{proof}

The first property in Prop.~\ref{projection:prop}, a Galois connection, states
that we can conclude from $c$ that $e$ is in the distributed space of $I$
exactly when $e$ is $I$-group derivable from $c$. The second says that the
bigger the  group, the bigger the projection. The last property says that
whatever is $I$-join derivable is $I$-group derivable, although the opposite 
is not true as shown in Ex.~\ref{ds:counter-example}.

\subsection{Group Compactness.}
\label{ssec:group-compact}

Suppose that an \emph{infinite} group of agents $I$ can derive $e$ from $c$
(i.e.,  $c \cgeq \Dfunapp{I}{e}$).  A legitimate question is whether there
exists a \emph{finite} sub-group $J$ of agents from $I$ that can also derive
$e$ from $c$. The following theorem provides a positive answer to this
question given that $e$ is a compact element (see Section \ref{sec:back})
and $I$-join derivable from $c$.

\begin{theorem}[Group Compactness]
\label{thm:compact}
\index{Group compactness}
Let  $(\Dfun{I})_{I \subseteq G}$ be the distributed spaces of an scs
$({\Con},\cleq,(\sfun{i})_{i \in G}).$ Suppose that  $c \cgeq \Dfunapp{I}{e}.$
If $e$ is compact and $I$-join derivable from $c$ then there exists a finite
set $J \subseteq I$ such that $c \cgeq \Dfunapp{J}{e}$.
\end{theorem}

\begin{proof}
Suppose that $c \cgeq \Dfunapp{I}{e}.$  If $I$ is finite then take $J=I$. If
$I$ is not finite, since $e$ is $I$-join derivable from $c$ we have
$\pfunapp{I}{c} = \bigjoin S \cgeq e$ where $S = \{ \pfunapp{i}{c} \mid {i \in
I} \}$.  

Define $D_I = \{  \pfunapp{J}{c} \mid J \subseteq I \mbox{ and } J \mbox{ is
finite} \}$. Take any $\pfunapp{H}{c}, \pfunapp{K}{c} \in D_I$. Since $H$ and
$K$  are finite, their union $K\cup H$ must also be finite and included in
$I$. Hence $\pfunapp{H\cup K}{c} \in D_I$. Therefore, $D_I$ is a directed set.

Since $S = \{ \pfunapp{i}{c} \mid {i \in I} \} = \{ \pfunapp{\{i\}}{c} \mid {i
\in I} \}$ is included in $D_I$, we obtain $\bigjoin D_I \cgeq \bigjoin S
\cgeq e$. But $e$ is compact and $D_I$ directed hence there must be
$\pfunapp{J}{c} \in D_I$, with $J$  a finite set, such that $\pfunapp{J}{c}
\cgeq e.$ From Prop.\ref{projection:prop} (3) and Prop.\ref{projection:prop}
(1), we conclude  $c \cgeq \Dfunapp{J}{e}$ as  wanted.
\end{proof}

Let us illustrate Th.~\ref{thm:compact} with our recurrent example.

\hltext{
\begin{example}
\resp{We rewrote this example.}
Consider an scs $({\Con},\cleq, (\sfun{i})_{i \in G})$ where $G = \nmat$ and $\C$
is a countable set.
Let $d = \bigjoin_{i \in \nmat} \sfunapp{i}{a_i}$ for some
increasing chain $a_0 \cleq a_1 \cleq \cdots$ , and $b \in \C$ such that
$b \cleq \bigjoin_{i \in \nmat} {a_i}$.

Notice that $d \cgeq \Dfunapp{\nmat}{b}$ and $\pfunapp{\nmat}{d} \cgeq b$.
Hence $b$ is $\nmat$-join derivable from $d$.
If $b$ is compact, by Th.~\ref{thm:compact} there must be a finite
subset $J \subseteq \nmat$ such that  $d \cgeq \Dfunapp{J}{b}$.\qed
\end{example}
}

\subsection{Group-Compactness without I-join derivability}
\label{ssec:group-noncompact}

Let us assume $c \cgeq \Dfunapp{I}{e}$ as in Th.~\ref{thm:compact}. By Prop.~\ref{projection:prop}~(1), we know that $e$ is $I$-group derivable
from $c$ but not necessarily $I$-join derivable from $c$.
The problem in establishing group compactness in the \emph{absence of $I$-join
derivability} has to do with $d'$ in the infinite case in
Ex.~\ref{ds:counter-example}. We have $d' = \bigmeet_{i \in
\nmat}\sfunapp{i}{b'}$. Notice that we cannot guarantee that $b'$ is
$\nmat$-join derivable from $d'$ ($\pfunapp{\nmat}{d'} \cgeq b'$). One can verify that $d' \cgeq \Dfunapp{\nmat}{b'}$, i.e.,
$b'$ resides in the space of agent $i$ for some $i \in \nmat$. Then, $b'$ is
$I$-group derivable from $d'$  ($\Pfunapp{\nmat}{d'} \cgeq b'$). Nevertheless we cannot guarantee the existence of a finite $J \subset \nmat$ such that  that $d' \cgeq
\Dfunapp{J}{b'}$.  In fact, the existence of such a $J$ cannot be guaranteed even if
$e$ ($b'$ in Ex.~\ref{ds:counter-example}) is compact as stated in the next theorem.  

\begin{theorem}[Non-Compactness]
\label{thm:nocompact}
\index{Group non-compactness}
There exists an scs $({\Con},\cleq,(\sfun{i})_{i \in G})$ with
distributed spaces $(\Dfun{I})_{I \subseteq G}$ such that for some $c,e \in \Con$
and $I\subseteq G$: (1) $e$ is compact, (2) $c \cgeq \Dfunapp{I}{e}$ but
(3) there is no finite subset  $J \subseteq I$ with $c \cgeq \Dfunapp{J}{e}$.
\end{theorem}

\begin{proof}
Consider the scs $(\Con, \leq, (\sfun{n})_{n \in \nmat})$ (Fig.~\ref{fig:cons}) defined by
\[
\Con = \left\{0, \frac{1}{2} \right\} \cup
	\left\{ \frac{1}{2} + \frac{1}{2n}\ \Big| \ n \geq 1 \right\}
\quad \text{ and } \quad
\sfunapp{n}{x} =
\begin{cases}
  \frac{1}{2} + \frac{1}{2n} & x \geq \frac{1}{2}\\
  0 & x < \frac{1}{2}\\
\end{cases}
\]
where $\sfun{n}$ is a self-map on $\Con$ for every $n \geq 1$. For $n = 0$,
$\sfunapp{n}{x} = 0$ for every $x \in \Con$.

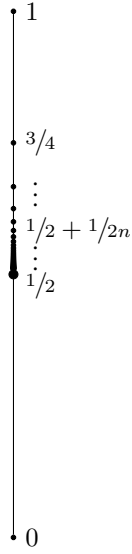
\begin{figure}[t]
\centering
\begin{tikzpicture}[scale=0.7]
\foreach \x in {1,...,50}{
\node at (1, 5 + 5/\x) [draw, circle,inner sep=0.6pt,fill] (\x){};
}
\node at (1,0)   [draw,circle,inner sep=0.6pt,fill,label={right:$0$}]  (0) {};
\node at (1,5)   [draw,circle,inner sep=1.2pt,fill]					           (n) {};
\node at (1,4.8) [circle,inner sep=0.6pt,label={right:$\sfrac{1}{2}$}] (12){};
\node also [label={right: $1$}]                             (1);
\node also [label={right: $\sfrac{3}{4}$}]                  (2);
\node also [label={right: $\ \vdots$}]						          (3);
\node also [label={right: $\sfrac{1}{2} + \sfrac{1}{2n}$}]  (6);
\node also [label={right: $\ \vdots$}]						          (11);
\draw (1) -- (0);
\end{tikzpicture}
\caption{Constraint system $(\Con,\leq)$ with $\Con = \left\{0, \frac{1}{2} \right\}
\cup \left\{ \frac{1}{2} + \frac{1}{2n} \mid \geq 1 \right\}$.}
\label{fig:cons}
\end{figure}

First we prove that for every $n \in \nmat$, $\sfun{n}$ is a space function.
From Def.~\ref{def:sfunc} we prove:

\begin{enumerate}
\item[a.] $\sfun{n}$ preserves 0 and binary joins.

Recall that for every $n \in \nmat$, $\sfunapp{n}{0} = 0$.
If $x < \sfrac{1}{2}$ and $y < \sfrac{1}{2}$, then $\sfunapp{n}{x \join y}
= 0 = \sfunapp{n}{x} \join \sfunapp{n}{y}$.
If either $x \geq \sfrac{1}{2}$ or $y \geq \sfrac{1}{2}$ hold, then
$\sfunapp{n}{x \join y} = \sfrac{1}{2} + \sfrac{1}{2n} = \sfunapp{n}{x}
\join \sfunapp{n}{y}$.

\item[b.] $\sfun{n}$ is continuous.

For $n=0$, it is immediate that $\sfun{n}$ is continuous.

Since $\Con$ is countable, from Prop.~\ref{prop:continuity} it suffices
to prove that $\sfun{n}$ preserves the join of increasing chains. For any $n
\geq 1$, consider any increasing chain $x_1 \leq x_2 \leq \cdots$ in $(\Con,
\leq)$.  To prove $\bigjoin_{i \geq 1} \sfunapp{n}{x_i} =
\sfunapp{n}{\bigjoin_{i \geq 1} x_i}$.

Let $J,K \subseteq \nmat \setminus \{0\}$ be two index sets such that: if $x_i
< \sfrac{1}{2}$ then $i \in J$ and if $x_i \geq \sfrac{1}{2}$ then $i \in K$.
Notice that $J \cap K = \emptyset$ and $J \cup K = \nmat \setminus \{0\}$.
Then using this and part (a), we have $\sfunapp{n}{\bigjoin_{i \geq 1} x_i} = \sfunapp{n}{\left(\bigjoin_{j \in J} x_j\right) \join \left(\bigjoin_{k \in K} x_k\right)} =
\sfunapp{n}{\bigjoin_{j \in J} x_j} \join \sfunapp{n}{\bigjoin_{k \in K} x_k} $. Also, $\bigjoin_{j \in J} x_j = 0$ and $\bigjoin_{k \in K} x_k \geq \sfrac{1}{2}$, therefore $\sfunapp{n}{\bigjoin_{j \in J} x_j} \join \sfunapp{n}{\bigjoin_{k \in K} x_k} = 0 \join (\sfrac{1}{2} + \sfrac{1}{2n}) = \bigjoin_{j \in J} 0 \join \bigjoin_{k \in K} (\sfrac{1}{2} + \sfrac{1}{2n}) = \bigjoin_{j \in J} \sfunapp{n}{x_j} \join \bigjoin_{k \in K} \sfunapp{n}{x_k} = \bigjoin_{i \geq 1} \sfunapp{n}{x_i}$. Then 
$\sfunapp{n}{\bigjoin_{i \geq 1} x_i} = \bigjoin_{i \geq 1} \sfunapp{n}{x_i}$ as wanted.
\end{enumerate}

To complete the proof we now show that given the above scs $(\Con, \leq, (\sfun{n})_{n \in \nmat})$ for $I = \nmat \setminus \{0\}$ and
$c = e = \sfrac{1}{2}$: (1) $e$ is compact,
(2) $c \geq \Dfunapp{I}{e}$ but
(3) $c \not\geq \Dfunapp{N}{e}$ for any finite set $N \subseteq I$.

\begin{enumerate}
\item $\sfrac{1}{2}$ is compact. This follows directly from the definition of our constraint system $(\Con,\leq)$.

\item $\sfrac{1}{2} \geq \Dfunapp{I}{\sfrac{1}{2}}$. By definition of $\Dfun{I}$, $\Dfunapp{I}{\sfrac{1}{2}} \leq
\sfunapp{n}{\sfrac{1}{2}}$ for every $n \in I$.
Then $\Dfunapp{I}{\sfrac{1}{2}} \leq \bigmeet_{n \geq 1}
\sfunapp{n}{\sfrac{1}{2}}$. Since for every $n \geq 1$, $\sfunapp{n}{\sfrac{1}{2}} = \sfrac{1}{2} +
\sfrac{1}{2n} \geq \sfrac{1}{2}$ then $\bigmeet_{n \geq 1}
\sfunapp{n}{\sfrac{1}{2}} = \sfrac{1}{2}$. Thus, $\sfrac{1}{2} \geq \Dfunapp{I}{\sfrac{1}{2}}$.

\item $\sfrac{1}{2} \not\geq \Dfunapp{N}{\sfrac{1}{2}}$ for any finite set
$N \subseteq I$. Notice that, for any finite set $N \subseteq I$,
$\Dfunapp{N}{\sfrac{1}{2}} = \sfunapp{m}{\sfrac{1}{2}} > \sfrac{1}{2}$,
where $m = \max(N)$. Hence, $\sfrac{1}{2} \not\geq \Dfunapp{N}{\sfrac{1}{2}}$.
\end{enumerate}
\end{proof}

\subsection{Distributed Spaces in Completely Distributive Lattices}
\label{DistributedSpaces:section}

In this section we present another characterization of distributed spaces for distributive lattices: For a group $I$, $\Dfunapp{I}{c}$ can be understood as the greatest
information below all possible combinations of information in the
spaces of the agents in $I$ that derive $c$.  We also provide compositionality properties capturing the intuition that just
like \emph{distributed information} of a  group  $I$ is the collective
information from all its members, it is also the collective information of its
subgroups.  We shall argue that the following results can be used to produce algorithms to efficiently compute $\Dfunapp{I}{c}$  for finite constraint systems.

We recall $J$-tuples, a general form of tuples that allows for an arbitrary index set $J$.

\begin{definition}[\cite{munkres-topology-1974}]
\label{def:j-tuples}  Let $J$ be an index set. Given a set $X$, a $J$-tuple of
elements of $X$ is a function $\func{\mathbf{x}}{J}{X}$. If $j \in J$, we
denote $\mathbf{x}(j)$ by $x_j$ and refer to it  as the $j$-th coordinate
of $\mathbf{x}$. The function $\mathbf{x}$ is denoted itself by $(x_j)_{j \in
J}$. We use $\tuplespace{X}{J}$ to denote the set of all $J$-tuples. 
\end{definition}
\index{$J$-tuples}
\index[operator]{$\tuplespace{X}{J}$, set of $J$-tuples $(x_j)_{j \in J}$}
\index[operator]{$(x_j)_{j \in J}$, $J$-tuple of elements $x_j \in X$ for each $j \in J$}

The next theorem is one of main results of this paper. It establishes that for completely distributed lattices,  $\Dfunapp{K}{c}$ is the greatest information below all possible combinations of  information in the spaces of the agents in $K$ that derive $c$.

\begin{theorem}
\label{thm:delta-ast}
Let  $(\Dfun{I})_{I \subseteq G}$ be the distributed spaces of an scs
$({\Con},\cleq,(\sfun{i})_{i \in G})$.
Suppose that $({\Con},\cleq)$ is completely distributive.
Let  $\dplusfun{K}:\Con \to \Con$, with $K\subseteq G$,
be the function defined as follows: 
\[ \dplusfunapp{K}{c} \defsymbol \bigmeet\left\{ \bigjoin_{k \in K}
\sfunapp{k}{a_k} \mid (a_k)_{k \in K} \in \tuplespace{\Con}{K} \text{ and } \bigjoin_{k \in K}
a_k \cgeq c \right\}.\]
Then $\Dfun{K} = \dplusfun{K}$.
\end{theorem}
\index{Completely distributive lattice}
\index[operator]{$\dplusfun{K}$}

The above theorem is presented in~\cite{guzman-reasondistknowldg-2019} for finite cs. Here we extend this for completely distributive lattices.

For the sake of the presentation, we give the proof of the above theorem in Section~\ref{ssec:proof-thms}. Nevertheless, we would like to mention that the central and non-obvious property used in the proof is that of $\dplusfun{K}$ being a \emph{continuous function}. The distributivity of $({\Con},\cleq)$ is crucial for this. In fact without it the equality $\Dfun{K} = \dplusfun{K}$ does not  necessarily hold as shown by the following counter-example.

\begin{figure}
\centering
\begin{subfigure}[t]{0.45\textwidth}
\centering
\begin{tikzpicture}[scale=0.45,>=stealth]
  \tikzstyle{every node}=[font=\scriptsize]
  \node[shape = circle, draw] (A) at (0,-4)   {$p \vee \neg p$};
  \node[shape = circle, draw] (B) at (-4,0)   {$\; p\; $};
  \node[shape = circle, draw] (C) at (4,0)    {$\neg p$};
  \node[shape = circle, draw] (D) at (0,4)    {$p \wedge \neg p$};

  \draw (A) to node {} (B);
  \draw (A) to node {} (C);
  \draw (B) to node {} (D);
  \draw (C) to node {} (D);

  \draw [blue,thick,->,bend left,out=20,in=160] (B) to node[fill=white] {$\sfun{1}$} (C);
  \draw [blue,thick,->,bend left,out=20,in=160] (C) to node[fill=white] {$\sfun{1}$} (B);
  \draw [blue,thick,->,loop left]               (A) to node             {$\sfun{1}$} (A);
  \draw [blue,thick,->,loop left]               (D) to node             {$\sfun{1}$} (D);

  \draw [red,thick,->,loop right]    (D) to node             {$\sfun{2}$} (D);
  \draw [red,thick,->,bend left]     (B) to node[fill=white] {$\sfun{2}$} (D);
  \draw [red,thick,->,loop right]    (A) to node             {$\sfun{2}$} (A);
  \draw [red,thick,->,loop below]    (C) to node             {$\sfun{2}$} (C);

  \draw [dotted,teal,loop below,->,thick]    (A) to node             {$\dapproxfun{I}$} (A);
  \draw [dotted,teal,->,thick]               (B) to node[fill=white] {$\dapproxfun{I}$} (C);
  \draw [dotted,teal,->,bend left,thick]     (C) to node[fill=white] {$\dapproxfun{I}$} (A);
  \draw [dotted,loop above,teal,->,thick]    (D) to node             {$\dapproxfun{I}$} (D);
\end{tikzpicture}
\caption{For $I = \{1,2\}$, $\dapproxfun{I}(c) = \bigmeet_{i \in I} \sfunapp{i}{c}$
is not a space function: $\dapproxfunapp{I}{p \join \neg p} \neq \dapproxfunapp{I}{p}
\join \dapproxfunapp{I}{\neg p}$.}
\label{fig:approx}
\end{subfigure}
~
\begin{subfigure}[t]{0.45\textwidth}
\centering
\begin{tikzpicture}[scale=0.4,>=stealth]
  \node [shape = circle, draw] (A) at (0,-6)   {$a$};
  \node [shape = circle, draw] (B) at (-6,0)   {$b$};
  \node [shape = circle, draw] (C) at (0,0)    {$c$};
  \node [shape = circle, draw] (D) at (6,0)    {$d$};
  \node [shape = circle, draw] (E) at (0,6)    {$e$};

  \draw (A) to node {} (B);
  \draw (A) to node {} (C);
  \draw (A) to node {} (D);
  \draw (B) to node {} (E);
  \draw (C) to node {} (E);
  \draw (D) to node {} (E);

  \draw [left,loop above,blue,->,dotted,thick]     (B) to node {} (B);
  \draw [above,blue,->,bend left,dotted,thick]     (C) to node {} (D);
  \draw [below,blue,->,bend left,dotted,thick]     (D) to node {} (C);
  \draw [below,loop below,blue,->,dotted,thick]    (A) to node {} (A);
  \draw [above,loop above,blue,->,dotted,thick]    (E) to node {} (E);
  \draw [above,loop right,red,->,dashed,thick]     (E) to node {} (E);
  \draw [left,loop right,red,->,dashed,thick]      (C) to node {} (C);
  \draw [above,red,->,bend left,dashed,thick]      (B) to node {} (E);
  \draw [right,loop above,red,->,dashed,thick]     (D) to node {} (D);
  \draw [below,loop left,red,->,dashed,thick]      (A) to node {} (A);

  \draw [above,loop right,teal,->,thick]  (A) to node {} (A);
  \draw [left,teal,->,bend right,thick]   (C) to node {} (A);
  \draw [right,teal,->,bend left,thick]   (D) to node {} (A);
  \draw [left,teal,->,bend right,thick]   (E) to node {} (A);
  \draw [below,loop below,teal,->,thick]  (B) to node {} (B);
\end{tikzpicture}
\caption{For $I = \{1, 2\}$,
{\color{blue} $\sfun{1}$} ({\color{blue}${\cdot}{\cdot}{\cdot}{\to}$}) and
{\color{red}  $\sfun{2}$} ({\color{red} $\dashrightarrow$}) are space functions. The function {\color{teal}
$\dplusfun{S}$} ({\color{teal}$\rightarrow$}) in Th.~\ref{thm:delta-ast}
is not a space function:
$\dplusfunapp{I}{b} \join \dplusfunapp{I}{c} = b \neq a =
\dplusfunapp{I}{b \join c}$.}
\label{ex:m3}
\end{subfigure}
\caption{Counter-examples over the four-element boolean algebra (a) and non-distributive lattice $\M_3$ (b).}
\label{fig:dapprox-nondistri}
\end{figure}
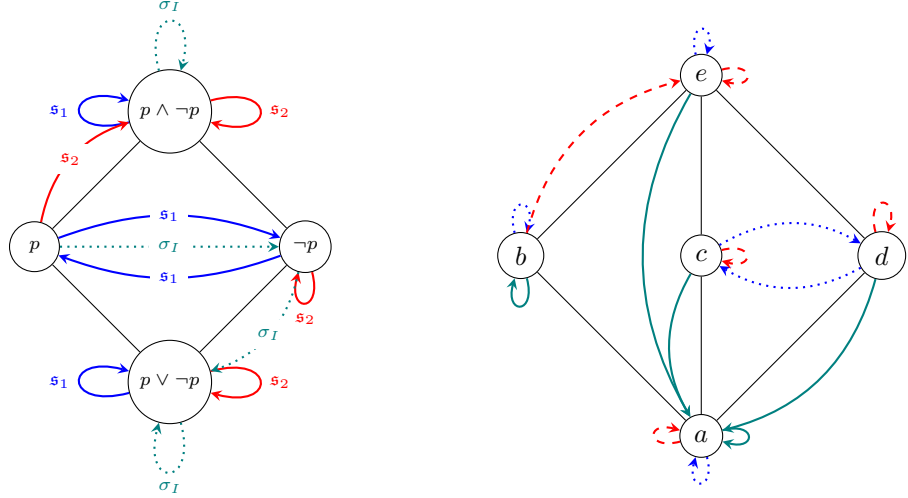

\begin{example}
Consider the lattice $\M_3$, which is not (completely) distributive, and the space functions
$\sfun{1}$ and $\sfun{2}$ in Fig.~\ref{ex:m3}.
We obtain $\dplusfunapp{I}{b \join c} = \dplusfunapp{I}{e} = a$ and
$\dplusfunapp{I}{b} \join \dplusfunapp{I}{c} = b \join a = b$. Then,
$\dplusfunapp{I}{b \join c} \neq \dplusfunapp{I}{b} \join \dplusfunapp{I}{c}$,
i.e., $\dplusfun{I}$ is not a space function.\qed
\end{example}

\index[operator]{$\M_3$, non-distributive lattice}

We can use Th.\ref{thm:delta-ast} to prove the following properties
characterizing the information of a group from that  of its subgroups. 

\begin{theorem}
\label{thm:comp-algo} 
Let $(\Dfun{I})_{I \subseteq G}$ be the distributed spaces of an scs
$({\Con},\cleq,(\sfun{i})_{i \in G})$. Suppose that $({\Con},\cleq)$ is
completely distributive. Let $I,J,K \subseteq G$ be such that $I = J
\cup K$. Then the following equalities hold:   
\begin{eqnarray}
  1.  \ \Dfunapp{I}{c} & = & \bigmeet\left\{ \Dfunapp{J}{a} \join 
  \Dfunapp{K}{b} \mid a,b \in \Con \ \mbox{ and } \ a \join b \cgeq c \right\}. \\
  2. \ \Dfunapp{I}{c}& = &  \bigmeet\left\{ \Dfunapp{J}{a} \join 
  \Dfunapp{K}{c \sop a } \mid a \in \Con \right\}. \\
  3.  \ \Dfunapp{I}{c}& = &  \bigmeet\left\{ \Dfunapp{J}{a} \join
  \Dfunapp{K}{c \sop a }  \mid a \in \Con \  \mbox{ and } \ a \cleq c \right\}.
\end{eqnarray}
\end{theorem}

We find it convenient to give the proof of Th.~\ref{thm:comp-algo} in Section~\ref{ssec:proof-thms}.  The properties in this theorem bear witness to the inherent
compositional nature of our notion of distributed space.
The first property in
Th.~\ref{thm:comp-algo} essentially  reformulates Th.~\ref{thm:delta-ast} in
terms of subgroups rather than  agents. It can be proven by replacing
$\Dfunapp{J}{a}$ and  $\Dfunapp{K}{b}$  by $\dplusfun{J}(a)$ and
$\dplusfun{K}(b)$, defined in Th.~\ref{thm:delta-ast} and using
distributivity of joins over meets. The second and third properties in
Th.~\ref{thm:comp-algo} are pleasant simplifications of the first one
using co-Heyting subtraction. These properties realize the intuition that by
joining the information $a$ and $c \sop a$ of their subgroups, the group $I$
can obtain $c$.  

In  Section~\ref{ssec:proof-thms} we use Th.\ref{thm:delta-ast} to prove Th.~\ref{thm:comp-algo}. 
We now  conclude this section with a brief discussion on how to use  Th.\ref{thm:delta-ast}
to solve a computational lattice problem. 

\paragraph{Computing  Distributed Information}

Let us assume that $\C$ is finite and distributive. We wish to compute
$\Dfun{I}$. Notice that under this finiteness assumption, space functions are
exactly those that preserve the join of finite sets, also known as
\emph{join-endomorphisms}~\cite{gratzer-latjoinend-1958}. Recall that
$\scal(\C)$ denotes the set of space functions (join-endomorphism in this
case) over $\C$.  

From Remark~\ref{rmk:ds-meet}, computing the distributed space $\Dfun{I}$ is
then equivalent to the following lattice problem:  \emph{Given a finite set
$S= \{ \sfun{i} \mid {i \in I} \}$ of join-endomorphisms over the finite
distributive lattice $\C$, find its meet $\bigmeet_{\scal(\C)}S$}.  Even  in
small lattices with four elements and two space functions, finding
$\Dfun{I}=\bigmeet_{\scal(\C)}S$ may not be immediate, e.g., consider $S=\{ \sfun{1},\sfun{2} \}$ in Fig.~\ref{fig:difficult-small}.

A \emph{naive approach} would be to compute each $\Dfunapp{I}{c}$ by taking
the point-wise meet construction $\dapproxfunapp{I}{c} \defsymbol\bigmeet\{ 
\sfun{i}(c) \mid i \in I  \}$. But this does not work in general since
$\bigmeet\{  \sfun{i} \mid i \in I  \}(c)$ is not necessarily
equal to $\dapproxfunapp{I}{c}$.
In fact $\dapproxfun{I} \fgeq \Dfun{I}$ but $\dapproxfun{I}$ may not even be a
space function as shown in Fig.~\ref{fig:approx}.

A \emph{brute force} solution to computing $\Dfunapp{I}{c}$ is to
generate the set $\{ f (c) \mid f \in \scal(\C) \mbox{ and } f \cleq  \sfun{i}
\mbox{ for all } i \in I\}$ and then take its join (see
Remark~\ref{rmk:ds-meet}). This approach works since for any set $S$ of
join-endomorphisms $(\bigjoinp{\scal(\C)} S)(c) = \bigjoinp{\C} \{ f(c) \mid f
\in S\}$. The problem, however, is that the number of join-endomorphisms over
a distributive lattice can be \emph{non-polynomial} in the size of the
lattice~\cite{quintero-ccjoinendo-2020}.

Nevertheless we can use Th.~\ref{thm:delta-ast} to obtain a worst-case
polynomial bound for computing $\Dfun{I}$. The next proposition shows this.

\begin{proposition}
\label{algo:prop}

Let $\C$ be a distributive lattice of size $n$. Let $S=\{ \sfun{i} \ | \  i
\in I \}$, where $I=\{1,\ldots, m\}$, be a set of join-endomorphisms over 
$\C$. Assuming that binary meets and joins over  $\C$ can be computed in O(1),
the meet $\Dfun{I}=\bigmeet_{\scal(\C)}S$ can be computed in $O(mn^3)$
worst-case time complexity.  
\end{proposition}

\begin{proof}
Let $\C$ be a distributive cs (lattice) of size $n$ and let $\sfun{j}$ and $\sfun{k}$ be space functions (join-endomorphisms) over $\C$.
From Th.~\ref{thm:delta-ast}, the value $\Dfunapp{K}{c}= (\sfun{j} \meetp{\scal(\C)} \sfun{k})(c) = \dplusfunapp{K}{c}$, with $K = \{j,k\}$, can be computed in $O(n^2)$ by performing $O(n^2)$ joins and $O(n^2)$ meets. Hence the function $\Dfun{K}$ can be computed in $O(n^3)$ whenever $|K| = 2.$

We now proceed by induction on the size of $S=\{ \sfun{i} \mid i \in I \}$ where $I=\{1,\ldots, m\}$. Suppose that $m=1$. We can then compute $\Dfun{\{{1}\}} = \sfun{1}$ in $O(n)$ and hence in $O(n^3)$. 
Assume that $\Dfun{\{{1},\ldots,{m-1} \}}$ can be computed in $O((m-1)n^3)$. From the associativity of the meet operation,
$\Dfun{I}= \Dfun{\{{1},\ldots,{m-1} \}} \ \meetp{\scal(\C)} \ \sfun{m}.$ Thus, we can  compute first  $\Dfun{\{{1},\ldots,{m-1} \}}$ in $O((m-1)n^3)$ and  then  
$\Dfun{K}$, with $\sfun{j}=\Dfun{\{{1},\ldots,{m-1} \}}$ and $\sfun{k}=\sfun{m}$,  in  $O(n^3)$. The total worst-case time complexity for computing  $\Dfun{I}$ is then in $O(mn^3).$
\end{proof}

\subsection{Summary of Section~\ref{sec:dist-info}}
\label{ssec:summ}

In this section we presented the main technical results of this paper.  We have formalized and developed the theory of the collective information of
a group of agents $I$ as the space function $\Dfun{I}$.   Intuitively, the space function $\Dfun{I}$ represents the smallest space that includes all the
local information of the agents in $I$.

We  first constructed the complete lattice $(\scal(\C),\fleq)$
(Lemma~\ref{lemma:closure-space}) where $\scal(\C)$ is the set of all space
functions defined on the complete lattice $(\C,\cleq)$. We then defined $\Dfun{I}$ as the greatest space function in $\scal(\C)$ below the space functions of agents in $I$ (Def.~\ref{ds:def}) and presented some of its basic compositional properties (Prop.~\ref{prop:comp}). We showed that $\Dfun{I}$ could also be alternative defined as the greatest group distribution candidate (gdc) (Th.~\ref{th:max-gdc}).   We illustrated in Ex.\ref{aumann:example} that $\Dfun{I}$ can be interpreted as \emph{Distributed Knowledge} in  Aumann structures, a representative model for epistemic group reasoning. 
\hltext{Furthermore,\resp{We added the next phrase.}
in Ex.~\ref{ex:kripke-ds} we showed that distributed knowledge in epistemic logic is also an instance of distributed information.}
 
We also defined agent, join and group projections (Def.~\ref{def:agent-proj},
Def~\ref{group-projection}). Group (agent) projection of a given $c \in \C$
represents the join of all the information that the group (agent) has in $c$.
Join projections are the join of individual agent projections. We stated that
group projections and distributed spaces form a Galois connection
(Prop.~\ref{projection:prop}). We then provided a \emph{group compactness}
result: Given an \emph{infinite} group $I$, we identified
\emph{join-derivability} (Def.~\ref{def:agent-proj}) as a condition under
which  $c \cgeq \Dfunapp{I}{e}$ implies $c \cgeq \Dfunapp{J}{e}$ for some
finite group $J \subseteq I$ (Th.~\ref{thm:compact}).  We then showed that
without this condition we cannot guarantee the existence of such finite set $J \subseteq I$ (Th.~\ref{thm:nocompact}).

Finally we showed that if $\C$ is completely-distributive, $\Dfunapp{I}{c}$ can be characterized as the greatest information below all
possible combinations of information in the spaces of the agents in $I$ (Th.~\ref{thm:delta-ast}) that derive $c$ and, more succinctly,  as the
combination of the information of its subgroups (Th.~\ref{thm:comp-algo}) that derive $c$. For the finite-case we briefly explained how Th.~\ref{thm:delta-ast} can be used to compute distributed space functions in polynomial time.

\section{Applications to Minkowski Addition and Mathematical Morphology}
\label{sec:app}

In this section we shall show that some fundamental operations from Mathematical Morphology (MM) have a counterpart in the theory we developed in the previous sections. In particular we shall show that \emph{distributed spaces}, the central notion of this paper, have a natural interpretation in MM. Furthermore, we shall use our results on distributed information to provide new constructions and results for MM.

\subsection{Modules}
\label{ssec:mm-modules}

We assume that the reader is familiar with basic concepts of abstract
algebra~\cite{dummit-modules-2003,ash-modules-2007}. To present the results in
this section uniformly  we shall use a fundamental structure from algebra,
namely, that of a \emph{module}.
Recall that a module $M$ over a \emph{ring} $R$ is  a generalization of the notion of \emph{vector space.} In a vector space the ring $R$ needs to be a \emph{field}.
We shall take the liberty of referring to the elements of $M$ and $R$ as
\emph{vectors} and \emph{scalars}, resp. The former will be written in
boldface to distinguish them from the latter.

More precisely, a \emph{module} $M$ over a \emph{ring}
$R$~\cite{ash-modules-2007}, also called an (left) \emph{$R$-module} $M$, is a
set that satisfies the following three conditions.
 It must be closed under addition and scalar multiplication: $\vect{u} +
\vect{v} \in M$ and $r\vect{u} \in M$ whenever $\vect{u}, \vect{v} \in M$ and
$r \in R$.  It must also form an \emph{abelian group} under addition:  $+$ is 
a commutative and associative  operator with $\vect{0}$ as additive identity
and with an additive inverse $\vect{-u}$ for every $\vect{u} \in M$. Finally,
it must also satisfy the following axioms for scalar multiplication: For every
$\vect{u}, \vect{v} \in M$ and every $r,s \in R$,  $r(s\vect{u})=
(rs)\vect{u}$, $r(\vect{u}+\vect{v})=r\vect{u}+r\vect{v}$, $(r+s)\vect{u}=
r\vect{u} + s\vect{u}$, and $1 \vect{u}=\vect{u}$ where  $1$ is the
\emph{multiplicative identity} of $R$.
\index[operator]{$R$-module $M$, module $M$ over a ring $R$}
\index{Module}
\index{Ring}
\index{Vector}

We shall use the following basic properties of modules. The additive identity
for any module $M$ and the additive inverse for every $\vect{u} \in M$ are
unique. If $R$ is a field then the $R$-module $M$ is a \emph{vector space}
over $R$.  If $M$ is a vector space over $R$ then $1$ is the only
multiplicative identity for $M$.

The following examples of modules are fundamental in Mathematical Morphology.
One of them is not a vector space; it justifies using modules rather than
vector spaces as the underlying structure. 

\begin{example}
\label{module:example} 
The set $R^n$ of all $n$-tuples of elements of a ring $R$ can be made into an $R$-module.
Given $\vect{u}= (p_1,\ldots, p_n)\in R^n, \vect{v}=(q_1,\ldots, q_n)\in R^n$
and $r \in R$, define $\vect{u} + \vect{v} = ( p_1+q_1,\ldots,
p_n+q_n )$ and $r\cdot \vect{v}= ( rp_1, \ldots, rp_n)$. The module additive
identity  $\vect{0}$ is $(0,\ldots,0)\in R^n$, where $0$ is the ring additive
identity, and the module inverse additive $\vect{-u}$ is $(-p_1,\ldots, -p_n)$
where $-p_i$ is the ring additive inverse of $p_i$. 
 
The \emph{Euclidean $n$-dimensional space} $\rmat^n$ is obtained by taking $R$
as the set of reals numbers $\rmat$ in the above example. Since $\rmat$ is
also a field, $\rmat^n$ is also a vector space.  The \emph{$n$-dimensional
grid} $\zmat^n$ is obtained by taking $R$ as the set of integers $\zmat$. This
is an example of a module that it is not a vector space since the ring $\zmat$
is not a field. \qed
\end{example}

\subsection{Minkowski Addition}
\label{ssec:mm-mink-add}

In geometry, vector addition is extended to addition of sets of vectors in an
operation known as \emph{Minkowski addition}. From now on we shall omit mentioning
the ring of the module when it is unimportant or clear from the context.

\begin{definition}[Minkowski Sum~\cite{schneider-minkowski-2013}] 
Let $M$ be a module and $A,B \subseteq M$. The \emph{Minkowski addition} of
$A$ and $B$ is defined thus \( A \oplus B = \{ \vect{u} + \vect{v} \mid
\vect{u} \in A \mbox{ and } \vect{v}  \in B \}. \)
\index{Minkowski addition}
\index[operator]{$\oplus$, Minkowski addition}
\end{definition}

It is easy to see that  $\oplus$ is associative and commutative, it has $\{
\vect{0} \}$ and $\emptyset$ as identity and absorbent elements, resp., and 
that it distributes over set union.

\begin{proposition}[\cite{schneider-minkowski-2013}]
\label{monoid:prop} 
Let $M$ be a module. Then $(\pcal(M),\oplus)$ is a \emph{commutative monoid
with zero element} $\emptyset$ and identity $\{ \vect{0} \}$. Furthermore,
$X\oplus(A \cup B)= (X\oplus A) \cup (X \oplus B)$ for every $X,A,B\subseteq
M$.
\end{proposition}

Recall that a  convex set is a set of points such that, given any two points
in that set, the line segment joining them lies entirely within that set.  It
is well-known that the distribution law $A \oplus (B \cap C)=(A  \cap B)
\oplus (A  \cap C)$ holds for convex sets~\cite{schneider-minkowski-2013}. 
Nevertheless, in general, $\oplus$ does not distribute over set intersection
as illustrated next.

\begin{example}
\label{ex:non-distri}
Take the Euclidean one-dimensional vector space $\rmat$ and let $X=\{0,1\}$,
$A=\{1\}$ and $B=\{2\}$. One can verify that $\emptyset=X\oplus(A \cap B) \neq
(X\oplus A) \cap (X \oplus B)=\{ 2 \}$.\qed
\end{example}

However, as part of the application results in this section we shall 
establish a pleasant new equation for $X\oplus(A \cap B).$ Namely, for every
module $M$ and for all $X,A,B \subseteq M$ we have  
\begin{equation}
\label{oplus-eq}
X\oplus(A \cap B)
= \bigcap_{Y \subseteq X} (Y \oplus A) \cup ( (X\setminus Y) \oplus B).
\end{equation}

The Minkowski sum has been applied in mathematical morphology as well as in
collision detection, robot motion planing, aggregation
theory~\cite{serra-mm-apps-2012,najman-mm-theory-app-2013,zelenyuk-aggregation-2015}.
In this section we focus on applications to mathematical morphology.

\subsection{Mathematical Morphology}
\label{ssec:mm-mat-morph}

Mathematical morphology (MM) is a theory developed for the analysis of
geometric structures~\cite{serra-intromm-1986} . It is founded upon, among
others, set theory, lattice theory, geometry, topology and probability.
Basically, this theory considers an arbitrary space $M$ where its objects are
transformed by two fundamental operations: \emph{dilation} and \emph{erosion}.
\index{Mathematical morphology}
\index{Dilation}

In~\cite{bloch-mm-2007} dilations and erosions are typically defined in terms
of Minkowski additions over the modules $\rmat^n$ or $\zmat^n$ given in
Ex.~\ref{module:example}. Here we generalize the definition
in~\cite{bloch-mm-2007} to arbitrary modules. 

\begin{definition}[Dilations and Erosions in Modules]
\label{def:dilation-set}
Let $M$ be a module. A \emph{dilation} by $S \subseteq M$ is a function 
$\func{\dilatn{S}}{\pcal(M)}{\pcal(M)}$ given by $\dilatnapp{S}{X} = X \oplus
S = \bigcup_{\vect{u} \in S} X \oplus \{\vect{u}\}$. An \emph{erosion} by
$S\subseteq M$ is a function $\func{\erosn{S}}{\pcal(M)}{\pcal(M)}$ given by
$\erosnapp{S}{X} =  X \ominus S$ where  $X \ominus S \defsymbol
\bigcap_{\vect{u} \in S} X \oplus \{-\vect{u}\}$.
\index{Dilation}
\index[operator]{$\dilatn{S}$, dilation by $S$}
\index{Erosion}
\index[operator]{$\erosn{S}$, erosion by $S$}
\end{definition}

\begin{figure}
    \centering
    \includegraphics[scale=0.3]{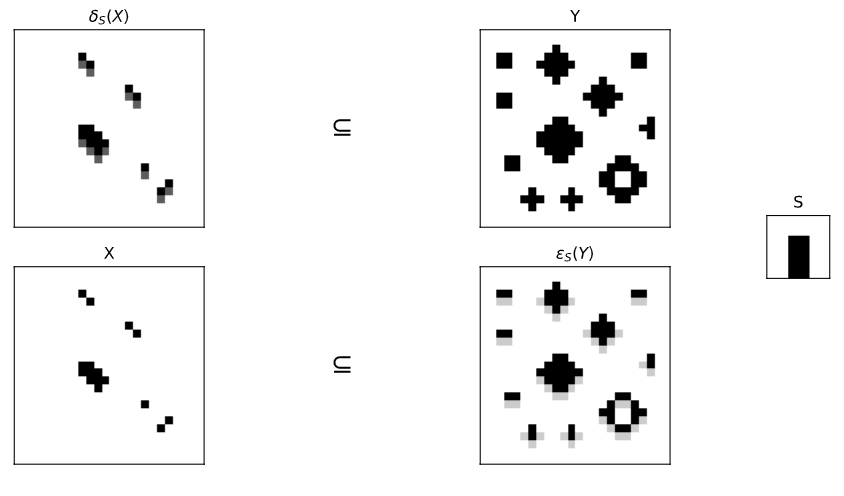}
    \caption{Example of Galois connection between dilations and erosions:
    $\dilatnapp{S}{X} \subseteq Y$ iff $X \subseteq \erosnapp{S}{Y}$.
    Here $X$ and $Y$ are images, $S$ is the structuring element (centered at 
    the origin), $\dilatnapp{S}{X}$ is a dilation and $\erosnapp{S}{Y}$ is an 
    erosion. Pixels added/removed by dilation/erosion are depicted in dark/light gray.}
    \label{fig:mm-galois}
\end{figure}

\begin{figure}[t]
	\centering
	\includegraphics[scale=0.3]{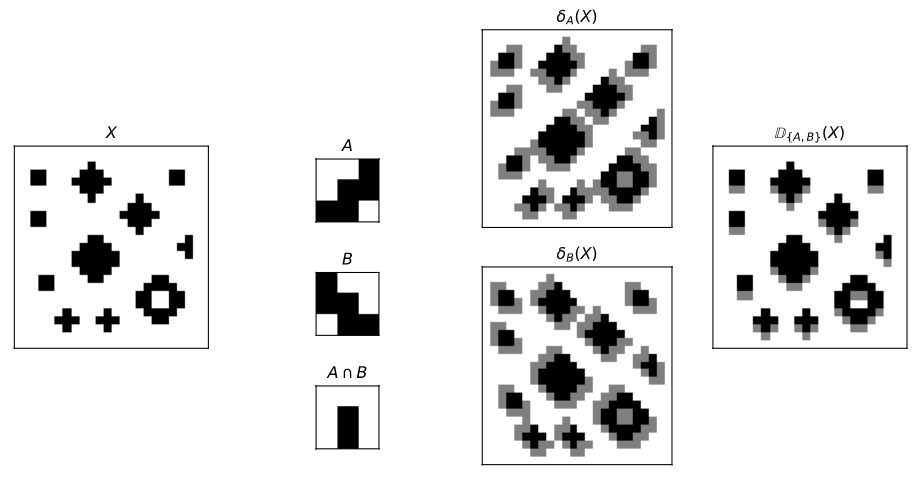}
	\caption{From left to right, image $X$, structuring elements $A$, $B$ and
	$A \cap B$, dilations $\dilatnapp{A}{X}$ and $\dilatnapp{B}{X}$, and 
	dilation $\Dfunapp{\{A,B\}}{X}$. Structuring elements are centered at 
	the origin. Pixels added by dilation are depicted in gray.}
	\label{fig:mm-dilations}
\end{figure}

In MM, a binary image $X$ is typically represented as a subset of the module
$M=\zmat^2$ where a pixel is activated if its corresponding coordinate (or
position vector) is in $X$. The \emph{translation} of a vector $\vect{u}$ by a
vector $\vect{v}$  is given by $ \vect{u} + \vect{v}$. The dilation
$\dilatnapp{S}{X}$ describes the interaction of $X$ with another image $S$
referred to as a \emph{structuring element} and typically  assumed to include
the center, i.e., $\vect{0}=(0,0) \in S.$ The dilated image 
$\dilatnapp{S}{X}$ ``inflates" the original one by including $X$ and adding
the pixels from the  translation of every $\vect{v}$ in $X$ by each $\vect{u}$
in $S$.  Intuitively, $\dilatnapp{S}{X}$ can be viewed as redrawing the image
$X$ with the brush $S$~\cite{bloch-mm-2007}. This is illustrated in
Fig.~\ref{fig:mm-galois} where an image $X$ is dilated by the structuring
element $S=\{(0,0),(0,-1)\}$, and in Fig.~\ref{fig:mm-dilations} where an
image $X$ is dilated by two different structuring elements
$A=\{(1,1),(0,0),(1,0),(-1,-1),(0,-1)\}$ and
$B=\{(-1,1),(-1,0),(0,0),(0,-1),(1,-1)\}$.
 
Furthermore, the erosion $\erosnapp{S}{X}$ in $\rmat^n$ or $\zmat^n$ can be
defined in terms of the translations of $S$ that are contained in $X$. The
next proposition states this result for modules.

\begin{proposition}[]
\label{prop:erosn-prop}

Let $M$ be a module and $S, X \subseteq M$. Then $\erosnapp{S}{X} = \{\vect{u}
\in M \mid S \oplus \{\vect{u}\} \subseteq X\}$.
\end{proposition}

\begin{proof}
Let $M$ be a module and $S, X \subseteq M$. From
Def.~\ref{def:dilation-set}, we will prove that $\bigcap_{\vect{u} \in
S} X \oplus \{-\vect{u}\} = \{\vect{u} \in M \mid S \oplus \{\vect{u}\}
\subseteq X\}$. Take any $\vect{v} \in \bigcap_{\vect{u} \in S} X \oplus
\{-\vect{u}\}$, then for every $\vect{u} \in S$, $\vect{v} = \vect{w} +
(-\vect{u})$ for some $\vect{w} \in X$. Since for every $\vect{u} \in S$,
$\vect{u} + \vect{v} \in S \oplus \{\vect{v}\}$ and $\vect{u} + \vect{v} =
\vect{w} \in X$ (see Prop.~\ref{monoid:prop}), we have $S \oplus \{\vect{v}\}
\subseteq X$. Therefore, $\vect{v} \in \{\vect{u} \in M \mid S \oplus
\{\vect{u}\} \subseteq X\}$.

Now, let $\vect{v} \in \{\vect{u} \in M \mid S \oplus \{\vect{u}\} \subseteq
X\}$, then $S \oplus \{\vect{v}\} \subseteq X$. Notice that for any $\vect{w}
\in S$, $\vect{w}+\vect{v} \in S \oplus \{\vect{v}\} \subseteq X$ and
therefore $\vect{v} = (\vect{w}+\vect{v}) + (-\vect{w}) \in X \oplus
\{-\vect{w}\}$ for every $\vect{w} \in S$. Then $\vect{v} \in
\bigcap_{\vect{u} \in S} X \oplus \{-\vect{u}\}$.
\end{proof}

Assume that $S$ includes the center. The above proposition tells us that  an erosion $\erosnapp{S}{X}$ reduces the image $X$ by erasing the pixels in
$X$ whose translation by some element of $S$ is not within $X$. This can also be
easily seen from the fact that erosions satisfy the equation
$\erosnapp{S}{X}=\{ \vect{u} \in M  \mid \mbox{ for each } \vect{v} \in S:
\vect{u} +  \vect{v} \in X  \}$ which follows directly from Prop.~\ref{prop:erosn-prop}.
Fig.~\ref{fig:mm-galois} illustrates the erosion of an image $Y$ by the
structuring element $S$. 

\subsection{Dilations as Space Functions}
\label{ssec:dilation-sfunc}

We now state that the power set with the usual order and the set of all
dilations over it form a \emph{(completely distributive) spatial constraint system}; i.e.,
an scs whose underlying lattice is completely distributive. 

\begin{theorem}[]
\label{thm:dilation-sf}
Let $M$ be a module. Then $(\pcal(M),\subseteq,(\dilatn{S})_{S \subseteq M})$
is a completely distributive scs.
\end{theorem}

\begin{proof}
The power set of any set ordered by inclusion is a completely distributive
lattice with join $\join = \cup$ and meet $\meet =
\cap$~\cite{davey2002introduction}. Hence $(\pcal(M),\subseteq)$ is a
completely distributive cs.

It remains to prove that for any $S \in \pcal(M)$, the dilation $\dilatn{S}$
is a space function. From Prop.~\ref{prop:join-preservation} part (2) it suffices to show that $\dilatnapp{S}{\bigcup_i A_i} = \bigcup_i \dilatnapp{S}{A_i}$ for
every arbitrary union $\bigcup_i A_i \in \pcal(M)$. This follows from the following equations:
\begin{align*}
 \dilatnapp{S}{\bigcup_i A_i}
=&\ \left\{\vect{x}+\vect{e} \mid \vect{x} \in \bigcup_i A_i \text{ and }
 \vect{e} \in S \right\}\\
=&\ \{\vect{x}+\vect{e} \mid \vect{x} \in A_i \text{ for some } i  \text{ and }
 \vect{e} \in S \}\\
=&\ \{\vect{x}+\vect{e} \mid \vect{x}+\vect{e} \in \dilatnapp{S}{A_i}
 \text{ for some } i \}\\
=&\ \bigcup_i \dilatnapp{S}{A_i}\\
\end{align*}

Thus, $(\pcal(M),\subseteq,(\dilatn{S})_{S \subseteq M})$ is a completely distributive scs.

\end{proof}

The above theorem states that dilations are space functions. Erosions, on the
other hand, are space projections (see Def.~\ref{def:agent-proj}).

\begin{proposition}
\label{prop:erosn-proj}

Let $M$ be a module. For every $S \subseteq M$, the function $\erosn{S}$ is
the $S$-projection in the scs $(\pcal(M),\subseteq,(\dilatn{S})_{S \subseteq M})$.
\end{proposition}

\begin{proof}
Let $M$ be a module and $S \subseteq M$. To prove that $\erosn{S}$ is an
$S$-projection we show that dilations and erosions form a Galois connection,
i.e., for every $X, Y \subseteq \pcal(M)$, $\dilatnapp{S}{X} \subseteq Y$ iff
$X \subseteq \erosnapp{S}{Y}$. It is known that a Galois connection determines
each function uniquely. Therefore from Prop.~\ref{projection:prop}~(1)
it follows that the erosion $\erosn{S}$ must then be a projection. 

Pick arbitrary $X, Y, S\subseteq \pcal(M)$. We have $\dilatnapp{S}{X} =
\bigcup_{\vect{v} \in S} X \oplus \{\vect{v}\} \subseteq Y$ iff for every
$\vect{u} \in S$, $X \oplus \{\vect{u}\} \subseteq Y$. Furthermore, with the
help of the monoid laws for $\oplus$ (Prop.~\ref{monoid:prop}) we can show that
for every $\vect{u} \in S$:
\begin{align*}
 &X \oplus \{\vect{u}\} \subseteq Y\\
\textit{iff}\ &\expl{\text{Property of }\subseteq}\\
&(X \oplus \{\vect{u}\}) \cup Y = Y\\
\textit{iff}\ &\expl{Z = Z \oplus \{\vect{0}\} \text{ and } \{\vect{0}\} = \{-\vect{u}\} \oplus \{\vect{u}\}}\\
&(X \oplus \{\vect{u}\}) \cup (Y \oplus \{-\vect{u}\} \oplus
\{\vect{u}\}) = Y\\
\textit{iff}\ &\expl{X\oplus(A \cup B)= (X\oplus A) \cup (X \oplus B)}\\
&(X \cup (Y \oplus \{-\vect{u}\})) \oplus \{\vect{u}\} = Y\\
\textit{iff}\ &\expl{\text{Adding } \{-\vect{u}\} \text{ by } \oplus}\\
&(X \cup (Y \oplus \{-\vect{u}\})) \oplus \{\vect{u}\} \oplus
\{-\vect{u}\} = Y \oplus \{-\vect{u}\}\\
\textit{iff}\ &\expl{\{-\vect{u}\} \oplus \{\vect{u}\} = \{\vect{0}\} \text{ and } Z \oplus \{\vect{0}\} = Z}\\
&X \cup (Y \oplus \{-\vect{u}\}) = Y \oplus \{-\vect{u}\}\\
\textit{iff}\ &\expl{\text{Property of }\subseteq}\\
&X \subseteq Y \oplus \{-\vect{u}\}.\\
\end{align*}
Clearly for every $\vect{u} \in S,$ $X \subseteq Y \oplus \{-\vect{u}\}$
iff $X \subseteq \bigcap_{\vect{v} \in S} Y \oplus \{-\vect{v}\} =
\erosnapp{S}{Y}$. 
We have then established that $\dilatnapp{S}{X} \subseteq Y$ iff
$X \subseteq \erosnapp{S}{Y}$ as wanted.
\end{proof}

In the proof of the above proposition, we show that dilations and erosions
form a Galois connection. This is a known fact in the MM community for
$\rmat^n$ or $\zmat^n$ space~\cite{ronse-mmclat-1990}. Here we proved a more
general version of it for modules.

\begin{figure}[t]
\centering
\includegraphics[scale=0.3]{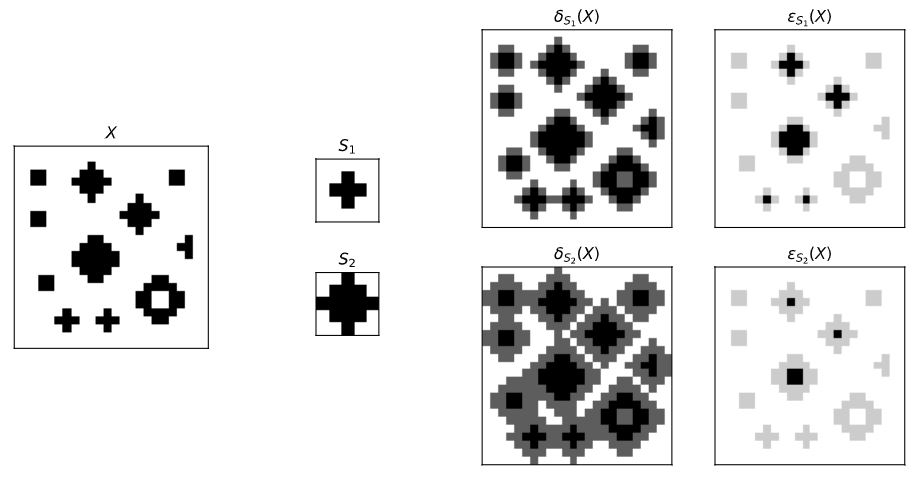}
\caption{From left to right, image $X$, structuring elements $S_1$ and $S_2$,
dilations $\dilatnapp{S_1}{X}$ and $\dilatnapp{S_2}{X}$, and  erosions
$\erosnapp{S_1}{X}$ and $\erosnapp{S_2}{X}$. Structuring elements are centered
at the origin. Pixels added/removed by dilation/erosion are depicted in
dark/light grey.} 
\label{fig:mm-dilations-2}
\end{figure}

\hltext{
\begin{remark}[An Epistemic Interpretation of MM operations]
\label{dilation-epistemic:remark}
\resp{We extended the following remark according to {\bf N46}. Also we added Fig.~\ref{fig:mm-dilations-2}}
Since we have shown that dilation is a space function, and erosion is a space
projection, we can now think of these functions in terms of information
available to an agent. Thus, for example in Fig.~\ref{fig:mm-galois}, when $Y$
is the actual state, $\erosnapp{S}{Y}$ shows what $S$ perceives. The
interpretation of $\dilatn{S}$ is more complicated: in fact, we interpret $X$ as
the information available to, or perceived by, agent $S$, when $\dilatnapp{S}{X}$
is the real situation. 

We notice that with the interpretation of dilations as visual perception, an
agent with perfect vision has the structuring element which is a single pixel
at the origin, and if structuring elements $S_1 \subseteq S_2$, then $S_2$
represents vision which is more blurry than the vision represented by $S_1$,
as seen in Fig.~\ref{fig:mm-dilations-2}. Thus, the agent only perceives some
of the real information, and some parts of the actual state of the world are
hidden from the agent. Specifically, we can consider this as an agent with
blurred vision who does not perceive some of the edges of objects. In the
instance of Fig.~\ref{fig:mm-dilations-2}, there are two agents, agent 1,
corresponding to structuring element $S_1$, with slightly blurred vision, and
agent 2 with corresponding structuring element $S_2$, with extremely blurred
vision.
Neither agent can perceive the edges of the objects.
So, when $\dilatnapp{S_1}{X}$ is the true state of affairs, $S_1$ perceives
the information in $X$; in effect, losing one pixel at the edge of each
object. On the other hand, when $\dilatnapp{S_2}{X}$ is the real situation,
$S_2$ perceives it as $X$, because this agent loses two pixels from the edge
of every object.

\noindent Thus, given an agent $S$, we may think of $\dilatnapp{S}{X}$ as the situation
where agent $S$ has information $X$, because $X$ is what the agent perceives
when $\dilatnapp{S}{X}$ is the real situation. 

In contrast, in Fig.~\ref{fig:mm-dilations-2} we note that if $X$ represents
the real situation, then agent 1 perceives $\erosnapp{S_1}{X}$, losing one
pixel at the edge of each object. Similarly, agent 2 perceives
$\erosnapp{S_2}{X}$ losing two pixels at the edge of each object.

In Fig.~\ref{fig:mm-galois}, the agent's vision is blurred in the vertical
direction: they only perceive a pixel if it also has another pixel below it,
but they are unable to perceive the bottom edge of any object.
\qed
\end{remark}
}

One may wonder if every space function over $(\pcal(M),\subseteq)$ is a
dilation $\dilatn{S}$ for some $S\subseteq M$.
Prop.~\ref{prop:non-dilations} answers this question negatively. First,
we need to introduce a new family of functions over modules.

\begin{definition}[Scale Function]
\label{def:scale-fun}
Let $M$ be a module over a ring $R$. Given $r\in R$, a scale by $r$ is a
function $\func{\scale{r}}{\pcal(M)}{\pcal(M)}$ defined as $\scaleapp{r}{X} =
\{ r \vect{u} \mid \vect{u} \in X \}$.
\index{Scale function}
\index[operator]{$\scale{r}$, scale by $r$}
\end{definition} 
 
It is easy to see that $\scale{r}$ is a space function over $(\pcal(M),\subseteq)$.
 
\begin{proposition}
Let $M$ be a module over a ring $R$. Then for every $r \in R$, 
$\scale{r}$ is a space function over $(\pcal(M),\subseteq)$.
\end{proposition}
\begin{proof}
Let $M$ be a module over a ring $R$. From
Prop.~\ref{prop:join-preservation}~(2), it suffices to show that given $r \in
R$ and an arbitrary $\bigcup_i A_i \in \pcal(M)$, $\scaleapp{r}{\bigcup_i A_i}
= \bigcup_i \scaleapp{r}{A_i}$.
Indeed, we have
\(
\scaleapp{r}{\bigcup_i A_i}
=\{r \vect{u} \mid \vect{u} \in \bigcup_i A_i \}
=\{r \vect{u} \mid \vect{u} \in A_i \text{ for some } i \}
=\{r \vect{u} \mid r \vect{u} \in \scaleapp{r}{A_i} \text{ for some } i \}
=\bigcup_i \scaleapp{r}{A_i}
\) 
as wanted. 
\end{proof}

The following proposition gives us a necessary and sufficient condition for a
scale function to be a dilation.  In particular, it tells us that we can have  infinitely
many space functions that \emph{are not dilations} by some structuring element if
the underlying module is, for example, the Euclidean vector space $\rmat^n$
or the grid $\zmat^n$ (see Ex.~\ref{module:example}).
 
\begin{proposition}
\label{prop:non-dilations}
Let $M$ be a module over a ring $R$. Then for each $r \in R$,
$\scale{r}=\dilatn{S}$ for some $S\subseteq M$ if and only if $r$ is a
multiplicative identity for $M$.
\end{proposition}

\begin{proof}
Suppose that $r$ is a multiplicative identity for $M$. Take $S = \{
\vect{0}\}$. Clearly $\scale{r} = \dilatn{\{ \vect{0} \}}$. For the other
direction we proceed by contradiction. Let us suppose that $r$ is not a
multiplicative identity for $M$ but that there exists $S$ such that $\scale{r} =
\dilatn{S}$. By applying both space functions to $\{ \vect{0}\}$, we obtain
$\scaleapp{r}{\{\vect{0}\}} = \{ r \vect{0} \} = \{ \vect{0} \} =
\dilatnapp{S}{\{\vect{0}\}}$. Then for every $\vect{v} \in S$,  $\vect{0} +
\vect{v} = \vect{0}$, hence $\vect{v} = \vect{0}$. Thus $S = \{ \vect{0} \}$.
It follows that for every $\vect{u} \in M$, $\scaleapp{r}{\{\vect{u}\}} =  \{ r
\vect{u} \} = \dilatnapp{\{ \vect{0} \}}{\{\vect{u}\}}=\{ \vect{u}\}$. This
implies that for every $\vect{u} \in M$, $r\vect{u} = \vect{u}$, thus $r$ is a
multiplicative identity for $M$, a contradiction. 

\end{proof}

\subsection{The Distributed Spaces and Dilations}
\label{ssec:dspace-dilation}

We have shown that dilations are space functions while erosions are space
projections. However, the main construction of this paper is that of
\emph{distributed spaces}: The greatest space function below a given set of
space functions. The problem we shall address is the following: Given two
dilations $\dilatn{A}$ and $\dilatn{B}$, \emph{find} the greatest space function
$\Dfun{\{A,B\}}$ below them. Let us consider some issues regarding this
question. 

Recall that join and meet operations of the power set $(\pcal(M),\subseteq)$
are set union and intersection. Notice that simply taking the point-wise
greatest lower bound does not work, i.e., in general, the equation
$\Dfun{\{A,B\}}(X) = \dilatn{A}(X) \cap \dilatn{B}(X)$ \emph{does not hold}.

\begin{example}
Consider Ex.~\ref{ex:non-distri}  with  the Euclidean
one-dimensional vector space $M=\rmat$, $X=\{0,1\}$, $A=\{1\}$ and
$B=\{2\}$. Let $f(Y) = \dilatn{A}(Y) \cap \dilatn{B}(Y)$ for every $Y\subseteq
M.$ One can verify that $f(\{0 \} \cup \{1\}) = \{2\} \neq \emptyset =
f(\{0\}) \cup f(\{1\})$, hence $f$ is  not even a space function.\qed
\end{example}

Furthermore, notice Prop.~\ref{prop:non-dilations}
tells us that there are space functions over $(\pcal(M),\subseteq)$ that are not
dilations. Thus, in principle it is not clear if $\Dfun{\{A,B\}}$ is itself a
dilation and if it is, we would like to identify what its structuring element
should be.  

The main result of this section, given next, addresses the above issues. 

\begin{theorem}[Distributed Spaces as Dilations]
\label{thm:ds-intersec}
Suppose that $M$ is a module. Let $(\Dfun{S})_{S \subseteq M}$ be the
distributed spaces of the scs $(\pcal(M),\subseteq,(\dilatn{S})_{S \subseteq
M})$. Then $\Dfun{\{A,B\}} = \dilatn{A \cap B}$ for every $A,B \subseteq M$.
\end{theorem}

Therefore, the greatest space function below two dilations is a dilation by
the intersection of their structuring elements. According to our intuition about dilations, the theorem tells us that given $\dilatn{A}$ and
$\dilatn{B}$, the dilation $\Dfun{\{A,B\}}$ applied to an image $X$ can be
intuitively described as the image obtained by re-drawing $X$ with (the brush)
$A \cap B$. This is illustrated in Fig.~\ref{fig:mm-dilations} by showing
$\Dfun{\{A,B\}}$ applied to an image $X$ \footnote{The result of $\Dfun{\{A,B\}}(X)$ was computed using the $O(n^2)$ procedure
mentioned in the beginning of the proof of Prop.~\ref{algo:prop} from Section~\ref{DistributedSpaces:section}.}.

We now devote the final part of this section to illustrate the application of the theory developed in previous sections to prove the above result. 

\subsection{Application: Proof of Theorem~\ref{thm:ds-intersec}}
\label{ssec:proof-thm-ds-intersec}

We wish to prove that $\Dfun{\{A,B\}} = \dilatn{A \cap B}$ for $A,B \subseteq
M$. From Def.~\ref{def:dilation-set}, we have $\dilatnapp{A \cap B}{X} = X
\oplus (A \cap B)$ and, from Th.~\ref{thm:comp-algo}, we have
$\Dfun{\{A,B\}}(X)=\bigcap_{Y\subseteq X}(Y \oplus A) \cup ((X\setminus Y)
\oplus B )$. Therefore, it suffices to show that
\[
X \oplus (A \cap B)
=\bigcap_{Y\subseteq X}(Y \oplus A) \cup ((X\setminus Y) \oplus B )
\]
for all $X \subseteq M$. Recall that the above equality is the distributivity
equation for the Minkowski addition discussed in Eq.~\ref{oplus-eq}. It
is important also to recall the equation $X\oplus(A \cap B) = (X\oplus A) \cap
(X \oplus B)$ \emph{does not hold} in general. Nevertheless, it does if $X$ is
a singleton set as shown next.

Let us  consider the \emph{singleton case}: Suppose that $X$
is an arbitrary set of the form $\{ \vect{v} \}$. We obtain the following equations:
\begin{align*}
\Dfun{\{A,B\}}(\{ \vect{v} \}) = & \ \bigcap_{Y\subseteq X}(Y \oplus A) \cup ((X\setminus Y) \oplus B )\\ 
=&\ ( (\emptyset \oplus A) \cup (\{ \vect{v} \} \oplus B)) \cap ((\{ \vect{v} \}
 \oplus A) \cup (\emptyset \oplus B))\\
=&\ (\{ \vect{v} \} \oplus B)  \cap (\{ \vect{v} \} \oplus A) \\
=&\ \{ \vect{v} + \vect{w} \mid \vect{w} \in B   \} \cap \{  \vect{v} + \vect{w}
 \mid \vect{w} \in A \}\\
=&\ \{ \vect{v} + \vect{w} \mid \vect{w} \in A \cap B \}\\
=&\ \dilatnapp{A \cap B}{\{ \vect{v} \}}.
\end{align*}

Now for the general case we will use the \emph{continuity of space functions}.
Suppose that $X$ is an arbitrary set.  From Def.~\ref{ds:def} and
Th.~\ref{thm:dilation-sf} we know that $\Dfun{\{A,B\}}$ and $\dilatn{A \cap
B}$ are \emph{space functions}. Furthermore from
Prop.~\ref{prop:join-preservation} it follows that space functions preserve
arbitrary joins. Thus, with the help of this preservation of arbitrary joins
(unions) and the singleton case above, we can obtain the desired result in a
simple way; namely $\Dfunapp{\{A,B\}}{X} = $
\[
\ \Dfunapp{\{A,B\}}{\bigcup_{\vect{v} \in X} \{ \vect{v} \}}
= \bigcup_{\vect{v} \in X} \Dfunapp{\{A,B\}}{\{ \vect{v} \}}
= \bigcup_{\vect{v} \in X} \dilatnapp{A\cap B}{\{ \vect{v} \}}
= \dilatnapp{A\cap B}{\bigcup_{\vect{v} \in X}\{ \vect{v} \}}
= \dilatnapp{A\cap B}{X}.
\]\qed

\subsection{Summary of Section~\ref{sec:app}}
\label{ssec:sum-app}

In this section we provided an application of the theory developed in
Section~\ref{sec:dist-info} to geometry and Mathematical Morphology. First, we
recalled the notion of Minkowski addition of sets in vector spaces as an
operation in the algebraic structure of modules
(Section~\ref{ssec:mm-modules}) and showed some of its basic properties
(Sections~\ref{ssec:mm-mink-add} and~\ref{ssec:mm-mat-morph}). We then proved
that given a module $M$, the structure $(\pcal(M),\subseteq,(\dilatn{S})_{S
\subseteq M})$ is an scs where dilations are space functions
(Section~\ref{ssec:dilation-sfunc}). Furthermore, we showed that erosions are
space projections (Prop.\ref{prop:erosn-proj}).  Since space functions and
projections can also be viewed as the information a given agent sees, we then
gave a natural epistemic interpretation of these MM operations as an agent's
perception of a given image (Remark \ref{dilation-epistemic:remark}).

In Sec.~\ref{ssec:dspace-dilation} we proved that given two dilations
$\dilatn{A}$ and $\dilatn{B}$, the \emph{distributed space function} of the
group $\{A,B\}$ (i.e., $\Dfun{\{A,B\}}$) corresponds to the dilation
$\dilatn{A \cap B}$.  Finally, in Section~\ref{ssec:proof-thm-ds-intersec} we
used the theory developed in previous sections to prove
Th.~\ref{thm:ds-intersec} and a novel law for $X \oplus (A \cap B)$: i.e., 
\(
X \oplus (A \cap B)
=\bigcap_{Y\subseteq X}(Y \oplus A) \cup ((X\setminus Y) \oplus B )
\).

\section{Proofs of Section~\ref{sec:dist-info}}
\label{ssec:proof-thms}

In this section we present proofs of some results from Sec.~\ref{sec:dist-info}.

\hltext{
\subsection{Proof of Proposition~\ref{prop:aumann-dist}}
\label{ssec:proof-aumann-dist}
\resp{The next proof is new, associated to {\bf N43}.}
Let $\acal = (S, \pcal_1, \dots, \pcal_n)$  be an Aumann structure and let 
$\mathbf{C}(\acal)$ be its induced scs defined in Ex.\ref{aumann:example}.
Then $\Dfun{I} = \Gfun{I}$ for every $I \subseteq G= \{ 1,\ldots, n\}$.

\begin{proof}
Let $I \subseteq G$. We shall prove that
(i) $\Gfun{I} \in \Cs$,
(ii) $\Dfunapp{I}{c} \cgeq \Gop{I}{c}$ and,
(iii) $\Dfunapp{I}{c} \cleq \Gop{I}{c}$.

\begin{enumerate}[(i)]
\item Recall that in $\C(\acal)$ joins are intersections
(Sec.~\ref{ssec:ds-aumann}). From Prop.~\ref{prop:join-preservation}~(2),
it suffices to show that given an arbitrary $\bigcap_j A_j \in \C$,
$\Gop{I}{\bigcap_j A_j} = \bigcap_j \Gop{I}{A_j}$.
Indeed, we have \(\Gop{I}{\bigcap_j A_j}=
\{s \mid \bigcap_{i \in I} \pcal_i(s) \subseteq \bigcap_j A_j \}
=\{s \mid \bigcap_{i \in I} \pcal_i(s) \subseteq A_j \text{ for every } j \}
=\bigcap_j \{s \mid \bigcap_{i \in I} \pcal_i(s) \subseteq A_j \}
=\bigcap_j \Gop{I}{A_j}
\) 
as wanted.
\item By definition of $\Afun{i}$ and $\Gfun{I}$ (see
Sec.~\ref{ssec:ds-aumann}), we know that for every $c \in \C$, $\Kop{i}{c}
\subseteq \Gop{I}{c}$. This implies $\Gop{I}{c} \cleq \Kop{i}{c}$. Then
$\Gop{I}{c}$ is a lower bound in $\Cs$ of the set of all the $\Kop{i}{c}$. Thus, by
Remark~\ref{rmk:ds-meet} $\Dfunapp{I}{c} \cgeq \Gop{I}{c}$ for every $c \in
\C$.

\item Now, let $t \in \Gop{I}{c}$. Then, by definition of $\Gfun{I}$, we have $\bigcap_{i \in I} \pcal_i(t) \subseteq c$. From Th.~\ref{thm:delta-ast},
\[
\Dfunapp{I}{c} = \bigcup\left\{ \bigcap_{i \in I} \Kop{i}{e_i} \mid \bigcap_{i \in I} e_i \subseteq c \right\}.
\]
Take $e_i = \pcal_i(t)$ for every $i \in I$. By assumption, $\bigcap_{i \in I}
e_i \subseteq c$ and, by definition of $\Afun{i}$, we know that $t \in
\Kop{i}{e_i}$. Therefore, $t \in \bigcap_{i \in I}\Kop{i}{e_i}$, i.e., $t \in
\Dfunapp{I}{c}$ for every $c \in \C$. Thus $\Dfunapp{I}{c} \cleq
\Gop{I}{c}$ for every $c \in \C$, as wanted.
\end{enumerate}
\end{proof}
}
\hltext{
\subsection{Proof of Proposition~\ref{prop:kripke-dist}}
\label{ssec:proof-kripke-dist}
\resp{This proof is new.}
Let $\kbold(\scal_n(\Phi)) = ({\Con},\sqsubseteq,\Kfun{1},\ldots,\Kfun{n})$ be an scs where $\scal_n(\Phi)$ is a non-empty set of $n$-agent Kripke structures over
$\Phi$ as in Def.~\ref{def:kripke-scs}. Then $\Dfun{I} = \Dkfun{I}$ for every $I \subseteq G= \{1,\ldots,n\}$.

\begin{proof}
Let $I \subseteq G$. We shall prove that
(i) $\Dkfun{I} \in \Cs$,
(ii) $\Dfunapp{I}{c} \cgeq \Dkfunapp{I}{c}$ and,
(iii) $\Dfunapp{I}{c} \cleq \Dkfunapp{I}{c}$.

\begin{enumerate}[(i)]
\item Recall that in $\kbold(\scal_n(\Phi))$ joins are intersections
(Def.~\ref{def:kripke-scs}). From Prop.~\ref{prop:join-preservation}~(2), it
suffices to show that given an arbitrary $\bigcap_j A_j \in \C$,
$\Dkfunapp{I}{\bigcap_j A_j} = \bigcap_j \Dkfunapp{I}{A_j}$.

Indeed, we have
\(\Dkfunapp{I}{\bigcap_j A_j}=
\{ (M,s) \in \Delta \mid \forall t: (s,t) \in \bigcap_{i \in I} \rcal_i  \text{ implies } (M,t) \in \bigcap_j A_j \}
=\{ (M,s) \in \Delta \mid \forall t: (s,t) \in \bigcap_{i \in I} \rcal_i  \text{ implies } (M,t) \in A_j \text{ for every } j \}
=\bigcap_j \{ (M,s) \in \Delta \mid \forall t: (s,t) \in \bigcap_{i \in I} \rcal_i  \mbox{ implies } (M,t) \in A_j \}
=\bigcap_j \Dkfunapp{I}{A_j}
\) 
as wanted.
\item Recall that in $\kbold(\scal_n(\Phi))$, for every $i \in
\{1,\ldots,n\}$, $\sfun{i}=\Kfun{i}$ (Prop.~\ref{prop:kripke-scs}). By
definition of $\Kfun{i}$ and $\Dkfun{I}$ (see Def.~\ref{def:kripke-scs} and
Ex.~\ref{ex:kripke-ds}), we know that for every $X \in \C$, $\Kfunapp{i}{X}
\subseteq \Dkfunapp{I}{X}$. This implies $\Dkfunapp{I}{X} \cleq
\Kfunapp{i}{X}$. Then $\Dkfunapp{I}{X}$ is a lower bound in $\Cs$ of the set
of all the $\Kfunapp{i}{X}$. Thus, by Remark~\ref{rmk:ds-meet} $\Dfunapp{I}{X}
\cgeq \Dkfunapp{I}{X}$ for every $X \in \C$.

\item Now, let $(M,s) \in \Dkfunapp{I}{X}$. Then, by definition of
$\Dkfun{I}$, we have for all $t$, $(s,t) \in \bigcap_{i \in I} \rcal_i$
implies $(M,t) \in X$. From Th.~\ref{thm:delta-ast},
\[
\Dfunapp{I}{X} = \bigcup\left\{ \bigcap_{i \in I} \Kfunapp{i}{X_i} \mid \bigcap_{i \in I} X_i \subseteq X \right\}.
\]
Take $X_i = \{(M,t) \mid (s,t) \in \bigcap_{i \in I} \rcal_i\}$ for every $i \in I$.
It is clear that $\bigcap_{i \in I} X_i \subseteq X$.
Notice that by definition of $\Kfun{i}$, we know that $(M,s) \in \Kfunapp{i}{X_i}$.
Therefore, $(M,s) \in \bigcap_{i \in I}\Kop{i}{X_i}$, i.e., $(M,s) \in
\Dfunapp{I}{X}$ for every $X \in \C$. Thus $\Dfunapp{I}{X} \cleq
\Dkfunapp{I}{X}$ for every $X \in \C$, as wanted.
\end{enumerate}
\end{proof}
}

\subsection{Proofs of Section~\ref{DistributedSpaces:section}}

This section is devoted to the proofs of the compositionality properties of
distributed spaces given in Section~\ref{DistributedSpaces:section}.
To simplify our notation, we define sets of $J$-tuples whose join derive a given constraint $c$.

\begin{definition}
\label{def:tuples-set}
Let $(\Con, \cleq)$ be a cs and $J$ some index set. For every $c \in \Con$,
let $\tupleset{c}{J} = \{ (a_j)_{j \in J} \in \tuplespace{\Con}{J} \mid
\bigjoin_{j \in J} a_j \cgeq c\}$. For simplicity, we use $\tuplesetc{c}$
instead of $\tupleset{c}{J}$ when no confusion arises.
\index[operator]{$\tupleset{c}{J}$, set of $J$-tuples s.t. the join of its elements derives $c$}
\end{definition}

\subsection{Proof of Theorem~\ref{thm:delta-ast}}
\label{sssec:proof-delta-ast}

The function $\dplusfun{K}$ is defined in Th.~\ref{thm:delta-ast}. Here we find it convenient to use the following simplified version.

\begin{proposition}
\label{def:dplus}
$\dplusfunapp{K}{c} = \bigmeet\{ \bigjoin_{k \in K} \sfunapp{k}{a_k}
\mid (a_k)_{k \in K} \in \tuplesetc{c} \}$.
\index[operator]{$\dplusfun{K}$}
\end{proposition}

The following is an immediate consequence of the above definition. 
\begin{proposition}
\label{prop:dplus-monotonic} The function $\dplusfun{K}$ is monotonic.
\end{proposition}
\begin{proof}
Let $c \cgeq d$. We have $\{ \bigjoin_{k \in
K}\sfunapp{k}{a_k} \mid (a_k)_{k \in K} \in \tuplesetc{c} \} \subseteq \{ \bigjoin_{k \in K}
\sfunapp{k}{a_k} \mid (a_k)_{k \in K} \in \tuplesetc{d} \}$.
Thus $\dplusfunapp{K}{c} \cgeq \dplusfunapp{K}{d}$.
\end{proof}

Next lemma states that $\dplusfun{K}$ is a space function. As pointed out
in Section~\ref{DistributedSpaces:section}, the proof of continuity of $\dplusfun{K}$  uses the assumption of  $({\Con},\cleq)$ being completely distributive.

\begin{lemma}
\label{lemma:dplus-sfun}
Let $({\Con},\cleq,(\sfun{i})_{i \in G})$ be an scs.
Suppose that $({\Con},\cleq)$ is completely distributive.
Then, for any $K \subseteq G$, $\dplusfun{K}$ is a space function.
\end{lemma}

\begin{proof}
Let $({\Con},\cleq,(\sfun{i})_{i \in G})$ be an scs and assume
$({\Con},\cleq)$ to be completely distributive. To show that
$\dplusfun{K}$ is a space function we prove:
(i) it satisfies S.1 and S.2 in Def.~\ref{def:sfunc}, and
(ii) it is continuous.
\begin{enumerate}[(i)]
\item $\dplusfun{K}$ \emph{satisfies} S.1 \emph{and} S.2.

For S.1, one can verify that $\dplusfunapp{K}{\true} = \true$.

To prove S.2, it suffices to show that $\dplusfunapp{K}{c \join d} \cleq
\dplusfunapp{K}{c} \join \dplusfunapp{K}{d}$. The other direction follows by
monotonicity (Prop.~\ref{prop:dplus-monotonic}). Consider the
following derivation:

\begin{align*}
&
\dplusfunapp{K}{c} \join \dplusfunapp{K}{d}\\
=\
&
\expl{\text{Definition of } \dplusfun{K}}\\
&
\dplusfunapp{K}{c} \join \bigmeet\left\{ \bigjoin_{k \in K}\sfunapp{k}{b_k} \mid (b_k)_{k \in K} \in \tuplesetc{d} \right\}\\
=\
&
\expl{\join \text{ distributes over } \meet}\\
&
\bigmeet\left\{ \dplusfunapp{K}{c} \join \bigjoin_{k \in K}\sfunapp{k}{b_k} \mid (b_k)_{k \in K} \in \tuplesetc{d} \right\}\\
=\
&
\expl{\text{Definition of } \dplusfun{K}}\\
&
\bigmeet\left\{ \bigmeet\left\{ \bigjoin_{k \in K}\sfunapp{k}{a_k} \mid (a_k)_{k \in K} \in \tuplesetc{c} \right\} \join \bigjoin_{k \in K}\sfunapp{k}{b_k} \mid (b_k)_{k \in K} \in \tuplesetc{d} \right\}\\
=\
&
\expl{\join \text{ distributes over } \meet}\\
&
\bigmeet\left\{ \bigmeet\left\{ \bigjoin_{k \in K}\sfunapp{k}{a_k} \join \bigjoin_{k \in K} \sfunapp{k}{b_k} \mid (a_k)_{k \in K} \in \tuplesetc{c} \right\} \mid (b_k)_{k \in K} \in \tuplesetc{d} \right\}\\
=\
&
\expl{ \text{Associativity of } \meet }\\
&
\bigmeet\left\{ \bigjoin_{k \in K}\sfunapp{k}{a_k} \join \bigjoin_{k \in K} \sfunapp{k}{b_k} \mid (a_k)_{k \in K} \in \tuplesetc{c} \text{ and } (b_k)_{k \in K} \in \tuplesetc{d}\right\}\\
=\
&
\expl{ \text{Associativity of } \join }\\
&
\bigmeet\left\{ \bigjoin_{k \in K}\left(\sfunapp{k}{a_k} \join \sfunapp{k}{b_k}\right)\mid (a_k)_{k \in K} \in \tuplesetc{c} \text{ and } (b_k)_{k \in K} \in \tuplesetc{d}\right\}\\
=\
&
\expl{ \sfun{k} \text{ is a space function}}\\
&
\bigmeet\left\{\bigjoin_{k \in K} \sfunapp{k}{a_k \join b_k} \mid
(a_k)_{k \in K} \in \tuplesetc{c} \text{ and } (b_k)_{k \in K} \in \tuplesetc{d} \right\}\\
\cgeq\
&
\expl{c_k = a_k \join b_k;\ (\bigjoin_{k \in K} c_k \ = \ \bigjoin_{k \in K}(a_k \join b_k) \
\cgeq\ c \join d)  \text{ implies } (c_k)_{k \in K} \in \tuplesetc{c \join d}}\\
&
\bigmeet\left\{\bigjoin_{k \in K} \sfunapp{i}{c_k} \mid 
(c_k)_{k \in K} \in \tuplesetc{c \join d} \right\}\\
=\
&
\expl{\text{Definition of } \dplusfun{K}}\\
&
\dplusfunapp{K}{c \join d}
\end{align*}

\item $\dplusfun{K}$ \emph{is continuous.}

Let $D$ be any directed set on $\Con$, we will prove $\dplusfunapp{K}{\bigjoin
D} = \bigjoin \{\dplusfunapp{K}{d} \mid d \in D \}$. We proceed with
$\dplusfunapp{K}{\bigjoin D} \cleq \bigjoin \left\{\dplusfunapp{K}{d} \mid d
\in D \right\}$.
The other direction follows by monotonicity.

By definition of $\dplusfun{K}$, $\bigjoin \left\{\dplusfunapp{K}{d} \mid d
\in D \right\} = \bigjoin \{\bigmeet\{ \bigjoin_{k \in K}\sfunapp{k}{a_k} \mid
(a_k)_{k \in K} \in \tuplesetc{d} \} \mid d \in D \}$.  Since $(\Con,\cleq)$
is completely distributive (see Def.~\ref{def:c-lat}), for the subset $\{\bigjoin_{k \in K}\sfunapp{k}{a_k}\}_{d \in D, (a_k)_{k \in
K} \in \tuplesetc{d}}$ of $\Con$, we have
\[
\bigjoin_{d \in D} \left\{\bigmeet_{(a_k)_{k {\in} K} \in \tuplesetc{d}}
\left\{ \bigjoin_{k \in K}\sfunapp{k}{a_k} \right\} \right\}
= \bigmeet_{f \in F} \left\{\bigjoin_{d \in D}
\left\{ \bigjoin_{k \in K}\sfunapp{k}{f_k(d)} \right\} \right\}
\]
where $F$ is the class of \emph{choice functions} $f$ choosing for each $d \in
D$ some index $f(d) \in \tuplesetc{d}$. Recall that $f_k(d)$ is the $k$-th element of $K$-tuple $f(d)$.
We can rewrite the right-hand side of
the above equality using $\join$ properties and the fact that $\sfun{k}$
preserves arbitrary joins (see Prop.~\ref{prop:join-preservation}). Then
we obtain
\[
\bigjoin_{d \in D} \left\{\bigmeet_{(a_k)_{k {\in} K} \in \tuplesetc{d}}
\left\{ \bigjoin_{k \in K}\sfunapp{k}{a_k} \right\} \right\}
=\bigmeet_{f \in F} \left\{\bigjoin_{k \in K}
 \sfunapp{k}{\bigjoin_{d \in D} f_k(d)} \right\}. 
\]
We now show that for every $f \in F$,
$\left(\bigjoin_{d \in D} f_k(d)\right)_{k \in K} \in \tuplesetc{\join D}$.
Notice that for every $d \in D$, $\bigjoin_{k \in K} f_k(d) \cgeq d$.
We have
\[
\bigjoin_{k \in K}\left( \bigjoin_{d \in D} f_k(d)\right)
=\bigjoin_{d \in D}\left( \bigjoin_{k \in K} f_k(d)\right)
\cgeq \bigjoin_{d \in D} d
=\bigjoin D.
\]
Therefore $\left( \bigjoin_{d \in D} f_k(d)\right)_{k \in K} \in
\tuplesetc{\join D}$ (see Def.~\ref{def:tuples-set}).

Then, for every $f \in F$, the element $\bigjoin_{k \in K}
\sfunapp{k}{\bigjoin_{d \in D} f_k(d)} \in \{ \bigjoin_{k \in K}
\sfunapp{k}{a_k} \mid (a_k)_{k \in K} \in \tuplesetc{\join D} \}$, this
implies
\[
\left\{\bigjoin_{k \in K}
\sfunapp{k}{\bigjoin_{d \in D} f_k(d)}\ \Big|\ f \in F\right\}
\subseteq \left\{ \bigjoin_{k \in K} \sfunapp{k}{a_k}\ \Big|\ (a_k)_{k \in K}
\in \tuplesetc{\join D} \right\}.
\]
Consequently,
\[
\bigmeet_{f \in F} \left\{\bigjoin_{k \in K} \sfunapp{k}{\bigjoin_{d \in D}
f_k(d)} \right\}
\cgeq \bigmeet \left\{ \bigjoin_{k \in K}\sfunapp{k}{a_k}\ \Big|\
(a_k)_{k \in K} \in \tuplesetc{\join D} \right\} = \dplusfunapp{K}{\bigjoin D}.
\]

Thus, we conclude $\dplusfun{K}$ is continuous.

\end{enumerate}

\end{proof}

Finally we prove Th.~\ref{thm:delta-ast}. Its statement now is 
simplified due to Prop.~\ref{def:dplus}.

\begin{theorem*}[\ref{thm:delta-ast}]
Let $(\Dfun{I})_{I \subseteq G}$ be the distributed spaces of an scs
$({\Con},\cleq,(\sfun{i})_{i \in G})$.
Suppose that $({\Con},\cleq)$ is a completely distributive lattice.
Then $\Dfun{K} = \dplusfun{K}$.
\index[operator]{$\dplusfun{K}$}
\end{theorem*}

\begin{proof}
Let $(\Dfun{I})_{I \subseteq G}$ be the distributed spaces of an scs
$({\Con},\cleq,(\sfun{i})_{i \in G})$ and assume $({\Con},\cleq)$ to be
completely distributive.

We divide the proof in two parts:
I. $\dplusfun{K} \fleq \Dfun{K}$ and
II. $\Dfun{K} \fleq \dplusfun{K}$.

\begin{enumerate}[I.]
\item $\dplusfun{K} \fleq \Dfun{K}$.

Recall that from Def.~\ref{ds:def}, $\Dfun{K} = \max\{ h \in
\sfunspace{\C} \mid h \fleq \sfun{k} \mbox{ for all } k \in K\}$.
Thus we have to prove:
(i) $\dplusfun{K} \in \sfunspace{\C}$ and
(ii) $\dplusfun{K} \fleq \sfun{k}$ for every $k \in K$.

From Lemma~\ref{lemma:dplus-sfun} we have (i). For part (ii),
let $c \in \Con$ and $S = \{ \bigjoin_{k \in K}\sfunapp{k}{a_k}
\mid (a_k)_{k \in K} \in \tuplesetc{c} \}$.
From definition of $\dplusfun{K}$, for every $k \in K$, the element
$\sfunapp{k}{c} = \sfunapp{k}{c} \join \bigjoin_{j \in K \setminus \{k\}}
\sfunapp{j}{\true} \in S$.
Then for every $c \in \Con$, $\dplusfunapp{K}{c} = \bigmeet S \cleq \sfunapp{k}{c}$.
Therefore for every $k \in K$, $\dplusfun{K} \fleq \sfun{k}$.
Then, $\dplusfun{K} \fleq \Dfun{K}$ holds.

\item $\Dfun{K} \fleq \dplusfun{K}$.

Let $c \in \Con$ and $S = \{ \bigjoin_{k \in K}\sfunapp{k}{a_k}
\mid (a_k)_{k \in K} \in \tuplesetc{c} \}$.
Notice that, for any $(a_k)_{k \in K} \in \tuplesetc{c}$, $\bigjoin_{k \in K}
\Dfunapp{K}{a_k} \cleq \bigjoin_{k \in K}\sfunapp{k}{a_k}$.
Since $\Dfun{K}$ is a space function and monotonic, we know that
\[
\bigjoin_{k \in K}\Dfunapp{K}{a_k} = \Dfunapp{K}{\bigjoin_{k \in K} a_k} \cgeq \Dfunapp{K}{c}.
\]
Thus, for every $(a_k)_{k \in K} \in \tuplesetc{c}$, we have $\Dfunapp{K}{c}
\cleq \bigjoin_{k \in K}\sfunapp{k}{a_k}$, i.e.,
$\Dfunapp{K}{c}$ is a lower bound of $S$.
Then for every $c \in \Con$, $\Dfunapp{K}{c} \cleq \bigmeet S
= \dplusfunapp{K}{c}$. Therefore $\Dfun{K} \fleq \dplusfun{K}$ as wanted.
\end{enumerate}
Thus, from (I) and (II), we conclude Th.~\ref{thm:delta-ast}, i.e.,
$\Dfun{K} = \dplusfun{K}$.
\end{proof}

\subsection{Proof of Theorem~\ref{thm:comp-algo}}
\label{sssec:proof-comp-algo}

We now prove the compositional properties of distributed spaces introduced
in Th.~\ref{thm:comp-algo}.

\begin{theorem*}[\ref{thm:comp-algo}]
Let $(\Dfun{I})_{I \subseteq G}$ be the distributed spaces of an scs
$({\Con},\cleq,(\sfun{i})_{i \in G})$. Suppose that $({\Con},\cleq)$ is
completely distributive and let $I,J,K \subseteq G$ be such that
$I = J \cup K$. Then
\begin{enumerate}
\item $\Dfunapp{I}{c} = \bigmeet \left\{ \Dfunapp{J}{a} \join 
\Dfunapp{K}{b} \mid a,b \in \Con \mbox{ and } a \join b \cgeq c \right\}$.
\item $\Dfunapp{I}{c} = \bigmeet \left\{ \Dfunapp{J}{a} \join 
\Dfunapp{K}{c \sop a} \mid a \in \Con \right\}$.
\item $\Dfunapp{I}{c} =  \bigmeet \left\{ \Dfunapp{J}{a} \join
\Dfunapp{K}{c \sop a} \mid a \in \Con \mbox{ and } a \cleq c \right\}$.
\end{enumerate}
\end{theorem*}

We present the proof of each item separately.

\begin{pot}[\ref{thm:comp-algo} (1)]
Let $(\Dfun{I})_{I \subseteq G}$ be the distributed spaces of an scs
$({\Con},\cleq,(\sfun{i})_{i \in G})$ where $({\Con},\cleq)$ is
completely distributive. Let $I,J,K \subseteq G$ be such that $I = J \cup K$.

\begin{itemize}
\item $\Dfunapp{I}{c} \cleq \bigmeet\{ \Dfunapp{J}{a} \join \Dfunapp{K}{b} \mid
a,b \in \Con \mbox{ and } a \join b \cgeq c \}$.

Let $a,b,c \in \Con$ such that
$a \join b \cgeq c$. Since $(\Dfun{I})_{I \subseteq G}$ is a gdc (see
Th.~\ref{th:max-gdc}) and, both $J \subseteq I$ and $K \subseteq I$ hold,
for every $e \in \Con$, we have $\Dfunapp{I}{e} \cleq \Dfunapp{J}{e}$ and
$\Dfunapp{I}{e} \cleq \Dfunapp{K}{e}$. Then $\Dfunapp{I}{a} \join \Dfunapp{I}{b}
\cleq \Dfunapp{J}{a} \join \Dfunapp{K}{b}$. From the fact that $\Dfun{I}$ is a
space function and hence monotonic we conclude $\Dfunapp{I}{c}  \cleq \Dfunapp{I}{a} \join
\Dfunapp{I}{b}$. Therefore $\Dfunapp{I}{c} \cleq \Dfunapp{J}{a} \join
\Dfunapp{K}{b}$. Thus,
$\Dfunapp{I}{c} \cleq \bigmeet\{ \Dfunapp{J}{a} \join \Dfunapp{K}{b} \mid a,b
\in \Con \mbox{ and } a \join b \cgeq c \}$.

\item $\Dfunapp{I}{c} \cgeq \bigmeet\{ \Dfunapp{J}{a} \join \Dfunapp{K}{b} \mid
a,b \in \Con \mbox{ and } a \join b \cgeq c \}$.

Let $c \in \Con$ and $S = \{ \bigjoin_{i \in I} \sfunapp{i}{c_i} \mid (c_i)_{i
\in I} \in \tupleset{c}{I} \}$. We first show the following claim:

\begin{claim*} For every  $\bigjoin_{i \in
I} \sfunapp{i}{c_i} \in S$, there are some $a,b \in \Con$ such that (i) $a
\join b \cgeq c$ and (ii) $\bigjoin_{i \in I} \sfunapp{i}{c_i} \cgeq
\Dfunapp{J}{a} \join \Dfunapp{K}{b}.$ \end{claim*}

Let $\bigjoin_{i \in I} \sfunapp{i}{c_i} \in S.$ Since $I = J \cup K$, we have
$\bigjoin_{i \in I} \sfunapp{i}{c_i} = \bigjoin_{j \in J} \sfunapp{j}{c_j}
\join \bigjoin_{k \in K} \sfunapp{k}{c_k}$. Let $a = \bigjoin_{j \in J} c_j$
and $b = \bigjoin_{k \in K} c_k$.
From Th.~\ref{thm:delta-ast}, we have
\[
\Dfunapp{J}{a} = \bigmeet \left\{ \bigjoin_{j \in J} \sfunapp{j}{a_j}\
\Big|\ (a_j)_{j \in J} \in \tupleset{a}{J} \right\} \text{ and }
\Dfunapp{K}{b} = \bigmeet\left\{ \bigjoin_{k \in K} \sfunapp{k}{b_k}\
\Big|\ (b_k)_{k \in K} \in \tupleset{b}{K} \right\}.
\]
Given that $(\Con,\cleq)$ is completely distributive and by associativity of
$\meet$, $\Dfunapp{J}{a} \join \Dfunapp{K}{b} = \bigmeet R$ where $R = \left\{
\bigjoin_{j \in J} \sfunapp{j}{a_j} \join \bigjoin_{k \in K} \sfunapp{k}{b_k}\
\big|\ (a_j)_{j \in J} \in \tupleset{a}{J} \text{ and } (b_k)_{k \in K} \in
\tupleset{b}{K} \right\}$.
Clearly $a \join b \cgeq c$ and $\bigjoin_{i \in I} \sfunapp{i}{c_i} \in R$.
Then $\bigjoin_{i \in I} \sfunapp{i}{c_i} \cgeq
\bigmeet R =  \Dfunapp{J}{a} \join \Dfunapp{K}{b}$. This shows (i) and (ii). 

From the above claim and Th.~\ref{thm:delta-ast}, we obtain
$\Dfunapp{I}{c} = \bigmeet S \cgeq \bigmeet\{ \Dfunapp{J}{a} \join
\Dfunapp{K}{b} \mid a,b \in \Con \mbox{ and } a \join b \cgeq c \}$ as wanted.\qed
\end{itemize}
\end{pot}

\begin{pot}[\ref{thm:comp-algo} (2)]
Let $(\Dfun{I})_{I \subseteq G}$ be the distributed spaces of an scs
$({\Con},\cleq,(\sfun{i})_{i \in G})$ where $({\Con},\cleq)$ is
completely distributive. Let $I,J,K \subseteq G$ be such that $I = J \cup K$.

Let $a,c \in \Con$. Recall that $c \sop a$ represents the least element $e \in
\Con$ such that  $a \join e \cgeq c$. Take any $b \in \Con$ such  that $a
\join b \cgeq c$. Then $b \cgeq c \sop a$ and since space functions are 
monotonic $\Dfunapp{J}{a} \join  \Dfunapp{K}{b} \cgeq \Dfunapp{J}{a} \join
\Dfunapp{K}{c \sop a}.$ From this it follows that  $\bigmeet ( S \cup \{
\Dfunapp{J}{a} \join \Dfunapp{K}{c \sop a}\ , \Dfunapp{J}{a} \join
\Dfunapp{K}{b} \}) = \bigmeet  ( S \cup \{ \Dfunapp{J}{a} \join 
\Dfunapp{K}{c \sop a} \})$ for any $S \subseteq \Con$.

From Th.~\ref{thm:comp-algo}~(1) and the above argument, we have 
\begin{align*}
\Dfunapp{I}{c}
=& \bigmeet\left\{ \Dfunapp{J}{a} \join  \Dfunapp{K}{b} \mid  a
\join b \cgeq c \right\}\\
=& \bigmeet \left(\left\{ \Dfunapp{J}{a} \join  \Dfunapp{K}{b} \mid  a
\join b \cgeq c \right\} \cup \left\{ \Dfunapp{J}{a} \join  \Dfunapp{K}{c \sop a} \mid a
\in \Con \right\}\right)\\
=& \bigmeet \left\{ \Dfunapp{J}{a} \join  \Dfunapp{K}{c \sop a} \mid a \in \Con \right\}.
\end{align*}
Thus $\Dfunapp{I}{c} = \bigmeet\{ \Dfunapp{J}{a} \join \Dfunapp{K}{c \sop a}
\mid a \in \Con \}$.\qed
\end{pot}

\begin{pot}[\ref{thm:comp-algo} (3)]
Let $(\Dfun{I})_{I \subseteq G}$ be the distributed spaces of an scs
$({\Con},\cleq,(\sfun{i})_{i \in G})$ where $({\Con},\cleq)$ is
completely distributive. Let $I,J,K \subseteq G$ be such that $I = J \cup K$.

Let $c \in \Con$ and take any $a' \not\cleq c$. It suffices to find $a \in
\Con$ such that $a \cleq c$ and $\Dfunapp{J}{a'} \join \Dfunapp{K}{c \sop a'}
\cgeq \Dfunapp{J}{a} \join \Dfunapp{K}{c \sop a}$ since then $\bigmeet ( S
\cup \{ \Dfunapp{J}{a} \join \Dfunapp{K}{c \sop a}\ , \Dfunapp{J}{a'} \join
\Dfunapp{K}{c \sop a'} \}) = \bigmeet ( S \cup \{ \Dfunapp{J}{a} \join
\Dfunapp{K}{c \sop a} \})$ for any $S \subseteq \Con$.

Given $a' \not\cleq c$ either (a) $a' \cg c$ or (b) $a'$ and $c$ are
incomparable w.r.t $\cleq$, written $a' \parallel c$.
\begin{itemize}
\item Suppose (a) holds.
Then take $a = c$ thus $c \sop a = true$. By monotonicity we have $
\Dfunapp{J}{a'} \join \Dfunapp{K}{c \sop a'} \cgeq  \Dfunapp{J}{a} \join 
\Dfunapp{K}{c \sop a}$ as wanted.

\item Suppose (b) holds, i.e., $a' \parallel c$.
Notice that $c \sop a' \cleq c$. By cases, assume $c \sop a' = c$. Then we can
take $a = \true$, and thus $c \sop a = c = c \sop a'$. By monotonicity we have $
\Dfunapp{J}{a'} \join \Dfunapp{K}{c \sop a'} \cgeq \Dfunapp{J}{a} \join 
\Dfunapp{K}{c \sop a}$ as wanted. Now suppose $c \sop a' \cl c$ holds. We can
build a poset $P = (\{ a' \join c, \ a', \ c, \ c \sop a', \ a' \meet (c \sop
a') \ \}, \cleq)$ which is a non-distributive sub-lattice of $({\Con},\cleq)$,
isomorphic to a lattice known as $\mathbf{N}_5$ (see Fig.~\ref{ex:n5}).
But this contradicts $({\Con},\cleq)$ to be distributive
(see~\cite{davey2002introduction}).
\end{itemize}
From Th.~\ref{thm:comp-algo}~(2) and the above argument, we have
\begin{align*}
\Dfunapp{I}{c} 
=& \bigmeet \left\{ \Dfunapp{J}{a} \join \Dfunapp{K}{c \sop a} \mid a
\in \Con \right\}\\
=& \bigmeet \left(\left\{ \Dfunapp{J}{a} \join  \Dfunapp{K}{c \sop a} \mid a
\cleq c \right\} \cup \left\{ \Dfunapp{J}{a} \join  \Dfunapp{K}{c \sop a} \mid a \not\cleq c \right\}\right)\\
=& \bigmeet \left\{ \Dfunapp{J}{a} \join  \Dfunapp{K}{c \sop a} \mid a \cleq c \right\}.
\end{align*}
Thus, $\Dfunapp{I}{c} = \bigmeet \{ \Dfunapp{J}{a} \join \Dfunapp{K}{c \sop a}
\mid a \cleq c \}$.\qed
\end{pot}

\begin{figure}
\centering
\begin{tikzpicture}[scale=0.8,inner sep=2pt]
\node [fill,circle,draw,label={below:$a' \meet (c \sop a')$}] (A) at (0,0.5) {};
\node [fill,circle,draw,label={left:$c \sop a'$}] 	  		  (B) at (-2,2)  {};
\node [fill,circle,draw,label={left:$c$}] 		  			  (C) at (-2,4)  {};
\node [fill,circle,draw,label={above:$a' \join c$}] 		  (D) at (0,5.5) {};
\node [fill,circle,draw,label={right:$a'$}] 		  		  (E) at (2,3)   {};

\draw (A) to node {} (B);
\draw (A) to node {} (E);
\draw (B) to node {} (C);
\draw (C) to node {} (D);
\draw (D) to node {} (E);
\end{tikzpicture}
\caption{Poset $(\{ a' \join c, a', c, c \sop a', a' \meet (c \sop a') \},
\cleq)$ isomorphic to lattice $\mathbf{N}_5$.}
\label{ex:n5}
\end{figure}
\index[operator]{$\mathbf{N}_5$, non-modular lattice}

\section{Conclusions and Related Work}
\label{sec:ds}

We have introduced an algebraic theory for reasoning about possibly infinite
groups of agents. We have also shown that the theory can be applied to other
domains such as geometry and mathematical morphology.

This paper is an extended version of our CONCUR'19 paper~\cite{guzman-reasondistknowldg-2019}.
With respect to that work, we have provided significant advances in both the
theoretical and practical implications of our notion of distributed spaces in
scs. On the theoretical side, we characterized spatial functions as maps
preserving arbitrary joins (Prop.~\ref{prop:join-preservation}). This allowed us
to delve into the interpretation of space functions of distributed spaces as
meaningful operations in the field of mathematical morphology. We also pursued
the study of conditions under which a piece of information derived by the
combined local information of an infinite group of agents could be
derived by some finite subgroup of those agents. We showed that this is the 
case when the information inferred by the group from a given supplied piece of
information, is itself compact, and derivable from the combination of what each
agent can infer from it in the local space (Th.~\ref{thm:compact}). We further
showed, however, that compactness does not hold in general without those
conditions (Th.~\ref{thm:nocompact}).

Furthermore, we showed a fundamental result that provides a way to compute, for
completely distributive lattices, the greatest information that can infer some
other given piece of information, and is below all possible combinations
of local informations deriving that piece in the spaces of some given group of
agents (Th.~\ref{thm:delta-ast}).
This result is presented in~\cite{guzman-reasondistknowldg-2019} for finite cs; in this work we extended it for completely distributive lattices. In~\cite{guzman-reasondistknowldg-2019} we had also stated some properties relating
the information of a group of agents w.r.t the information of subgroups of 
those agents but only for finite cs. In this paper we generalized these for completely distributive lattices (Th.~\ref{thm:comp-algo}).

Finally, we used the developed theory to investigate applications in
mathematical morphology (MM). In this domain, two fundamental operations,
dilation and erosion, provide ways to perform geometric transformations, in
particular within the realm of image processing based on so-called structuring
elements. We considered these MM operations, generalized with Minkowski
addition over modules, and used our theory to derive some interesting
distribution properties. We also gave the interpretation of maps in group
spatial constraint systems as MM operations over structuring elements and
showed that the maximum map under two given group distribution maps (seen as
dilations) corresponds to the dilation over the intersection of their
structuring elements. In so doing we provided a proof that erosion and
dilations for structuring elements that are modules form a Galois connection
(Prop.~\ref{prop:erosn-proj}). This allowed us to prove that the operation of
erosion corresponds to our defined operation of projection of information into
the spaces of the structural element (Def.~\ref{group-projection}). We also discussed
an interpretation of dilations and erosions as epistemic  as an agent's perception of a given image (Remark \ref{dilation-epistemic:remark}).

\subsection{Related Work}

Below we discuss closely related work divided into four categories.
\resp{We added related work as suggested by the reviewers
({\bf N4}, {\bf N8}, {\bf N11}, {\bf N47}).
Also, we reorganized the related work that we had before.}

\hltext{
\paragraph{Algebraic Epistemic/Spatial Reasoning}
There are two major branches of approaches to modal logic: Kripke semantics
and Algebraic semantics~\cite{blackburn}. The majority of modern work on
epistemic logic has taken the Kripke semantics approach, whereas algebraic
semantics are more common in the general study of modal logic. This is
probably explained by the fact that the study of epistemic logic is most
common in philosophy and computer science, with specific applications in both
fields, and Kripke semantics are often the most practical approach for these
applications. The Handbook of Epistemic Logic, a good survey of the recent
state of work on epistemic logic, does not include algebraic
semantics~\cite{handbook}. The correspondence between modal logic and boolean
algebras with operators, and lattice theory more generally, however, is well
known~\cite{bao,chagrov-modalogic-1997,blackburn}. Since concurrent constraint
programming (ccp) uses lattices as the underlying structure~\cite{vijay}, in
our first work adding epistemic information to ccp~\cite{knight:hal-00761116},
it was natural to take the algebraic approach to epistemic logic, rather than
the Kripke structure approach which is much more common in computer science
applications of epistemic logic and modal logic in general. We continued to
develop our theory of modal constraint systems~\cite{expressiveness,
guzman:hal-01675010, guzman:hal-01328188, guzman:hal-01257113, semantic,
guzman-reasondistknowldg-2019, quintero-ccjoinendo-2020}, and besides its
applicability to ccp, it has the added advantage of providing a more natural
method than Kripke semantics for talking about infinite groups of agents,
which is a focus of the present paper. 

Since we take the algebraic approach which is less common in epistemic logic,
there is not a great deal of recent work in epistemic logic which is closely
related to the present paper. As already mentioned, spatial constraint systems
(scs) were developed in~\cite{knight:hal-00761116} and are dual to (poly)modal
algebras~\cite{goldblatt:2000}. These algebras are boolean algebras that
preserve meets and top and, they characterize the minimal (multi)modal logic
$K_n$~\cite{popkorn-modalogic-1994}. McKinsey and Tarski extended the Stone
representation theorem for Boolean algebras  to modal algebras whose operators
are closure operators  to give topological semantics for  the epistemic modal
logic $S4$ \cite{mckinsey-topo-1944}. Epistemic constraint systems (ecs) are
scs where the space functions are closure operators and thus they are dual to
closure algebras~\cite{bergmann-s5-1956}. There is more recent work on the
relation of  several modal logics and topology, among many others $S5$,
$KD45$, and first-order extensions of modal logic
(see~\cite{parikh-topo-elog-2007}). 

The work in~\cite{ap1} generalizes the epistemic update from the Logic of
Epistemic Actions and Knowledge~\cite{bms}, to a general class of algebras. In
the future, it will be interesting to study whether these epistemic updates on
algebras apply to our systems. There is also work on proof theory for versions
of epistemic logic, taking the algebraic approach~\cite{pt}. 

Nevertheless none of the above-mentioned works deal with an algebraic
characterization of distributed knowledge/information/space. To our knowledge
this paper provides the first-algebraic characterization for distributed
information of infinitely many agents. 

\paragraph{Constraint Systems} The closest related work is that
of~\cite{knight:hal-00761116} (and its extended version~\cite{KPV:Journal})
which introduces spatial constraint systems (scs) for the semantics of a
spatial ccp language. Their work is confined to a finite number of agents and
to reasoning about agents individually rather than as groups. We added the
continuity requirement to the space functions of~\cite{knight:hal-00761116} to
be able to reason about possibly  infinite groups.
In~\cite{guzman:hal-01257113,guzman:hal-01328188,guzman:hal-01675010,haar:hal-01256984}
scs are used to reason about beliefs, lies and other epistemic utterances but
also restricted to a finite number of agents and individual, rather than
group, behaviour of agents.

\paragraph{Distributed Knowledge} Our  work is inspired by the epistemic
concept of distributed knowledge~\cite{fagin1995reasoning}. Knowledge in
distributed systems was discussed in~\cite{halpern1987using}, based on
interpreting distributed systems using Hintikka's notion of possible worlds.
In this definition of distributed knowledge, the system designer ascribes
knowledge to processors (agents) in each global state (a processor's local
state).  In~\cite{halpern1990knowledge} the authors present a general
framework to formalize the knowledge of a group of agents, in particular the
notion of distributed knowledge.  The authors consider distributed knowledge
as knowledge that is distributed among the agents belonging to a given group,
without any individual agent necessarily having this knowledge.

\paragraph{Infinitely many agents}
In~\cite{halpern2004reasoning} the authors study knowledge and common
knowledge in situations with infinitely many agents. The authors highlight the
importance of reasoning about  infinitely many agents in situations where the
number of agents is not known in advance.  Their work does not address
distributed knowledge but points out potential technical difficulties in their
future work.  In the realm of Economics and game theory, models of infinitely
many agents are used  to discover mass phenomena that do not necessarily occur
in the case of a fixed finite number of
agents~\cite{yeneng-inf-agents-2017,debreu-economy-1963,hildenbrand-economy-1974,mclean-core-2005}.
Also infinite sets of agents are used in game
theory~\cite{aumann-disagree-1976,geanakoplos-common-1994}. For example for
games played with two teams, we may want to specify that  everyone in a team 
knows that everyone in  the other team knows a given proposition, regardless
of the team size. This could be naturally specified with infinitely many
agents.

\paragraph{Mathematical Morphology and Topology} Complete lattices have been
used as a framework to define morphological operators specifically to study
grey-level images~\cite{ronse-mmclat-1990}. In this context, dilations and
erosions are defined as operators that preserve arbitrary suprema and infima,
resp. This proposal is a generalization of what we studied in
Section~\ref{sec:app} where we present dilations and erosions by some
\emph{structuring element}. As a novelty, we proposed the scs
$(\pcal(M),\subseteq,(\dilatn{S})_{S \subseteq M})$ where $M$ is a module and
$(\dilatn{S})_{S \subseteq M}$ are dilations defined on the cs
$(\pcal(M),\subseteq)$. It allowed us to apply our theoretical results to
prove MM properties, e.g., that dilations and erosions form a Galois
connection (Prop.~\ref{prop:erosn-proj}). Also, we provided the interpretation
of distributed information for images. Namely, we showed that given two
dilations $\dilatn{A}$ and $\dilatn{B}$, the greatest dilation below them is
exactly $\Dfun{\{A,B\}}$ which in turn equals $\dilatn{A \cap B}$
(Th.~\ref{thm:ds-intersec}). As a future work, we plan to explore these
results for grey-scale images.

We note that the $\dilatn{S}$ function bears a superficial resemblance to the
topological notion of closure, and $\erosn{S}$ resembles topological interior.
Topology has important connections with modal logic and S4 type epistemic
logic in particular~\cite{parikh-topo-elog-2007,aybuke-topo-2019}. However, in
the case of mathematical morphology, the transformations are different from 
topological operations, most importantly because the operators are not
necessarily idempotent,  whereas closure and interior operators must be
idempotent.

\paragraph{Other related work} 

In~\cite{quintero-ccjoinendo-2020} the authors investigate the cardinality of
the set $\mathcal{E}(L)$ of all join-endomorphisms of a given lattice $L$. (A
join-endomorphism is a self-map that preserves finite joins, hence it is a
space function without the continuity requirement.)  The authors also provide
efficient algorithms to compute the meet of a given set of join-endomorphisms.
In this paper, we briefly illustrated the use of Th.\ref{thm:delta-ast} to
derive a polynomial complexity bound for computing this meet. 

There are other  group phenomena that are closely related to the group
phenomena here studied. In particular, group
polarization~\cite{Esteban:94:Econometrica,alvim:hal-02410747}, from social
sciences, and group improvisation~\cite{Rueda2004}, from computer music. Group
polarization refers to the natural tendency of a group to make more extreme
decisions than their individuals. Group improvisation involves constraining
musical pattern variation choices of a participant according to choices made
by  others in the group. We plan to study these phenomena in future work by
building upon the present work.
}

\hltext{
\subsection{Future Work}
\resp{We added future work according to {\bf N7}, {\bf N17}, {\bf N45}, {\bf N50}, {\bf N57}.}
Some of the proofs of the main results of this work rely on the completeness
of the lattice of the underlying constraint system (Th.~\ref{thm:delta-ast}). 
As future work we would like to generalise our results to  the
directed-complete partial orders, the central semantic structure of domain
theory~\cite{gierz2003continuous}. In fact, Scott-continuity is a central
concept of our theory, hence this research direction seem promising. 
 
We plan to extend our work with a process calculus with dynamic
creation/removal of agents. The idea is to investigate situations where we
could prove that it is not possible to reach a state where $\Dfunapp{A}{e}$
holds, with $A$ as the set of all possible agents and $e$ is some sensible
information. This could be interpreted as saying that the agents will never be
able to derive $s$ by pooling or joining their own information. Along the same
lines, for meaningful scs such as Kripke scs with infinitely many agents,  we
would like to study,  the decision and complexity problem of whether  $c \cgeq
\Dfunapp{I}{e}$ given $c,e$ and infinite set $I$ under the conditions of our
compactness result (Th.\ref{thm:compact}).  Notice that our compactness
result, does not provide us with a bound on the size of the finite subset $J
\subset I$ such that $d \cgeq \Dfunapp{I}{e}$.
}

\section*{Acknowledgments}

We are grateful to the anonymous reviewers for their constructive comments
that help us improve both the presentation and contents of this paper. 

\bibliographystyle{elsarticle-num-names}
\bibliography{biblio}

\printindex

\printindex[operator]
\end{document}